\documentstyle[a4,epsf,12pt]{article}
\newcommand{\ewxy}[2]{\setlength{\epsfxsize}{#2}\epsfbox[10 60 640 570]{#1}}
\newcommand{\ltaeq}{$\raisebox{-.6ex}{$\stackrel{\textstyle{<}}{\sim}$}$}
\newcommand{\gtaeq}{$\raisebox{-.6ex}{$\stackrel{\textstyle{>}}{\sim}$}$}
\newcommand{\bra}[1]{\left< #1 \right|}

\newcommand{\ket}[1]{\left| #1 \right>}
\newcommand{\err}[2]{{\footnotesize {$\;\begin{array}{@{}l@{}}
                          +\makebox[1.3em][r]{#1} \\[-0.4em]
                          -\makebox[1.3em][r]{#2}
                        \end{array}$}}}

\begin{document}
\begin{titlepage}
\begin{flushright}
{\small
FSU-SCRI-96-13\\
GUTPA/96/2/8\\
OHSTPY-HEP-T-96-002\\
Edinburgh 96/1}
\end{flushright}

%prerint numbers OSU
%
%

\begin{center}
{\Large $B$ Spectroscopy from NRQCD with Dynamical Fermions}\\[5mm]

{\bf S.~Collins}\\
SCRI, Florida State University, Tallahassee, Fl 32306-4052, USA\\
and Edinburgh University, Edinburgh, Scotland EH9 3JZ.\\[3mm]

{\bf U.~M.~Heller and J.~H.~Sloan}\\
SCRI, Florida State University, Tallahassee, Fl 32306-4052, USA\\[3mm]

{\bf J.~Shigemitsu}\\
The Ohio State University, Columbus, Ohio 43210, USA\\[3mm]

{\bf A.~Ali~Khan\footnote{UKQCD Collaboration} and
C.~T.~H.~Davies${\vphantom{\Large A}}^1$}\\
University of Glasgow, Glasgow, Scotland G12 8QQ

\end{center}

\begin{abstract}
We present a lattice investigation, partially including the effects of
dynamical quarks, of the heavy-light mesons using NRQCD for the heavy
quark and the Wilson action for the light quark. We performed an
extensive calculation of the spectrum employing a multi-state,
multi-exponential fitting analysis which enabled us to extract the
$2S{-}1S$ splitting as well as the $^1P_1{-}\bar{S}$ and hyperfine
splittings.  The Wilson action introduces a large systematic error
into the calculation, and within this uncertainty we obtain agreement
with experiment. We performed a comprehensive calculation of the
heavy-light meson mass, investigating three methods and their range of
validity. The agreement found between these methods confirms that
Lorentz invariance can be restored at this order in NRQCD by a
constant shift to all masses. We calculated spectroscopic
quantities over a wide range in heavy quark mass and were able to
perform a detailed investigation of heavy quark symmetry around the
$b$ quark mass. In particular, we extracted the nonperturbative
coefficients of terms in the heavy quark expansion of the meson
binding energy.  We demonstrate the importance of using
tadpole-improved operators in such a calculation.
\end{abstract}
\vspace{2cm}
\thispagestyle{empty}
\end{titlepage}
\newpage
\section{Introduction}
Lattice field theory has an important role to play in the study of
$B$-Physics. Through this approach it is possible not only to perform
a first principles calculation of experimentally important quantities,
but also to test heavy quark symmetry predictions of Heavy Quark
Effective Theory~(HQET) and extract the nonperturbative coefficients
of the heavy quark expansions which are required to extract numerical
results from HQET.  Furthermore, comparison with the predictions of
various models for these coefficients provides a test of the
underlying assumptions of the models.

The first stage of a lattice study of $B$ mesons is the calculation of
the spectrum.  As the simplest quantity to extract, it is the
foundation for more complicated calculations of $B$ decay processes.  A
correct reproduction of the spectrum gives confidence in the
calculational procedure, as well as providing predictions for states
not yet measured experimentally.  In the spectrum, several physical
scales are important.  The gross structure of the $B$ spectrum is due
to excitations of the light quark and hence mass splittings related to
these, such as $P{-}S$ and $2S{-}1S$ are of
$O(\Lambda_{QCD})$. Conversely, the heavy quark is approximately a
static colour source and its spin affects the spectrum at the scale of
the hyperfine splitting, $O(\Lambda_{QCD}^2/M)$.

Thus, to reproduce the spectrum, the systematic errors arising from
both the heavy and light quarks must be rigorously controlled.
Previously, simulating the $b$ quark (as opposed to a static quark) on
the lattice presented a major problem since the $O(Ma)$ discretisation
errors of the standard relativistic fermionic action were $O(1)$ for a
$b$ quark on a typical lattice, $a^{-1}\sim 2.0$~GeV.  However, the
success of the non-relativistic QCD~(NRQCD) approach in calculations
of $\Upsilon$ spectroscopy~\cite{nrqcdups,junkpaul} shows that the $b$
quark can be simulated directly on lattices currently available.  In
NRQCD the heavy quark mass scale is removed and the action becomes an
expansion in the typical~(non-relativistic) heavy quark velocity,
which is a small parameter.  The precision of the calculation is
controlled simply by the order at which the expansion is
truncated. For heavy-light mesons, the terms included in the action at
each order in the heavy quark velocity is determined by the order of
the terms in $1/M$; this is discussed in more detail in
reference~\cite{arifa}.  Considering the almost spectator role of the
heavy quark in the heavy-light meson, we include only first order
i.e. $O(1/M)$ terms in the NRQCD action in this study.

For light quarks, recent studies have shown~\cite{scrispec,hugh} that
the $O(ma)$ discretisation errors of the Wilson action are significant
for a high statistics calculation on typical lattices even for small
bare masses~(i.e. the constituent mass seems to be the important
scale for $O(ma)$ effects).  Improved light fermion actions, for
example the tadpole-improved Clover action with errors of roughly
$O(m^2a^2)$, are needed.  However, in this initial calculation we have
employed the Wilson action and expect this to give rise to the
dominant systematic error.

Another uncertainty in connecting lattice calculations to the physical
world is the quenched approximation. In this study we have partially
included the effects of dynamical quarks. In addition, these
computations are performed in conjunction with a simulation using
$\beta=6.0$ quenched configurations~\cite{arifa} and in the future we
expect to be able to isolate and remove the full effects of quenching.

The organisation of this paper is as follows. The details of the
simulation are given in section~\ref{simdet} followed by a discussion
of the expected systematic errors. Particular attention has been paid
in our analysis to the fitting procedure and extraction of ground and
excited state energies. Our methods are outlined in section~\ref{fitanal} and
illustrated with a subset of the data. We use the results over a range
of heavy quark masses to investigate heavy quark symmetry in the
meson binding energy in section~\ref{hqsym}. The lowest order coefficients of
the heavy quark expansion $\bar{\Lambda}$ and $<-\vec{D}^2>$ are
extracted. The heavy-light meson mass is calculated in
section~\ref{mesonmass}; three methods are compared and their range of
validity investigated. In section~\ref{massplit} we turn to the mass
splittings, $B^*{-}B$, $B^{**}{-}B$ and $B_s{-}B$, and compare the
simulation results with experiment and the expectations of heavy quark
symmetry. The conclusions and a summary of the spectrum are given in
section~\ref{conc}.

\section{Simulation Details}
\label{simdet}
The simulations were performed using 100 $16^3\times32$ gauge
configurations at $\beta=5.6$ with two flavours of staggered dynamical
sea quarks with a bare quark mass of $am_{sea}=0.01$. These
configurations were generously made available by the \mbox{HEMCGC}
collaboration; more details can be found in~\cite{hemcgc}. We fixed
the configurations to the Coulomb gauge. The light quark propagators
were generated using the Wilson fermion action, without an $O(a)$
improvement term, at two values of the hopping parameter,
$\kappa=0.1585$ and $0.1600$.  The former corresponds to a quark mass
close to strange, where $\kappa_s=0.1577$ from $M_\phi$, while
$0.1600$ is somewhat lighter, with $\kappa_c=0.1610$.

In this simulation we truncate the NRQCD series at $O(1/M_0)$ and the
action takes the form:
\begin{equation}
S = Q^{\dagger}(D_t + H_0 + \delta H) Q
\end{equation}
where
\begin{equation}
H_0 = -\frac{\Delta^{(2)}}{2M_0}\hspace{0.5cm} \mbox{and} \hspace{0.5cm}
\delta H = -c_B\frac{\sigma\cdot B}{2M_0}.
\end{equation}
Tadpole improvement of the gauge links is used throughout, where
$u_0=0.876$ measured from the plaquette, and the hyperfine coefficient
is given the tree-level value $c_B=1$. We use the standard Clover-leaf
operator for the $B$ field. A constant mass term has
been omitted from the action for simplicity of the calculation. It has
no effect on mass splittings and amplitudes and a suitable energy
shift for all masses can be added post-simulation. This will be
discussed in further detail in section~\ref{mesonmass}. The heavy
quark propagators were computed using the evolution equation:
\begin{eqnarray}
G_1 & = & \left( 1-\frac{aH_0}{2n}\right)^n U_4^{\dagger}
\left(1-\frac{aH_0}{2n}\right)^n \delta_{\vec{x},0}\nonumber\\
G_{t+1} & = & \left(1-\frac{aH_0}{2n}\right)^n U_4^{\dagger}
\left(1-\frac{aH_0}{2n}\right)^n (1-a\delta H)G_t \hspace{0.5cm} (t>1)
\end{eqnarray}
In the static limit this reduces to
\begin{equation}
G_{t+1} - U_4^\dagger G_t = \delta_{x,0}
\end{equation}
The parameter $n$ is introduced to stabilise unphysical higher momentum
modes as $M_0$ is decreased and from perturbation theory one estimates
$n\gtaeq3/M_0$. This is only a rough guide and for a high statistics
simulation a more conservative lower limit on $n$ may be needed to
ensure a smooth transition between regions of different $n$ as $M_0$
decreases. This point will be illustrated in Section~\ref{esim}.

The heavy quark propagators were computed over a range of values of
$M_0$ to probe both the lower limit of NRQCD where the perturbative
couplings diverge and the theory breaks down, around $aM_0\ltaeq
0.6$~\cite{colin}, and the $aM_0\rightarrow\infty$ limit where
$signal/noise$ becomes poor as NRQCD tends to the static limit.  Note
that for heavy-light mesons there is no upper limit imposed on $M_0$ by
discretisation errors in the heavy quark motion,
$ap_Q\sim a\Lambda_{QCD}\sim 0.2 \ll 1$ independent of $M_0$~(where we
take $a\Lambda_{QCD}=a\Lambda_V=0.185$ for these configurations). This is
in contrast to heavy-heavy mesons where the kinetic energy, not the
momentum, of the heavy quark is approximately $\Lambda_{QCD}$ and
$ap_Q \sim \sqrt{a^2M\Lambda_{QCD}}$. Thus, discretisation errors naively
become
large when $ap_Q\sim1$ i.e. $aM\sim 1/a\Lambda_{QCD}\sim 5$.  We generated
heavy quark propagators at 11 values of ($aM_0$,$n$) corresponding to
(0.8,5), (1.0,4), (1.2,3), (1.7,2), (2.0,2), (2.5,2), (3.0,2),
(3.5,1), (4.0,1), (7.0,1) and (10.0,1), and the static limit. This
roughly corresponds to a range of meson masses from $M_B/2$ to $4M_B$
and is sufficient for a reasonable investigation of heavy quark
symmetry. However, it is not possible to simulate the $D$ meson on
this lattice using NRQCD; a larger lattice spacing is required,
$\beta^{n_f=2}\ltaeq 5.5$ or $\beta^{n_f=0}\ltaeq 5.85$.

A NRQCD heavy quark propagator describes the forward propagation of a
quark.  To increase statistics we applied the time reversal
transformation to both the gauge fields and the light quark
propagators and repeated the heavy quark evolution; this is equivalent
to calculating the propagation of the backwardly moving anti-particle.

Following the methods of~\cite{nrqcdups,eichten} meson correlation functions
corresponding to both $S$ and $P$ states were constructed from the
quark propagators using interpolating operators at the source
of the form:
\begin{equation}
O = \sum_{\vec{x}_1,\vec{x}_2} Q^{\dagger}\Gamma^{\dagger}
(\vec{x}_1-\vec{x}_2) q(\vec{x}_2),
\end{equation}
and $O^\dagger$ at the sink, where $\Gamma = \Omega
\phi(\vec{x}_1-\vec{x}_2)$ specifies the quantum numbers~($\Omega$) of
the state, detailed in table~\ref{ops}, and the smearing~($\phi$)
applied to the heavy quark.  Smearing functions motivated by the
hydrogen model were chosen to project out the ground and first excited
state; for S-states
\begin{eqnarray}
\phi_1(r) & = &   e^{-r/r_0} \nonumber\\
\phi_2(r) & = &  (1-r/(2r_0))e^{-r/(2r_0)}
\end{eqnarray}
where $r_0=3.0$. Similarly for P-states
\begin{eqnarray}
\phi_1(r) & = &  (-1-r/(2r_0)) e^{-r/(2r_0)} \nonumber\\
\phi_2(r) & = &  (r/(3r_0))^2 e^{-r/(3r_0)}\label{firstp}
\end{eqnarray}
It is sufficient to use the same radius, $r_0$, for all $M_0$ since
the heavy quark is almost a spectator in the heavy-light system and the
meson wavefunction is not expected to change significantly with $M_0$.
A delta function was also used for both $S$ and $P$ states.

% table of operators
%
\begin{table}
\begin{center}
\begin{tabular}{|c|c|}\hline
$^{2S+1}L_J$ ($J^{PC}$)& $\Omega$ \\\hline
 ${^1S}_0\;(0^{-+})$ &$\hat{I}$  \\
 ${^3S}_1\;(1^{--})$ & $\sigma_i $   \\
 ${^1P}_1\;(1^{+-})$ & $\Delta_i $  \\
 ${^3P}_0\;(0^{++})$ & $\sum_j  \Delta_j  \sigma_j$   \\
 ${^3P}_1\;(1^{++})$ & $ \Delta_i \sigma_j - \Delta_j \sigma_i $  \\
 ${^3P}_2\;(2^{++})$ & $ \Delta_i \sigma_i - \Delta_j \sigma_j$  \\
                     & $ \Delta_i \sigma_j + \Delta_j \sigma_i$  \\
                     & $\qquad$ ($i \neq j$)  \\\hline
\end{tabular}
\caption{The lattice operators, $\Omega$, and associated quantum numbers used
in the simulation. $\Omega$ is a $4\times 2$ matrix in spin space.
Note that since charge conjugation is not a good
quantum number for the heavy-light system there is mixing between the
$^1P_1$ and $^3P_1$ states.\label{ops}}
\end{center}
\end{table}

With the aim of performing a multi-state, multi-correlation function
analysis the heavy quark was smeared using all the source-sink
combinations of the smearing functions. The meson propagator
is then
\begin{equation}
C_{sc,sk}(\vec{p},t) = \sum_{\vec{y}_1, \vec{y}_2} Tr \left[
\gamma_5(L^{-1})^\dagger(\vec{y}_2)\gamma_5\Gamma^{(sk)\dagger}(\vec{y_1}
-\vec{y}_2)\tilde{G}_t(\vec{y}_1)\right]
e^{-i\frac{\vec{p}}{2}\cdot(\vec{y}_1 + \vec{y}_2)}
\end{equation}
where
\begin{equation}
\tilde{G}_t(\vec{y}_1) = \sum_{\vec{x}} G_t(\vec{y}-\vec{x})\Gamma^{(sc)}
(\vec{x})e^{i\frac{\vec{p}}{2}\cdot\vec{x}},
\end{equation}
and $sc,sk = 1,2,l$ corresponding to the ground state, first excited
state and delta function smearing functions respectively. $L^{-1}$ is
the light quark propagator. Finite momentum $^1S_0$ correlators were
generated for a subset of the data, $aM_0=0.8{-}4.0$, at $a|\vec{p}|=1$
and $\sqrt{2}$ in units of $\frac{\pi}{16}$. To improve statistics,
where appropriate, the correlators were averaged over different
polarisations and different momentum directions.

\section{Systematic Errors}

There are systematic errors associated with both the heavy and the
light quark in the heavy-light meson. The truncation of the NRQCD
series at $O(1/M)$ introduces an absolute systematic error of
$O(\Lambda_{QCD}(\Lambda_{QCD}/M)^2)$ associated with the heavy
quark. This corresponds to a relative error of $O(\Lambda_{QCD}/M)$ in
quantities of $O(\Lambda_{QCD}(\Lambda_{QCD}/M))$, naively estimated
to be $O(v)\sim 10\%$, and an $O((\Lambda_{QCD}/M)^2)\sim 1\%$ error
in quantities of $O(\Lambda_{QCD})$. The use of tree-level
coefficients for the relativistic corrections will introduce errors of
the form\\ \mbox{$O(\alpha_S \Lambda_{QCD}(\Lambda_{QCD}/M))$}, which
are of the same magnitude as those above for $\alpha_S\sim
0.1$. Similarly, the magnitude of the scaling violations due to using
the Wilson action for the light quark are $O(\Lambda_{QCD}^2a)$, or a
relative error of $O(\Lambda_{QCD}a)\approx 20\%$ for this ensemble.
Hence, the systematic errors associated with the light quark are
expected to dominate and the finite volume errors for a lattice of
size $\approx 1.6$~fm are unlikely to be larger than this for a $B$
meson. It is important to note that this is a high statistics
calculation with statistical errors of around $5\%$, much smaller than
the systematic errors.

An indication of the size of the light quark uncertainties in the
simulation can be found by comparing the lattice spacing obtained from
various quantities. Table~\ref{latspac} details the values of $a^{-1}$
extracted from various light spectroscopic quantities for this
ensemble, where the Wilson action was used for the light quarks. Also
included is the inverse lattice spacing obtained from the $\Upsilon$
spectrum using NRQCD~\cite{nrqcdups}. In this case all terms in the
NRQCD series were included up to $O(Mv^4)$ and the estimated
systematic error is $\sim 1\%$. Clearly there is marked lack of
agreement in the values obtained from observables dominated by
presumably similar typical physical scales indicating large light
quark uncertainties. In addition there is disagreement between the
scale obtained from light and heavy spectroscopy. With partial
inclusion of quark loops at finite sea quark mass, the running of the
strong coupling does not match that in the real world but should
provide an improvement on the quenched approximation. At present the
light quark uncertainties in the spectroscopy hide any indication of
this and the variation in the lattice spacing is similar to that found
for a lattice with a similar $a^{-1}$~($\beta\sim6.0$) in the quenched
approximation. Currently, the light quark spectroscopy is being
re-analysed using a multi-smearing, multi-state
analysis with Clover fermions~\cite{scrispec}.

A large uncertainty in $a^{-1}$ translates into similar uncertainties in
predictions for physical quantities. While it is useful to quantify
these by using a range of values for $a^{-1}$ to convert into physical
units, observables will exhibit systematic errors differently and the
resulting uncertainty may not necessarily be expressible by merely
using a range of values for $a^{-1}$. With this caveat we will take
$a^{-1} = 1.8-2.4$~GeV. The emphasis of our analysis will be on
determining the heavy quark mass dependence of various quantities and
the nonperturbative coefficients of the $O(1/M)$ terms.

With light quark propagators at only two values of $\kappa_l$ and with
a much worse signal to noise ratio for the data at $\kappa_l=0.1600$
compared to $\kappa_l=0.1585$, it is not possible to perform a
trustworthy chiral extrapolation. In addition, the light quark
uncertainties affect the determination of the bare strange quark
mass. $\kappa_s$ extracted from pseudoscalar mesons, using the lowest
order chiral mass dependence to find the `experimental mass' of the
pure $s\bar{s}$ pseudoscalar, differs from $\kappa_s$ obtained from
the vector meson $\phi$. Thus, most of the results
presented will be for $\kappa_l=0.1585$.

\begin{table}[hbt]
\begin{center}
\begin{tabular}{|c|c|}\hline
``force'' &    2055\\
$m_\rho$ &   2140 \\
$m_p$ & 1800\\
$\Upsilon$(NRQCD) & 2400\\\hline
\end{tabular}
\caption{$a^{-1}$ in MeV obtained from various observables
for this ensemble. The ``force'' is calculated from the static quark
potential using $r_0$. Statistical errors are 50 -- 100 MeV.}
\label{latspac}
\end{center}
\end{table}
\section{Analysis}
\label{fitanal}

With three smearing functions for the heavy quark in all source-sink
combinations we generated a $3\times3$ matrix of meson propagators,
$C_{ij}(t)$, for each state, where $i,j=1,2,l$. With several
independent determinations of the meson correlation function it is
possible to perform simultaneous multi-exponential fits and thus
rigorously constrain the ground state energy and amplitudes of the
meson; higher excited states can also be obtained. Using a subset of
the data involving correlation functions smeared at the source and
local at the sink, vector fits were performed using the fitting
function,
\begin{equation}
C_{il}(t) = \sum_{n=1}^{n_{exp}} b(i,n) e^{-E_n t} \hspace{1cm} i=1,2
\end{equation}
where $n_{exp}$ is the number of exponentials used in the fit.
Similarly,
\begin{equation}
C_{ij}(t) = \sum_{n=1}^{n_{exp}} a(i,n) a(j,n) e^{-E_n t} \hspace{1cm} i,j=1,2.
\end{equation}
was used to fit to the $2\times2$ matrix of correlators formed
from the correlators smeared using the ground and first excited
state smearing function.

The fitting procedure applied to extract the mass spectrum from these
correlation functions can be expressed as follows:
\begin{enumerate}
\item Choose the end of the fitting range, $t_{max}$, to be a timeslice
where the $noise/signal\gtaeq 3$.
\item Begin with a 1 exponential correlated fit.
\item Vary the beginning of the time range, $t_{min}$, over all possible
values i.e. $n_{cor}(t_{max}-t_{min}+1)-n_{par}>0$, where $n_{cor}$ is
the number of correlation functions and $n_{par}$ the number of
parameters in the fit.
\item Accept fits for which the quality of fit parameter $Q>0.1$.
\item In the case of multi-correlation function fits repeat~(3)
including 2 then 3 exponentials.
\end{enumerate}
For each set of fits with a fixed number of exponentials, the
criteria for the `best' fit is taken to be the earliest value of
$t_{min}$ where $Q$ has reached a plateau i.e. is stable to successive
increases in $t_{min}$.

With each addition of another exponential to the fit function the
range of $t_{min}$ with an acceptable $Q$ moves closer to the origin
and the extra exponential removes the excited state contamination from
the lower state.  A plateau is sought in the ground state fit
parameters obtained using $1$, $2$ and $3$ exponential fits.  To
achieve this and resolve the first, and higher, excited states there
must be a sufficiently different overlap with the ground and excited
states between the correlation functions with different smearing
combinations.  In addition, a signal for isolating the first excited
state must be a non-zero overlap of the state with the excited state
smearing function.
%amplitude for the first excited state.
In general, the highest excited state is likely to be untrustworthy,
i.e. a 3 exponential fit is needed to provide confidence in the first
excited state.

In order to perform correlated fits to quantities as a function of the
meson mass 100 bootstrap ensembles were generated for each correlated
fit to the propagators. However, we found that our statistics were
insufficient to be able to perform the bootstrap procedure above
$n_{exp}=1$. Thus, only $n_{exp}=1$ fits were used.  It was possible
to do this with confidence only because we were able to check
agreement with the multi-state, multi-smearing analysis of the
unbooted correlation functions.

In the construction of the correlation functions we are combining
light quark propagators which include the contribution of the
backwardly propagating anti-quark with the NRQCD heavy quark
propagating solely in the forward direction. This leads to a
unphysical contribution to the meson correlation function which should
be small at least through $t_{max} \approx L_t/2 = 16$. Examination
of the correlation functions indicates the contribution is small
for $t_{max}\ltaeq 24$.

\section{Results}
\subsection{Fitting results for $E_{sim}$}
\label{esim}
The simplest quantities to extract are the $^1S_0$ and $^3S_1$
energies, $E_{sim}^{PS}$ and $E_{sim}^{V}$ respectively, which
determine the exponential fall-off of the S-state correlators. With
the omission of the constant mass term in the NRQCD action $E_{sim}$
is not the meson mass, $M_2$; the calculation of this quantity is
discussed in the next section. However, physical mass splittings can be
obtained from differences in $E_{sim}$ for different mesons. In
particular, to have confidence in the results it is vital to ensure
the excited state contamination of $E_{sim}$ is minimal.  We use the
$aM_0=1.0$ and $\kappa_l=0.1585$ results to illustrate how we achieve
this with our fitting analysis.

\begin{figure}
\centerline{\ewxy{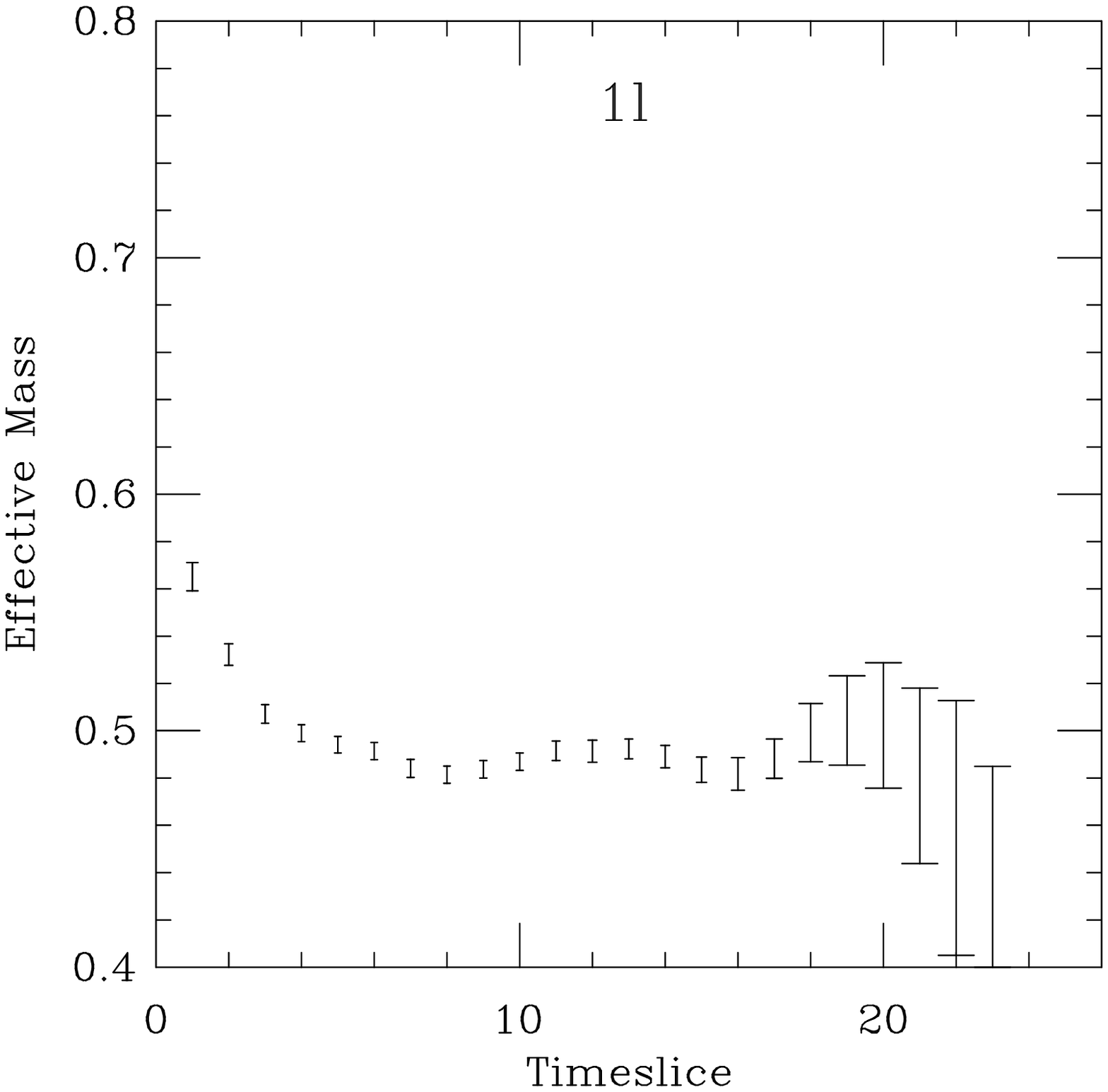}{80mm}
\ewxy{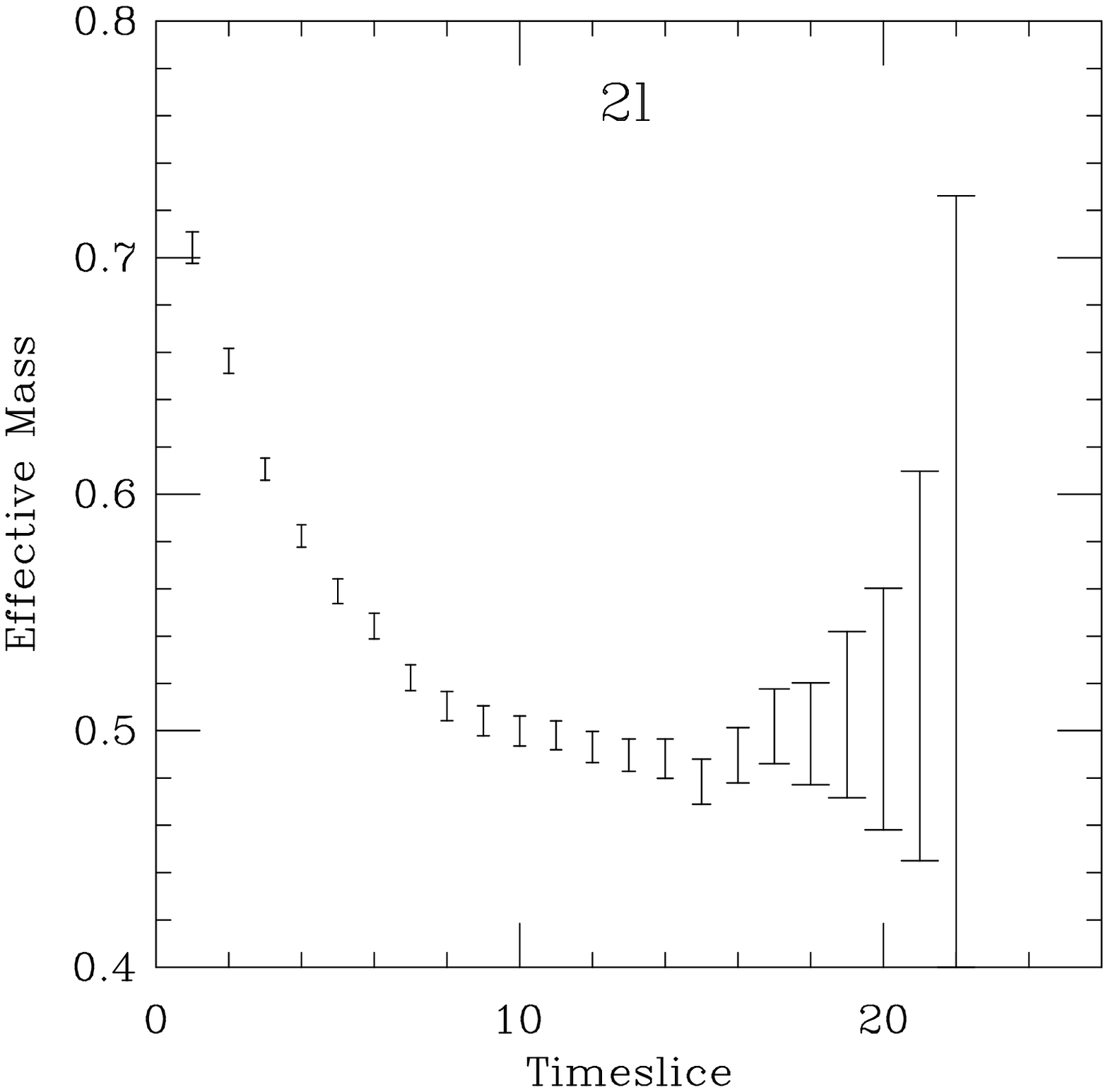}{80mm}
}
\vspace{0.5cm}
\centerline{\ewxy{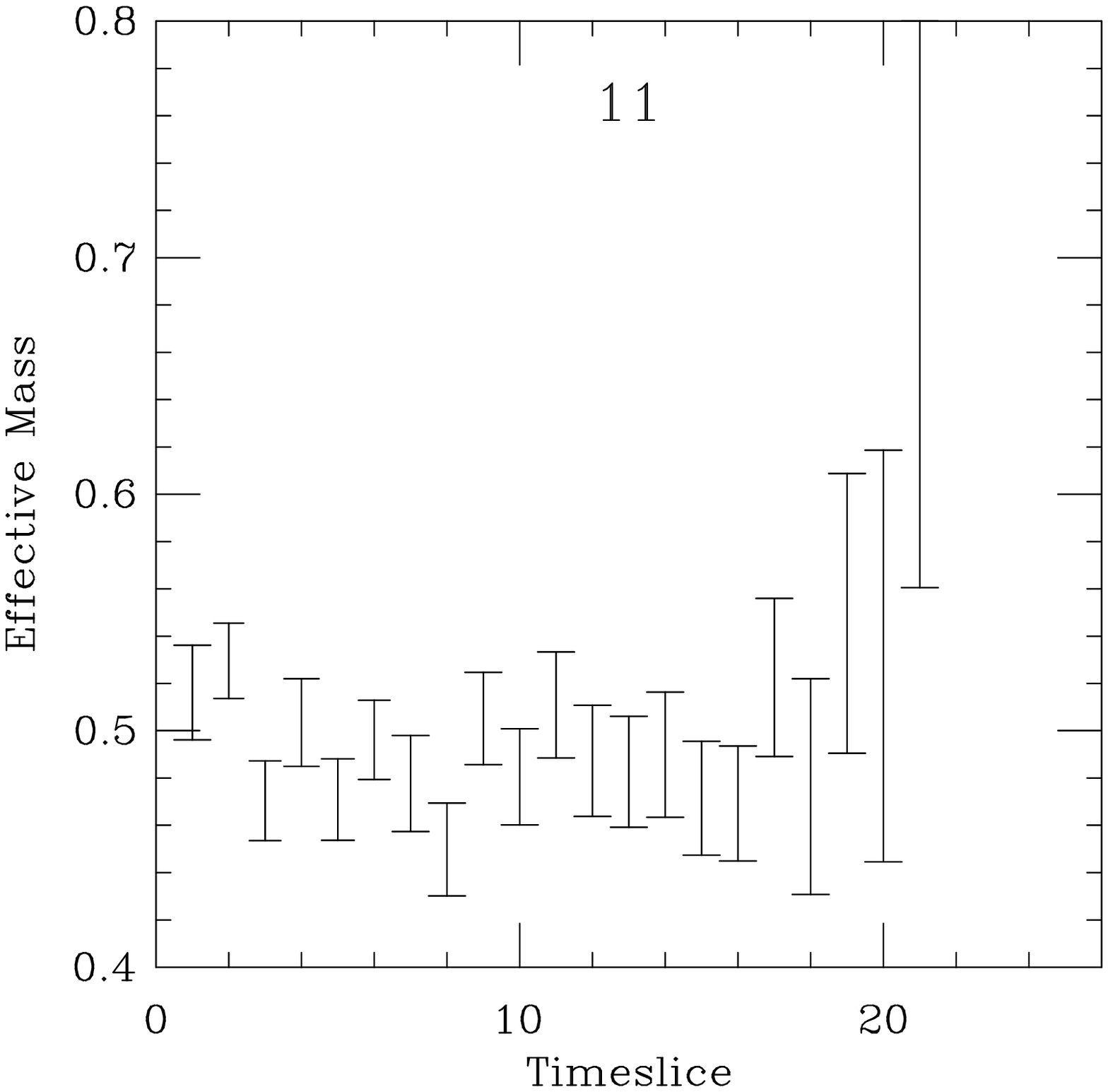}{80mm}
\ewxy{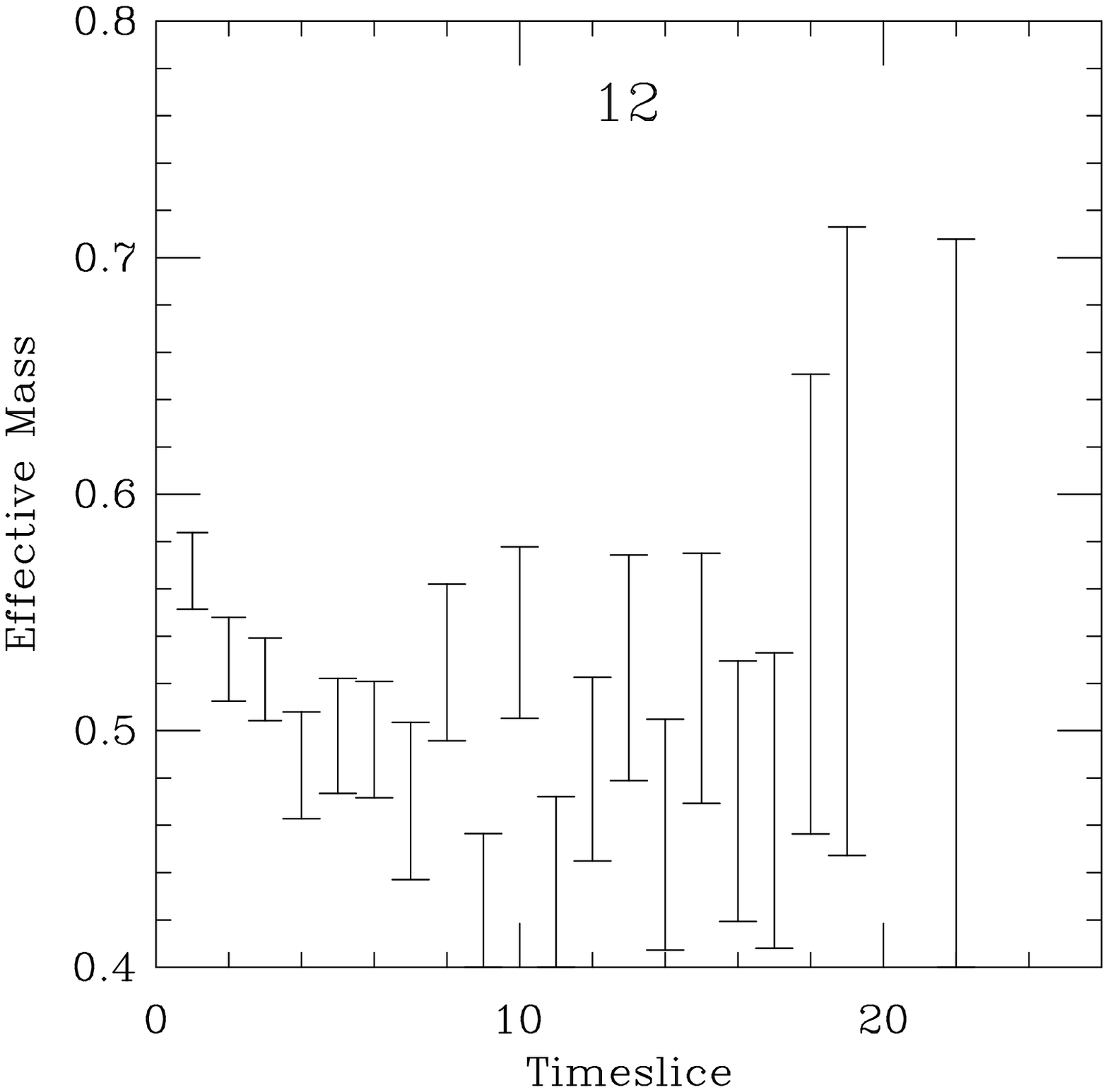}{80mm}
}
\vspace{0.5cm}
\centerline{\ewxy{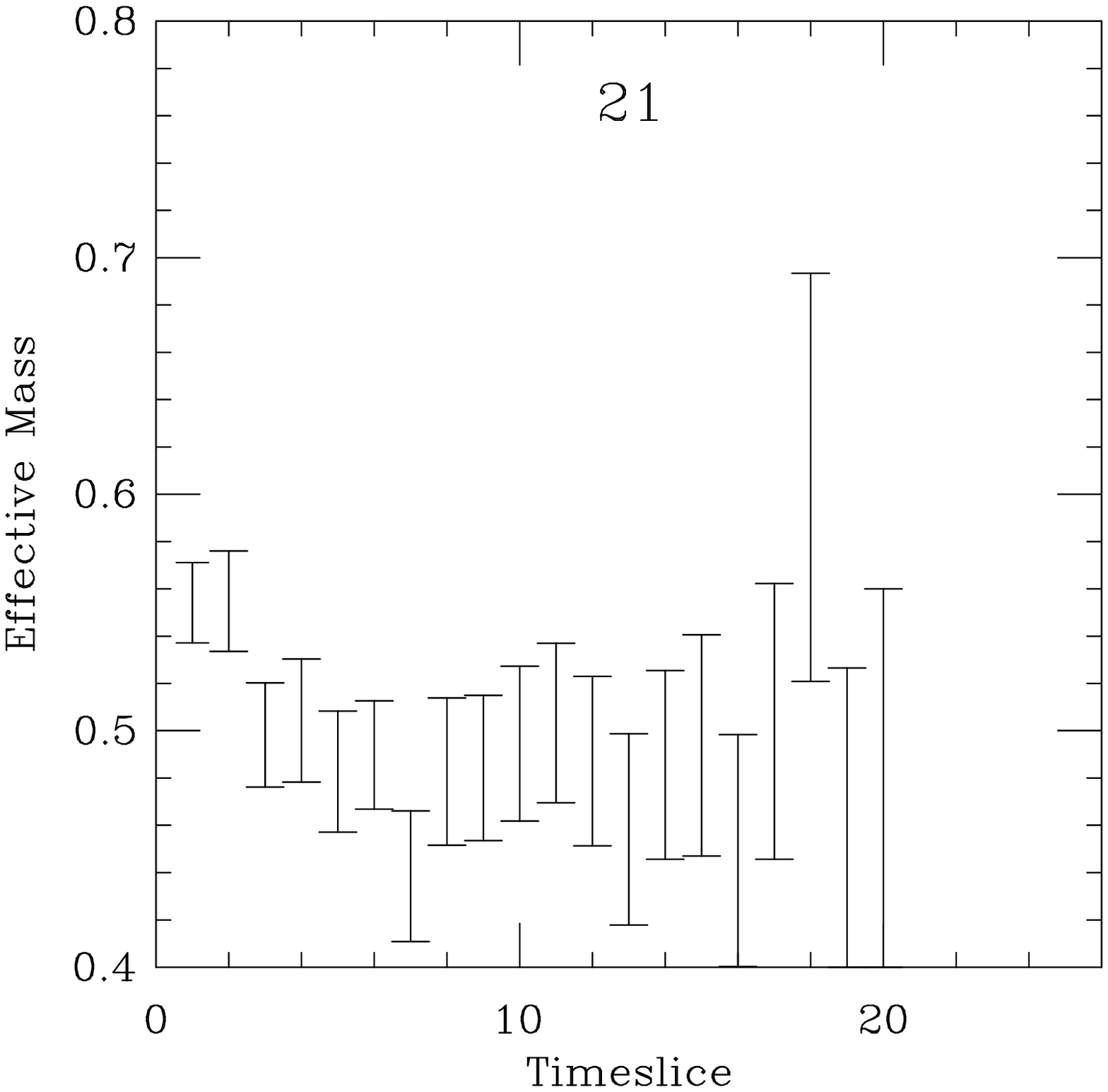}{80mm}
\ewxy{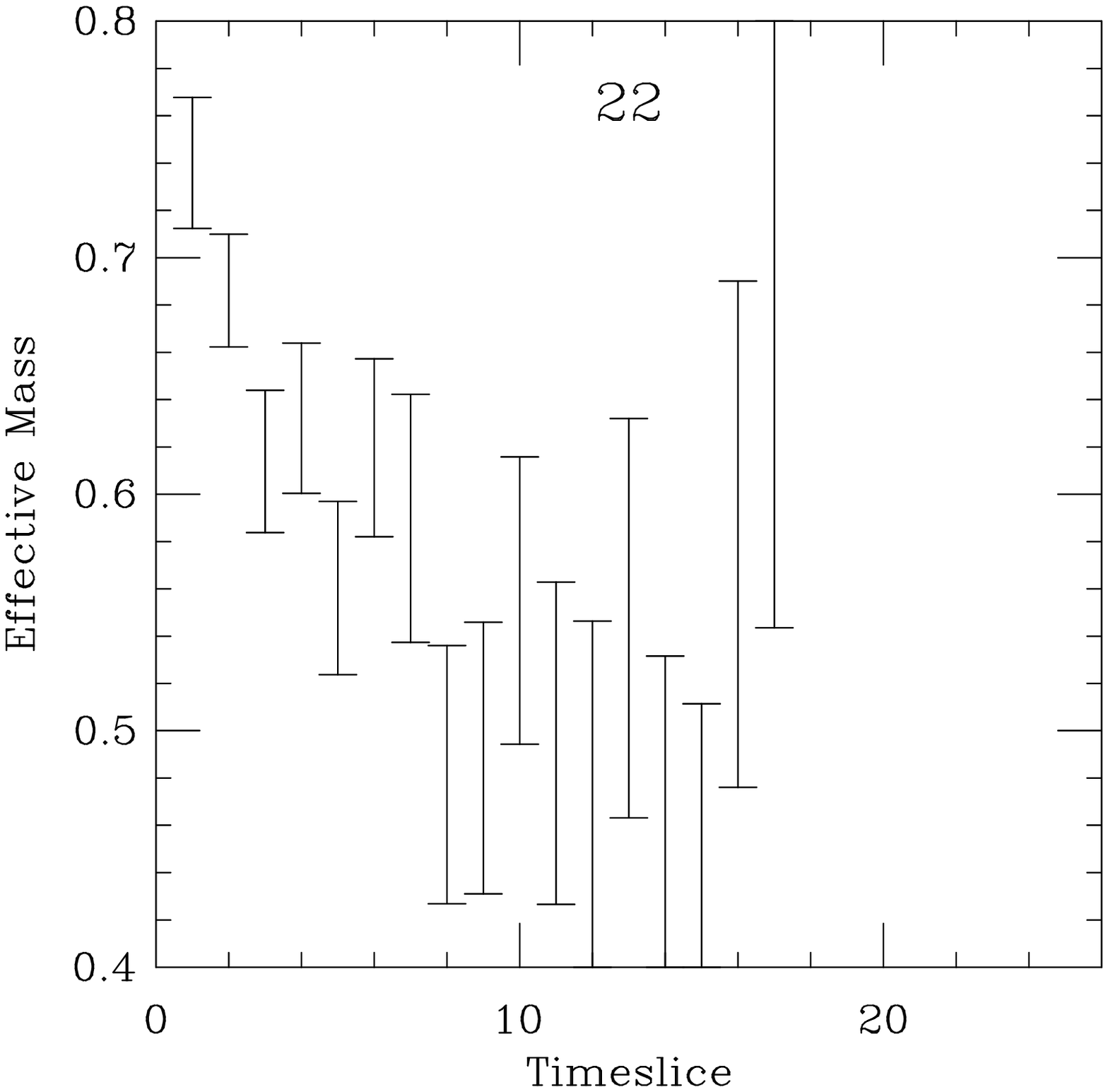}{80mm}
}
\caption{The effective masses of the $^1S_0$ meson
correlators for $aM_0 = 1.0$ and  $\kappa_l=0.1585$.
\label{meff_1s0_esim_1}
}
\end{figure}

\begin{figure}
\centerline{\ewxy{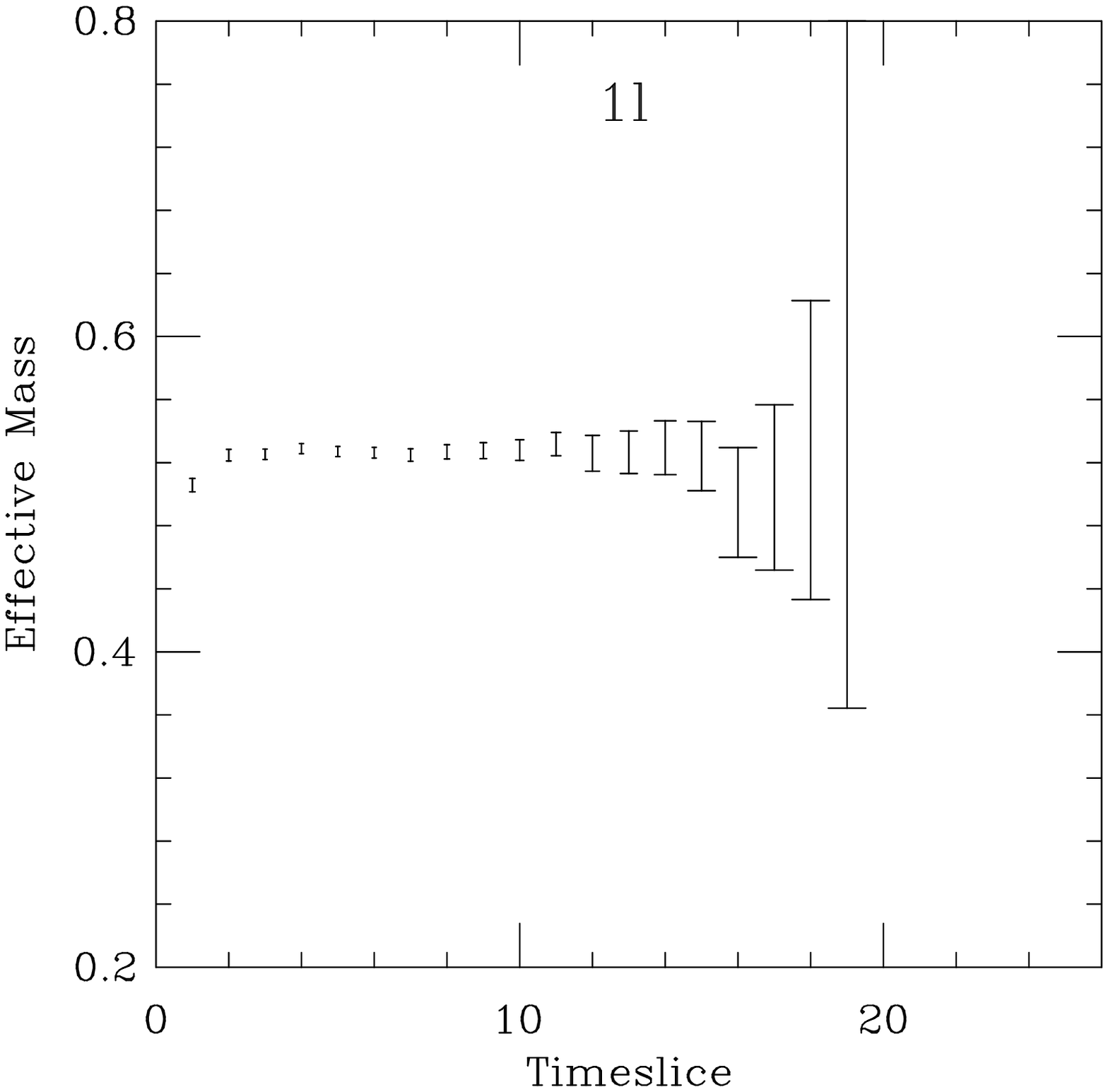}{80mm}
\ewxy{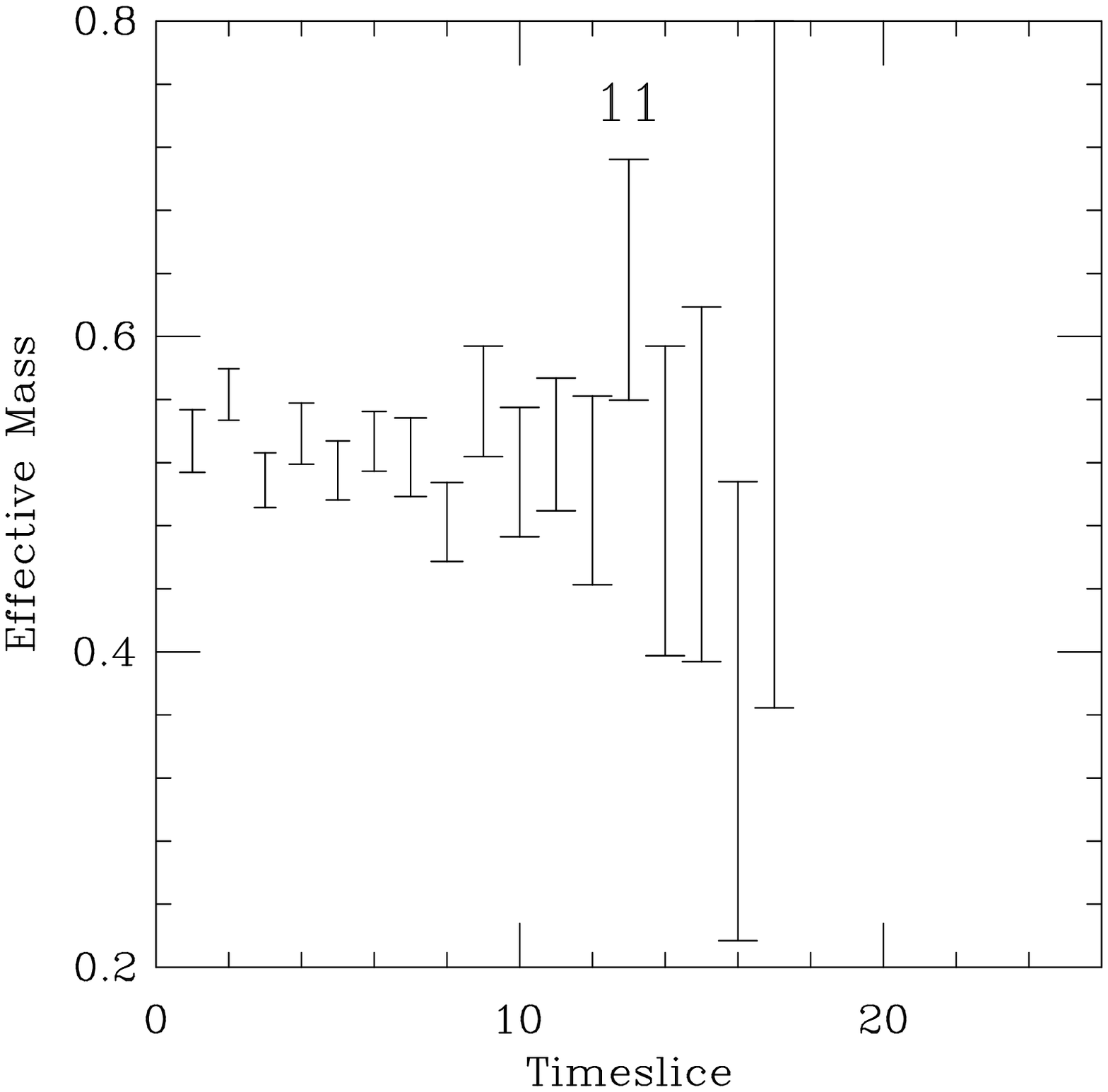}{80mm}
}
%\vspace{0.5cm}
%\centerline{\ewxy{b500.u01_10.0_1585.1s0.11.20_1_mcor_q.ps}{80mm}
%\ewxy{b500.u01_10.0_1585.1s0.11.20_1_mcor_param1.ps}{80mm}
%}
%
\caption{The effective masses of the $^1S_0$
meson correlators for $aM_0 = 10.0$ and $\kappa_l=0.1585$.
\label{meff_fit_10}
}
\end{figure}

Figure~\ref{meff_1s0_esim_1} shows the effective masses of all the
$^1S_0$ smeared correlation functions.  The correlators smeared with
the excited state smearing function have a larger contribution from
excited states compared to those smeared only with the ground state
smearing function, as required to separate ground state and excited
states.  Increasing $M_0$ from $1.0$ to $10.0$ corresponds to an
approximate $7$-fold increase in the meson mass. However, a
comparison with the effective mass of $C_{1l}$ for $aM_0=10.0$ in
figure~\ref{meff_fit_10} shows there is only a gradual change in the
meson wavefunction, as expected from a hydrogen-like picture where the
heavy quark plays a minor role.

% effective mass for M0=1.0 for 1l 2l
% 11 12  21 22.
%

\begin{figure}
\centerline{\ewxy{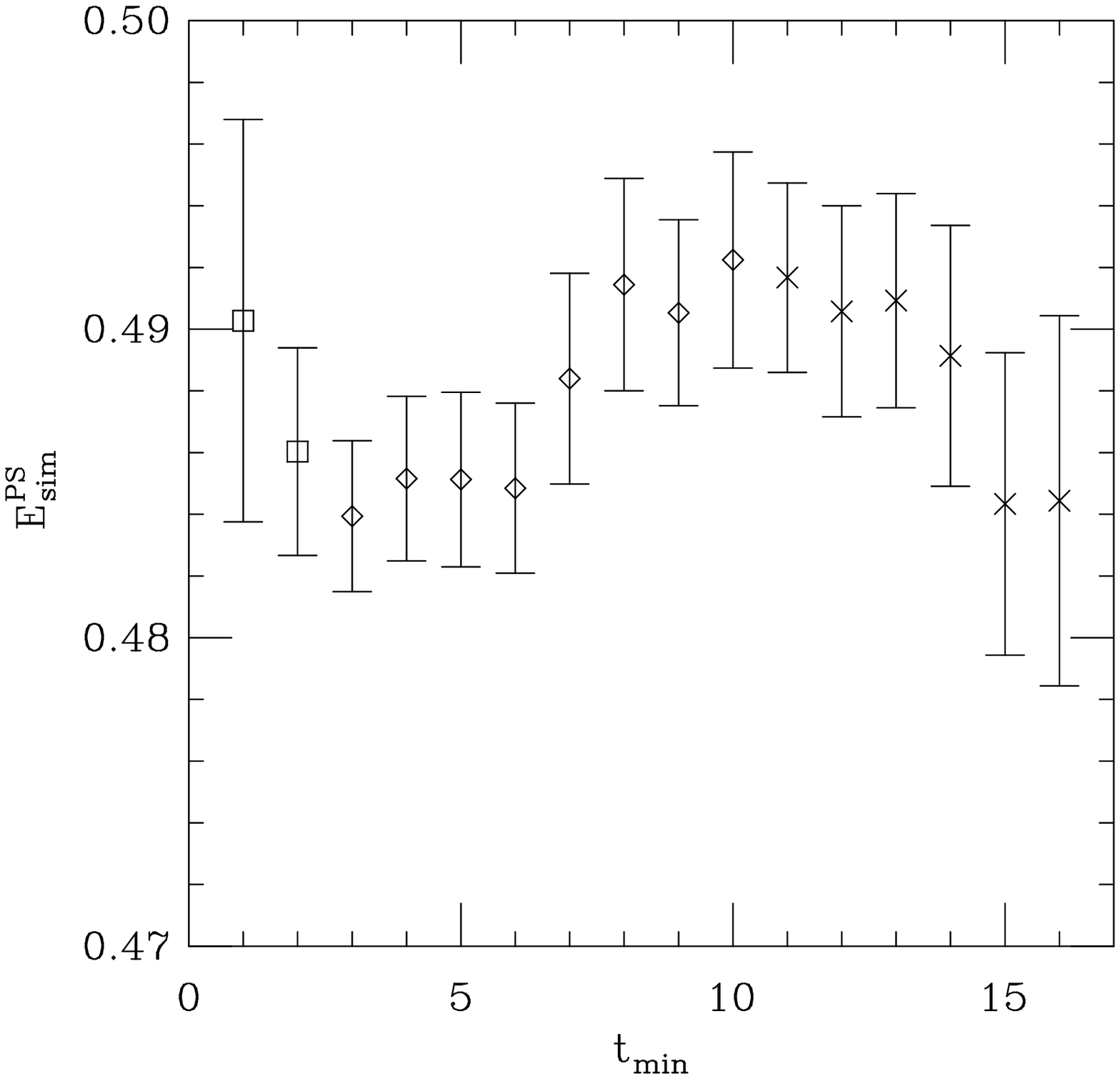}{80mm}
\ewxy{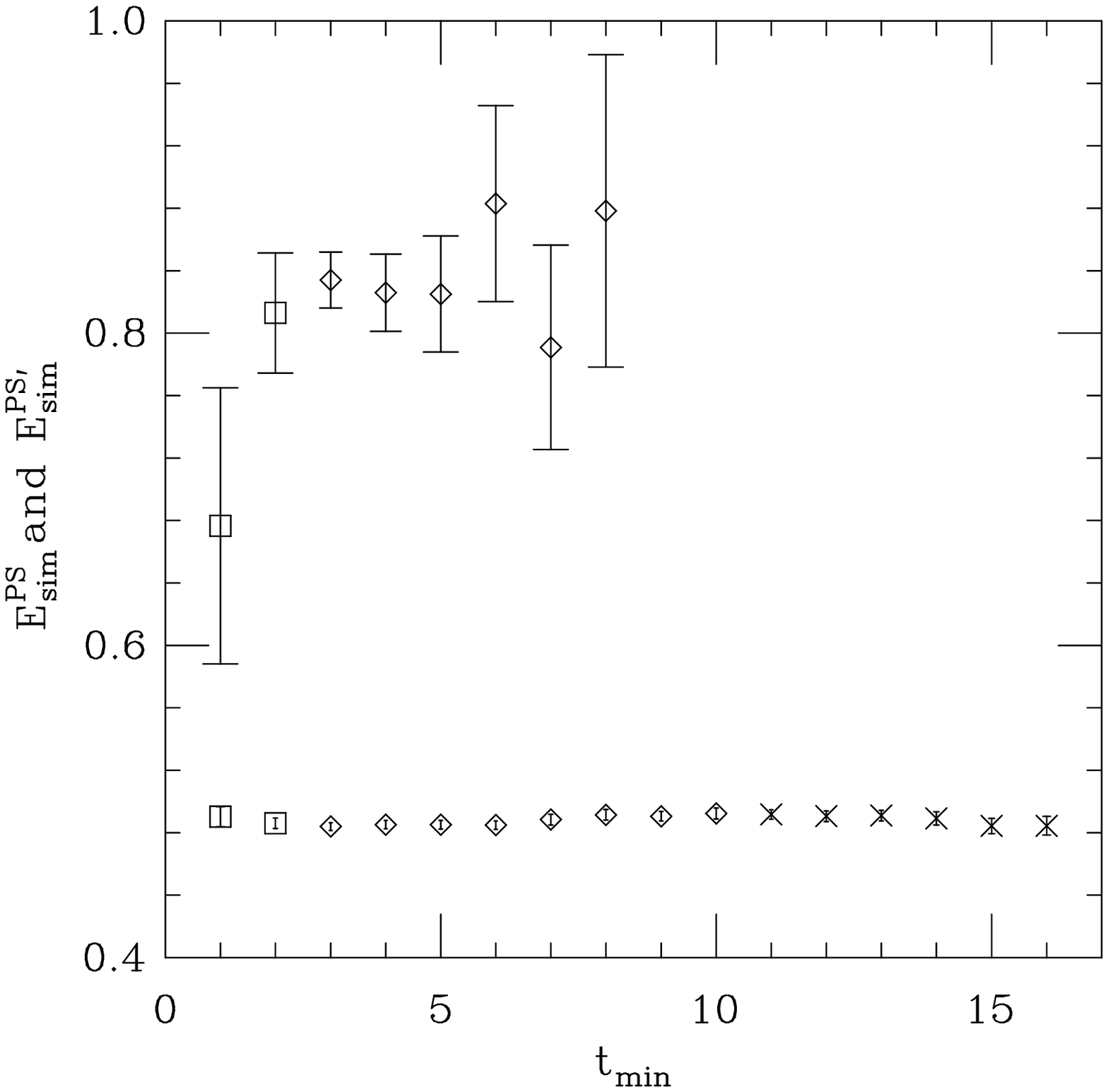}{80mm}
}
\vspace{0.5cm}
\centerline{\ewxy{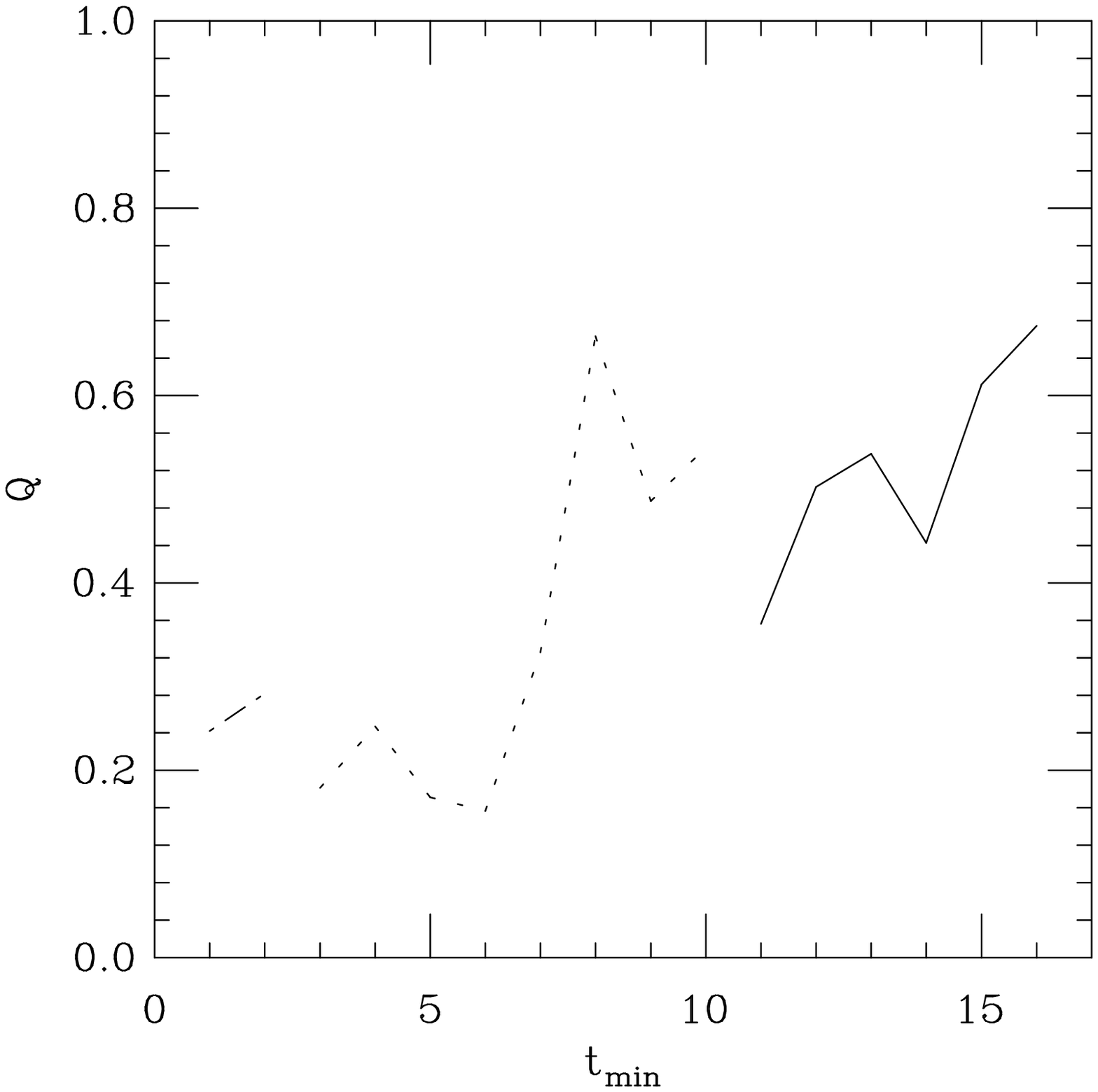}{80mm}
\ewxy{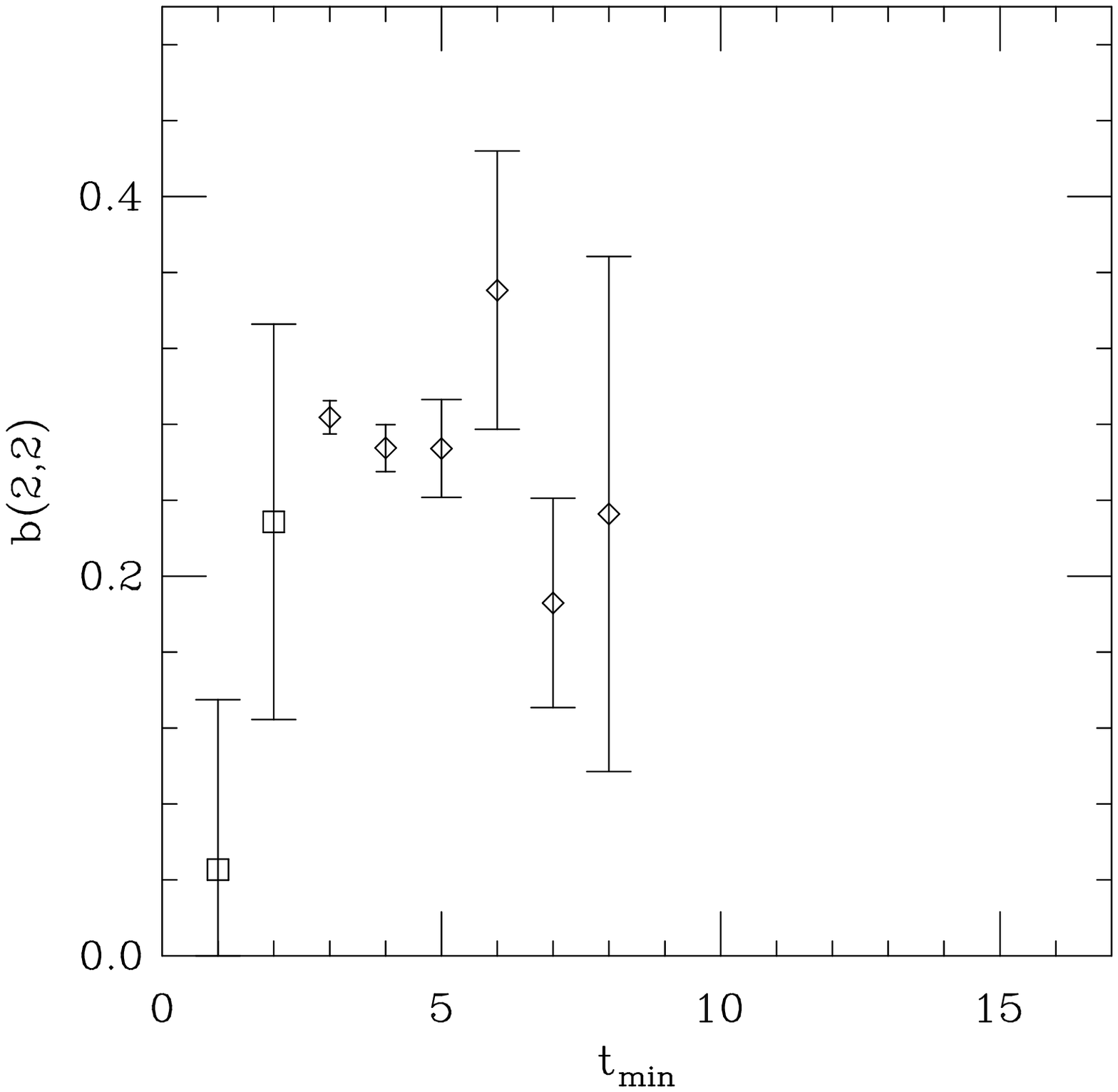}{80mm}
}
\caption{The variation of $^1S_0$ ground and first excited state energies
with $t_{min}$ extracted from vector fits to the $C_{1l}$ and $C_{2l}$
meson correlators for $aM_0 = 1.0$ and $\kappa_l=0.1585$. Fits
performed with $n_{exp}=3$ are shown as squares, diamonds represent
$n_{exp}=2$ and crosses $n_{exp}=1$.  Only fits for which $Q>0.1$ are
shown. The variation of $Q$ with $t_{min}$ is shown as the solid line
for $n_{exp}=1$ fits, the dotted line for $n_{exp}=2$ and the dashed
line for $n_{exp}=3$. $t_{max}$ is fixed at 20. Also presented is
$b(2,2)$, the overlap of the first excited state with the $C_{2l}$
correlator.
\label{fit_1l_2l_1}
}
\end{figure}

The $^1S_0$ ground state energy extracted from the $n_{exp}=1,2$ and
$3$ vector fits is presented in figure~\ref{fit_1l_2l_1} as a function
of $t_{min}$, with $t_{max}$ fixed at $20$. Fits are shown for which
$Q>0.1$ and for which all states included in the fit are resolved. For
example the $n_{exp}=2$ vector fit provides results for $E_{sim}$
consistent with $n_{exp}=1$ for $t_{min}>10$. The first excited state,
however, is no longer resolved as seen by an excited state amplitude
consistent with zero. A $n_{exp}=3$ fit can be performed using
$t_{min}=2$ and 1. This does not give sufficient additional
timeslices, however, to resolve the third excited state.

A noticeable feature is the $2\sigma$ `wiggle' in $E_{sim}$ which also
appears in the effective mass of the $C_{1l}(t)$ correlator.  There
are two possibilities, excited state contamination or finite
statistics.  If the former is responsible the contribution from
excited states is small enough that it cannot be resolved
using $n_{exp}=2$.
Thus, we quote the uncertainty in $E_{sim}$ due to
the fitting procedure to be $2\sigma$. This feature is seen in the
results for all values of $M_0$.

We found $E_{sim}'$ for the first excited state extracted using
$n_{exp}=2$ to be stable around $0.8$ and the corresponding amplitude
to be well determined for $t_{min}=2{-}6$; in this region there is a
significant excited state contribution to $C_{2l}$ while the ground
state dominates $C_{1l}$.  A $n_{exp}=3$ fit over a wider range in
$t_{min}$ is needed to provide confidence in this result. $C_{ll}$ was
not found to be useful in providing a larger overlap with the first or
second excited state since even higher excited states were also
present with significant contributions.

% plot of tmin and Q for M0=1.0
% for 1l 2l

\begin{figure}
\centerline{\ewxy{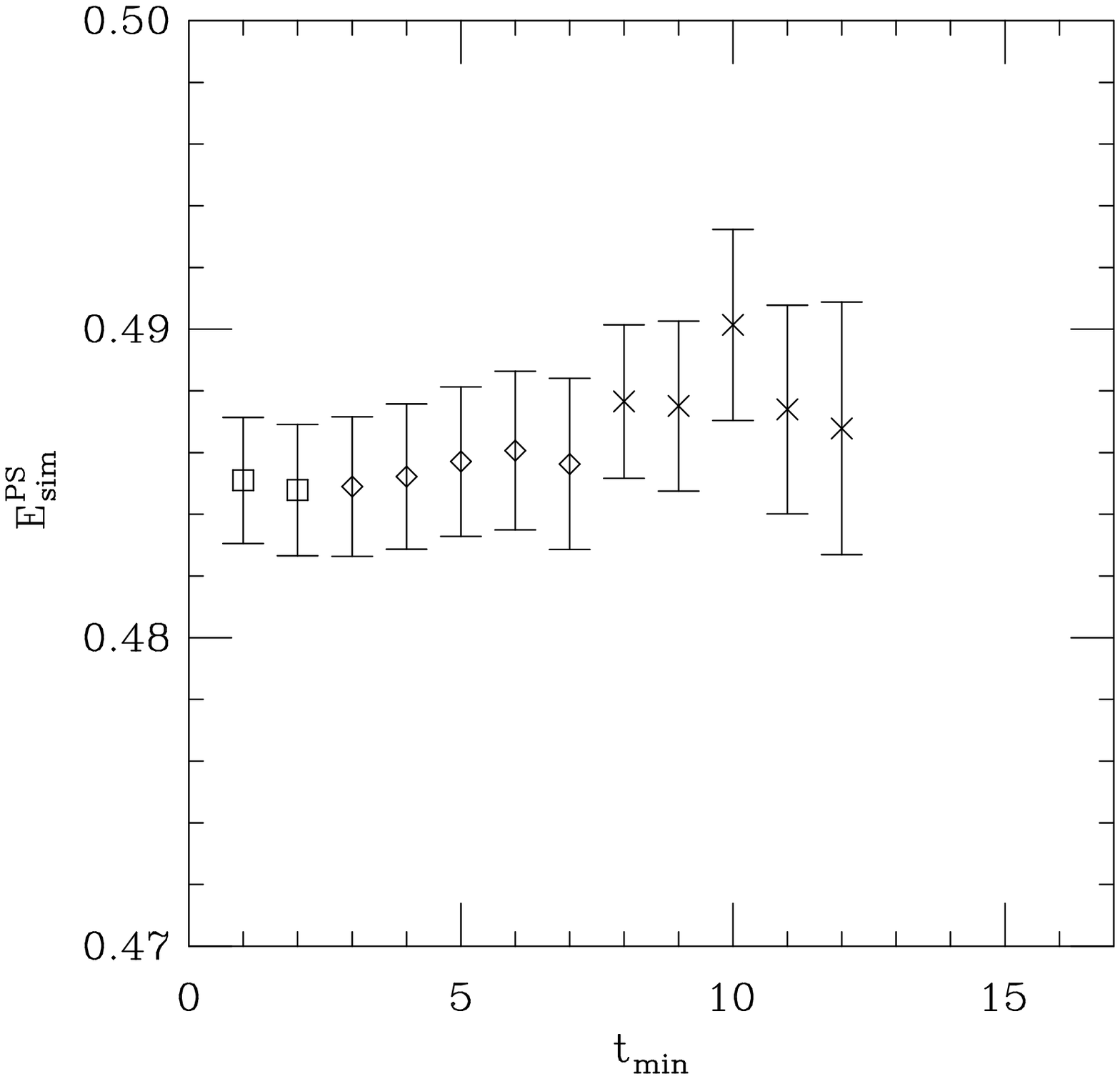}{80mm}
\ewxy{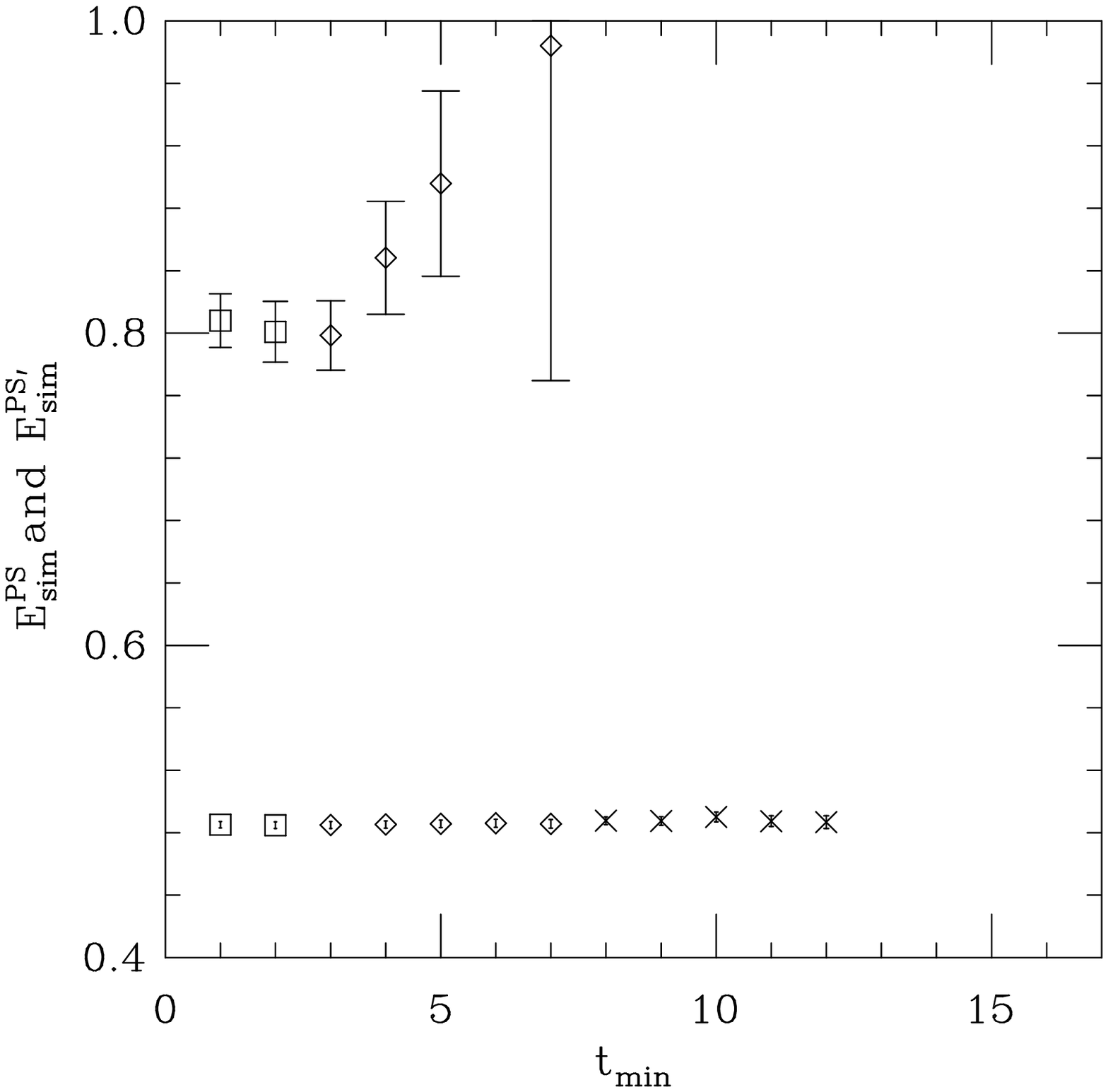}{80mm}
}
\vspace{0.5cm}
\centerline{\ewxy{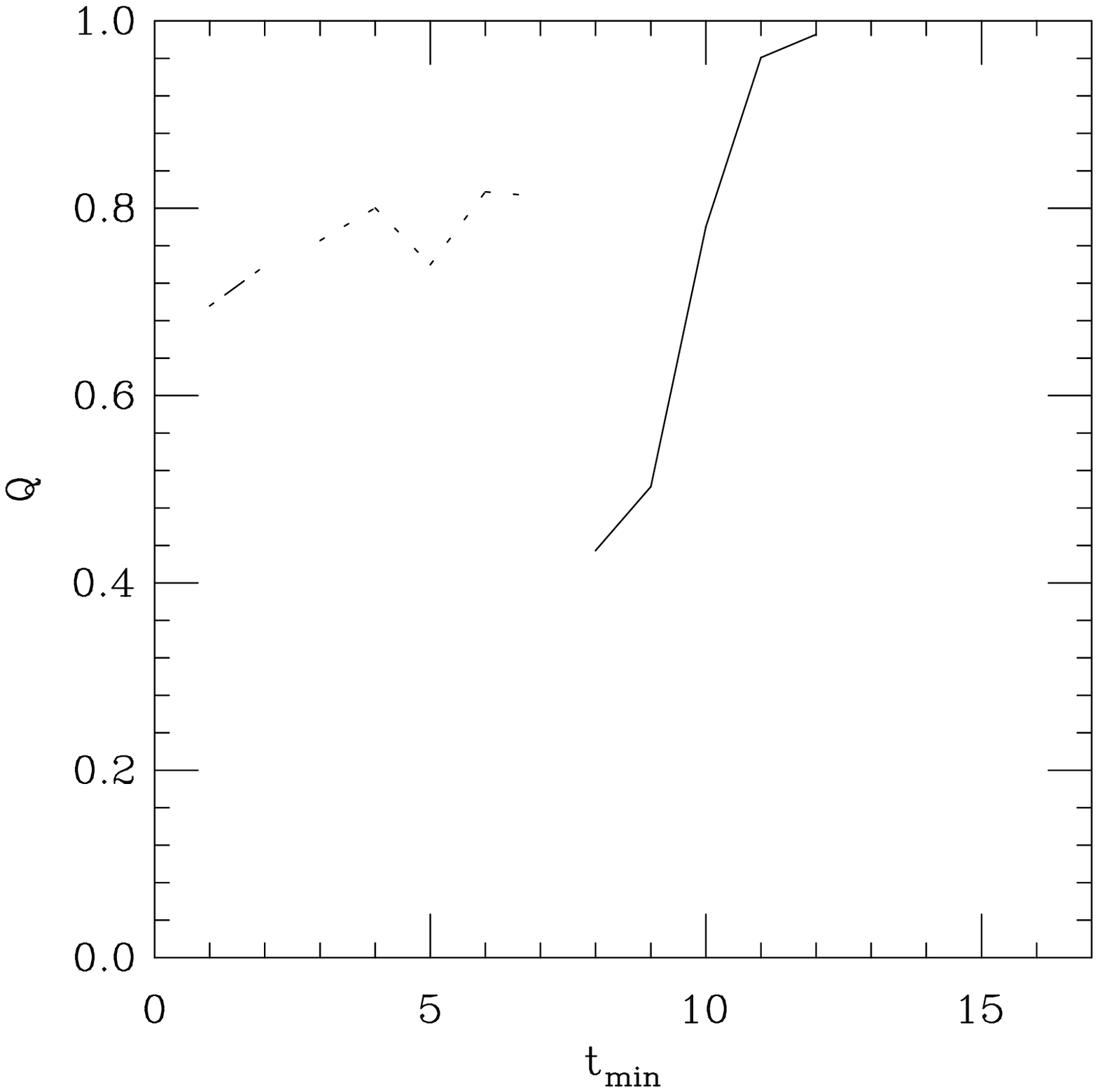}{80mm}
\ewxy{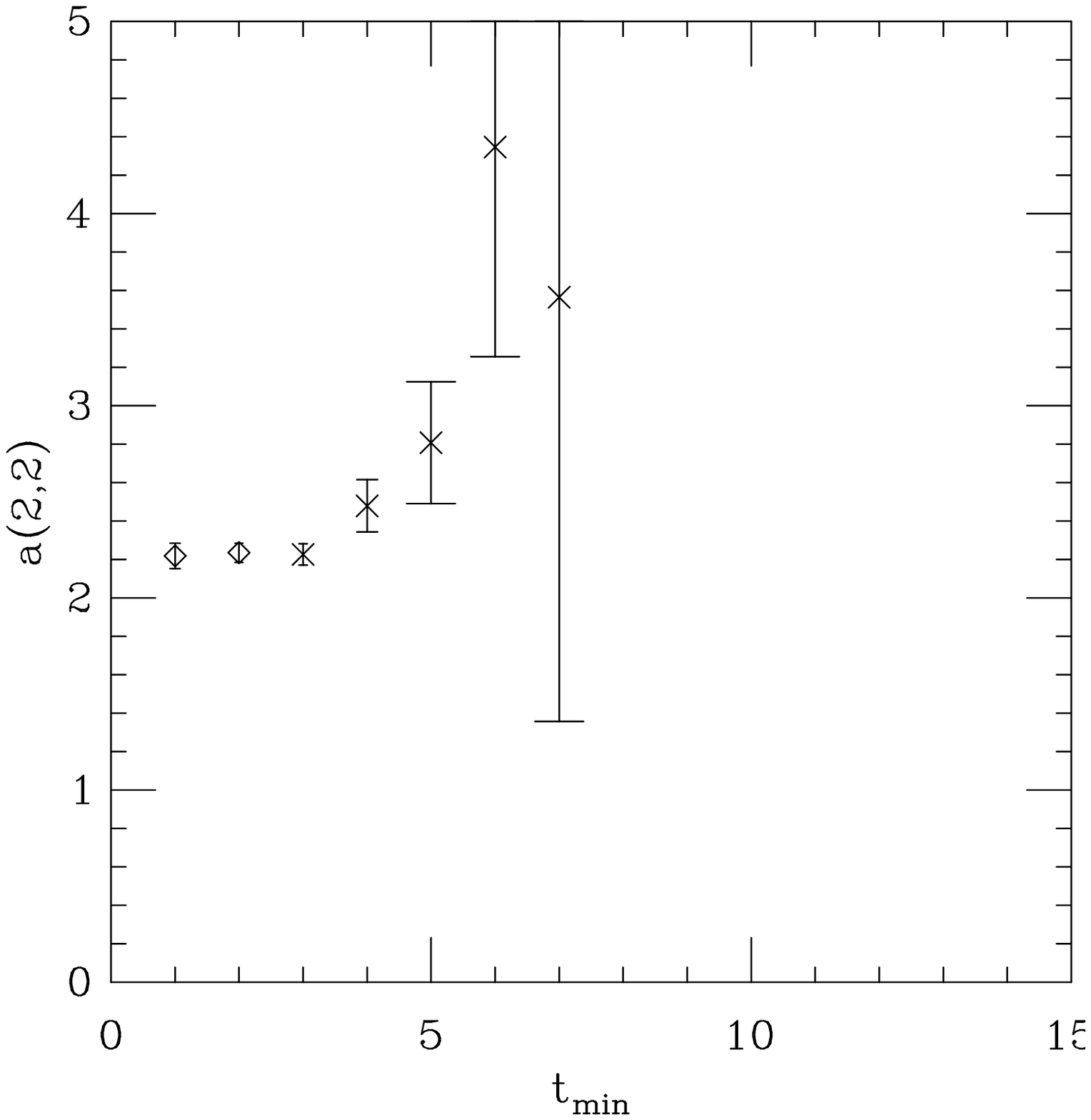}{80mm}
}
\caption{The variation of $^1S_0$ ground and first excited state energies
with $t_{min}$ extracted from matrix fits to the $C_{11}$, $C_{12}$,
$C_{21}$ and $C_{22}$ meson correlators for $aM_0 = 1.0$ and
$\kappa_l=0.1585$, where $t_{max} = 16$. The same symbols are used as
in the previous figure. Also presented is the overlap $a(2,2)$ of the
first excited state with the excited state smearing function.
\label{fit_mat_1}
}
\end{figure}

A similar picture is found in the results for the matrix fits
presented in figure~\ref{fit_mat_1}. $E_{sim}$ and $E_{sim}'$ are
consistent as $t_{min}$ and $n_{exp}$ are varied and in agreement with
those obtained using the vector fit. This is proof that we
have isolated the ground state and minimised excited state
contributions to $E_{sim}$.

% plot of tmin and Q for M0=1.0
% for 11 12 21 22
%

\begin{figure}
\centerline{\ewxy{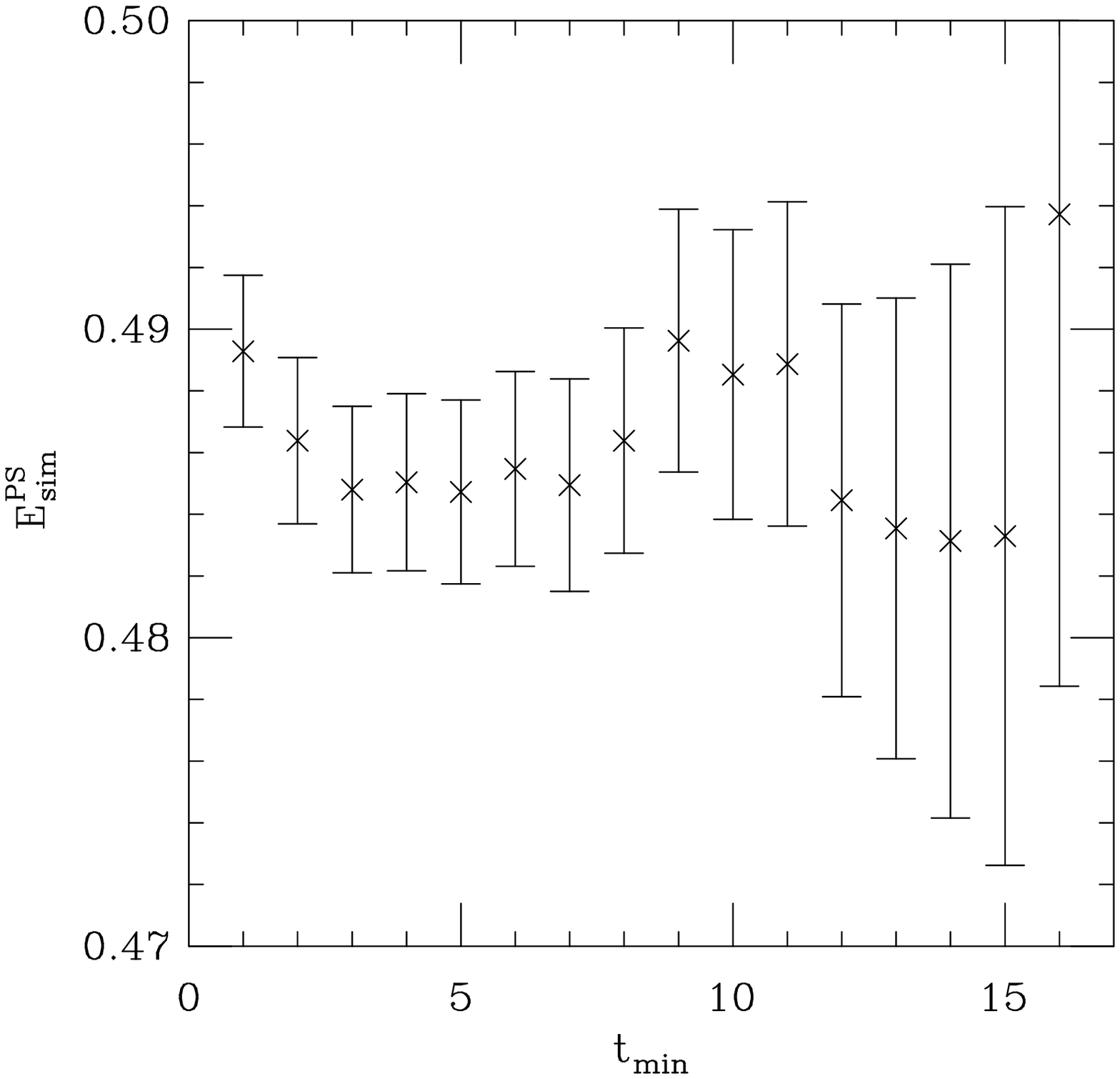}{80mm}
\ewxy{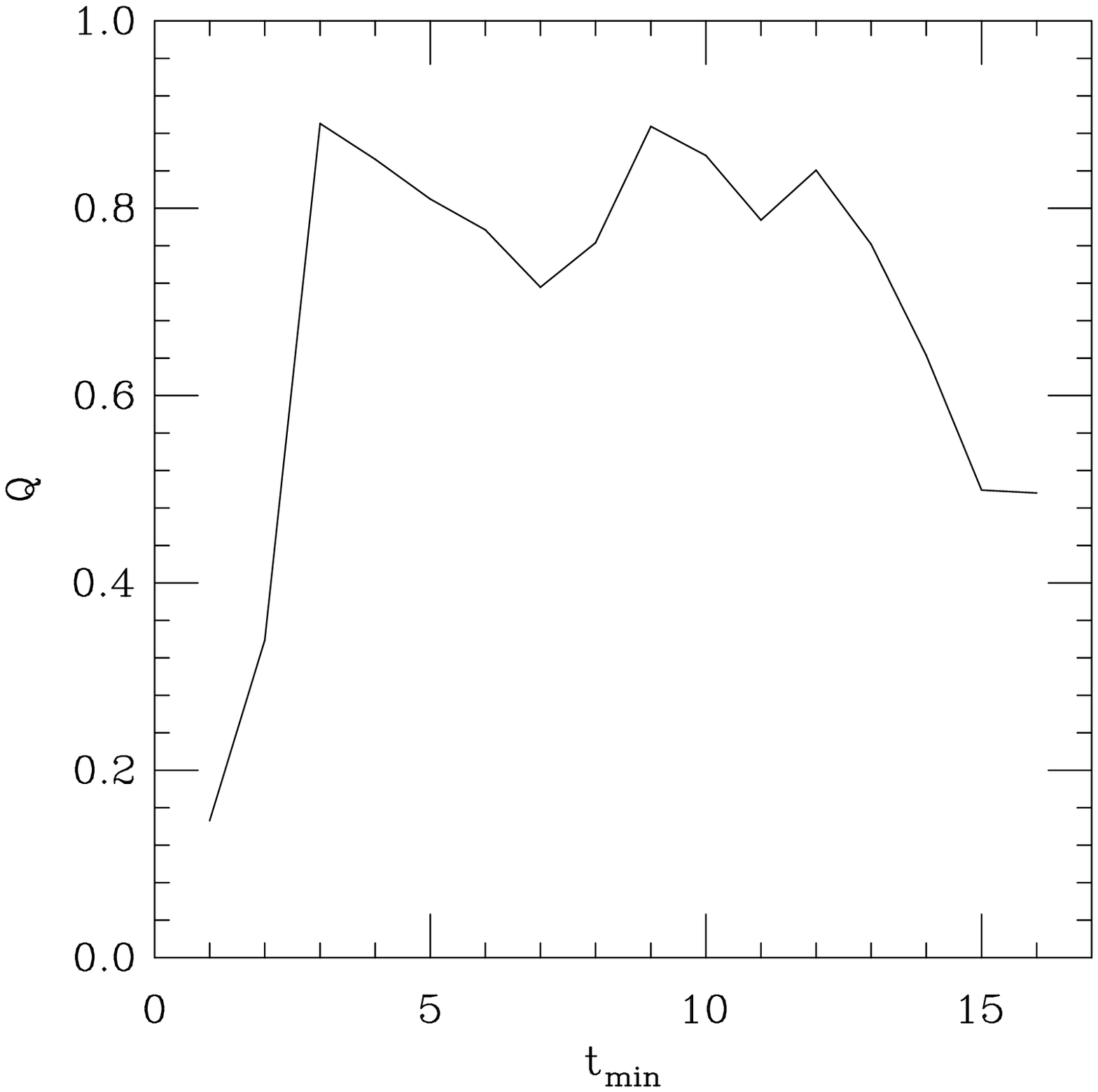}{80mm}
}
\caption{The variation of $E_{sim}^{PS}$ with $t_{min}$
extracted from a single exponential fit to  $C_{11}$
for $aM_0 = 1.0$ and $\kappa_l=0.1585$, where $t_{max} = 20$.
\label{fit_11}
}
\end{figure}

As mentioned previously, we were limited to bootstrapping only
$n_{exp}=1$ fits.  So we also performed $n_{exp}=1$ correlated fits to
the $C_{11}(t)$ correlation functions. The effective mass and values
of $E_{sim}$ obtained, with the corresponding values of $Q$, are shown
in figure~\ref{fit_11}. A plateau can be seen in the variation of
$E_{sim}$ with $t_{min}$ from $t_{min}=3$. Table~\ref{cmp_fits}
compares the best fits from all three fitting methods. The values for
the ground state using multi-exponential fits are consistent with the
$n_{exp}=1$ fit to $C_{11}(t)$.  The statistical error in the fits to
$C_{11}$ is larger. However, this will not affect quantitative results
of the analysis as the systematic errors are much larger than the
statistical errors.

% plot of tmin and Q for M0=1.0
% for 11
%
%

% table of fit parameters from various fits
% for 1.0.
\begin{table}
\begin{center}
\begin{tabular}{|c|c|c|c|c|c|}\hline
Data & $n_{exp}$ & fit range & Q & $aE_{sim}$ & $aE_{sim}'$  \\\hline
1l,2l &  2  &   3-20 &  0.2  & 0.484(2) & 0.83(2)\\\hline
11,12,21,22 & 2 & 3-16 & 0.8 & 0.485(2) & 0.80(2)\\\hline
11 & 1 & 3-20 & 0.9 & 0.485(3) & - \\\hline
\end{tabular}
\caption{A comparison of the energies extracted from
fits to 3 data sets of $^1S_0$ correlators for $aM_0 = 1.0$
and $\kappa_l=0.1585$\label{cmp_fits}}
\end{center}
\end{table}

The fitting range of $3{-}20$ was found to be optimal for the $^1S_0$
state for all $M_0$ at $\kappa_l=0.1585$, while for $^3S_1$, $4{-}20$
was used.  Table~\ref{tab:e_sim} details the corresponding values of
$E_{sim}^{PS}$ and $E_{sim}^V$. The first excited state energy
extracted using $n_{exp}=2$ correlated fits to the vector of smearing
functions with the fitting range $3{-}20$ for $aM_0=1.0$, 2.0 and 4.0
are presented in table~\ref{fit_1st}. The $E_{sim}'{-}E_{sim}$
splitting is of the order of $a\Lambda_{QCD}\sim 0.2$, and only weakly
dependent on the heavy quark mass, as expected for the excitation of a
light quark within the heavy-light meson.

% table of E_sim for 1s0 and 3s1.
%
\begin{table}
\begin{center}
\begin{tabular}{|c|c|c|c|}\hline
$aM_0$ & $aE_{sim}^{PS}$  & $aE_{sim}^V$ & $aE_{sim}^V$ {-}
$aE_{sim}^{PS}$\\\hline
0.8 & 0.472(3) &0.502(3) & 0.030(1)\\\hline
1.0 & 0.485(3) & 0.512(4)& 0.026(2)\\\hline
1.2 & 0.494(3) & 0.517(4)& 0.023(2)\\\hline
1.7 & 0.506(3) & 0.525(4)& 0.018(2) \\\hline
2.0 & 0.510(3) & 0.527(4)& 0.016(1)\\\hline
2.5 & 0.514(4) & 0.528(4)& 0.014(1)\\\hline
3.0 & 0.517(4) & 0.529(4)& 0.012(1)\\\hline
3.5 & 0.519(4) & 0.530(4)&0.010(1) \\\hline
4.0 & 0.520(4) & 0.530(4)& 0.009(1) \\\hline
7.0 & 0.522(4) & 0.529(5)&0.007(1) \\\hline
10.0 &0.522(4) & 0.528(5)& 0.005(1)\\\hline
$\infty$ & \multicolumn{2}{c|}{0.529(6)}& - \\\hline
\end{tabular}
\caption{Ground state energies extracted from $n_{exp}=1$ fits to $C_{11}$
 for $\kappa_l=0.1585$. Also shown are the hyperfine splittings
obtained from performing a single exponential fit to the ratio of the
$^3S_1$ and $^1S_0$ $C_{1l}$ correlators.\label{tab:e_sim}}
\end{center}
\end{table}

%
% table of excited state for 1s0 for 1.0,2.0,4.0
\begin{table}
\begin{center}
\begin{tabular}{|c|c|c|c|}\hline
$aM_0$ & $aE_{sim}^{PS}$ & $aE_{sim}^{PS'}$ & $a(2S{-}1S)$ \\\hline
1.0 &   0.485(3) &  0.83(2)  & 0.34(2) \\\hline
2.0 &   0.510(3) &  0.81(2)  & 0.30(2) \\\hline
4.0 &   0.520(4) &  0.77(1)  & 0.25(1) \\\hline
\end{tabular}
\caption{Ground and first excited state $^1S_0$ energies
obtained from $n_{exp}=2$ fits to $C_{1l}$ and $C_{2l}$ for
$\kappa_l=0.1585$. Also shown is the splitting between the ground and
first excited $^1S_0$ state.\label{fit_1st}}
\end{center}
\end{table}
%

% plot of effective masses for static
%
%
\begin{figure}
\centerline{\ewxy{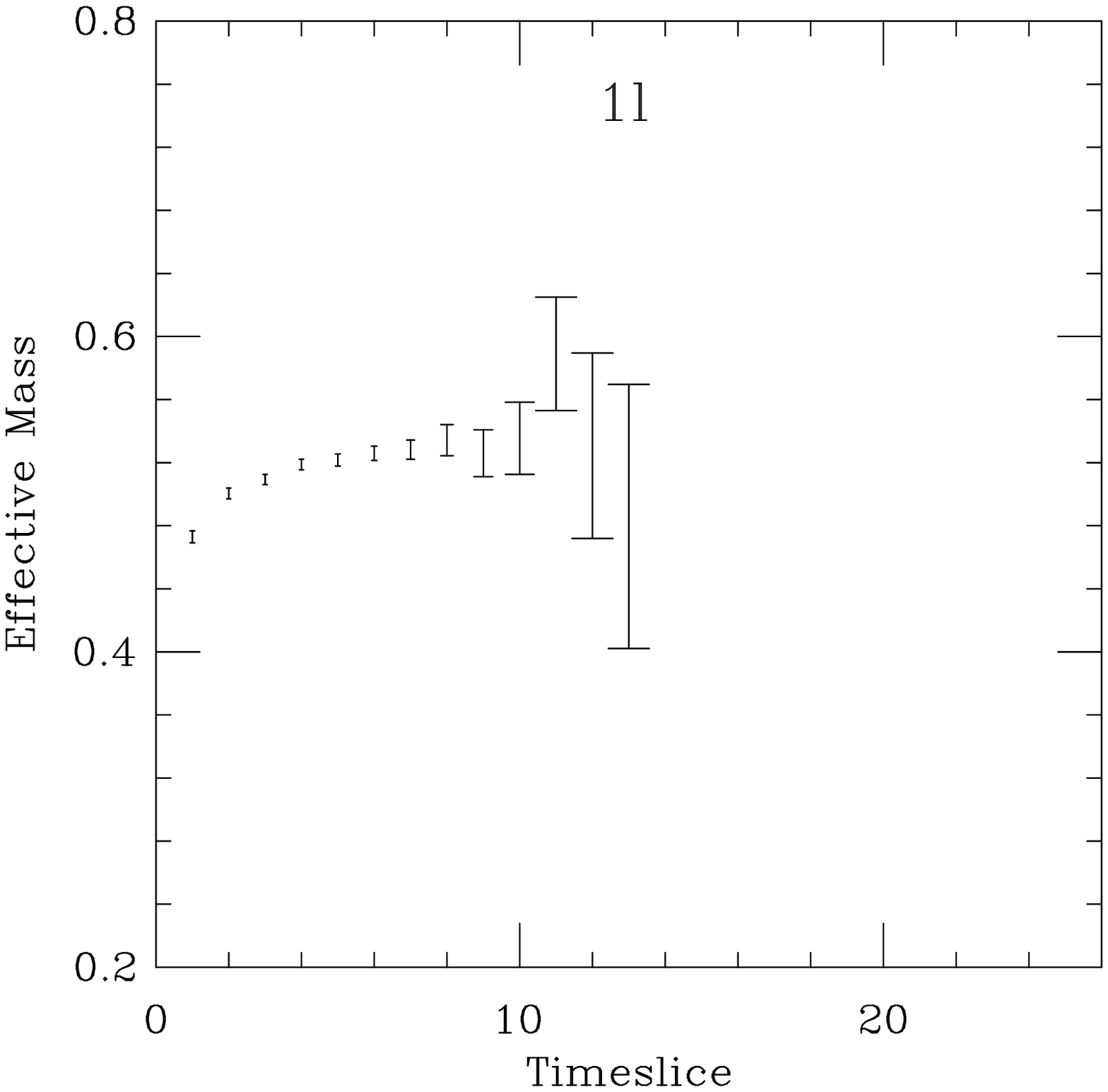}{80mm}
\ewxy{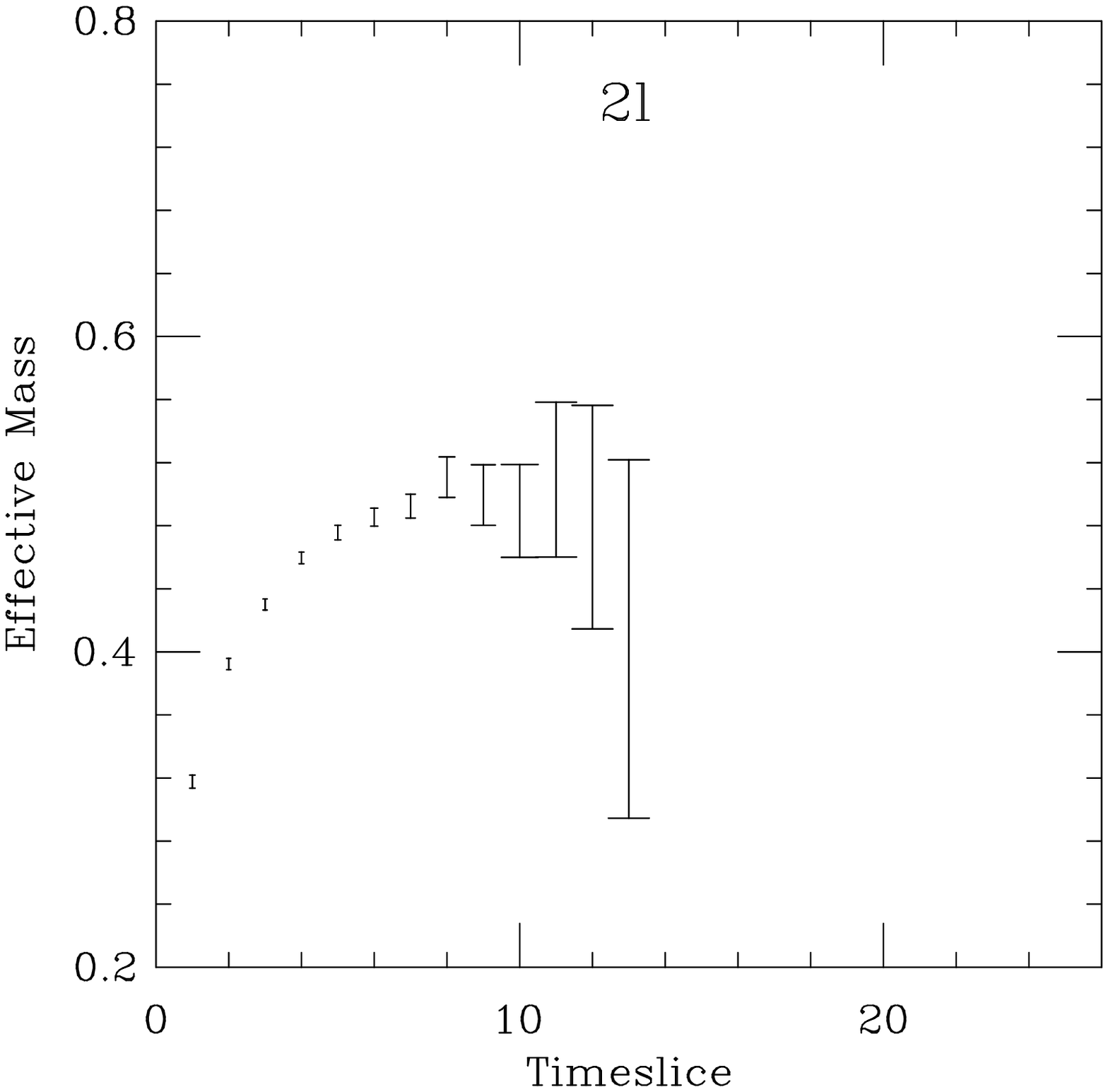}{80mm}
}
\vspace{0.5cm}
\centerline{\ewxy{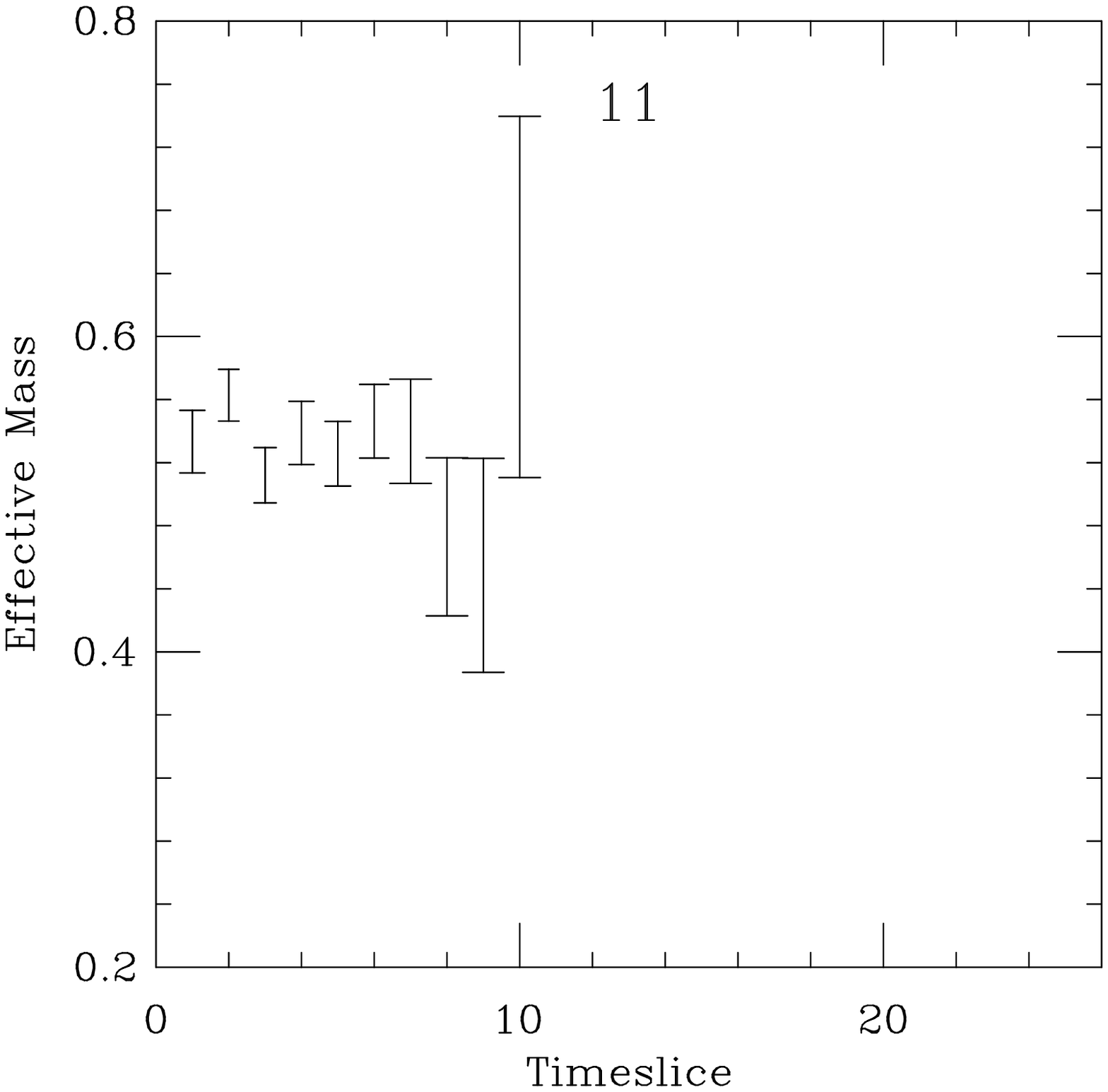}{80mm}
}
\caption{The effective masses of the $^1S_0$
meson correlators for $aM_0 = \infty$ and  $\kappa_l=0.1585$.
\label{meff_static}
}
\end{figure}

\begin{figure}
\centerline{\ewxy{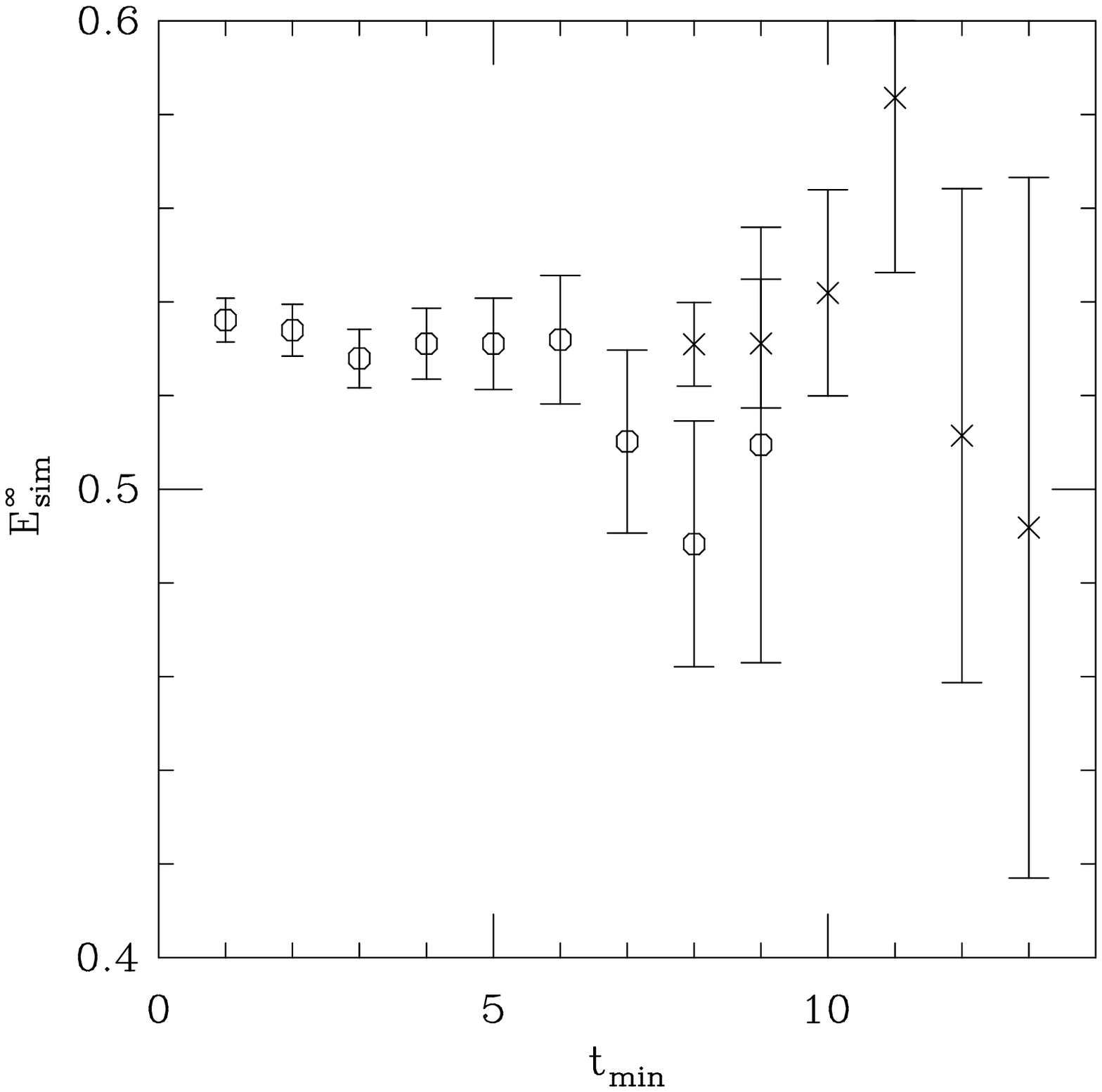}{80mm}
\ewxy{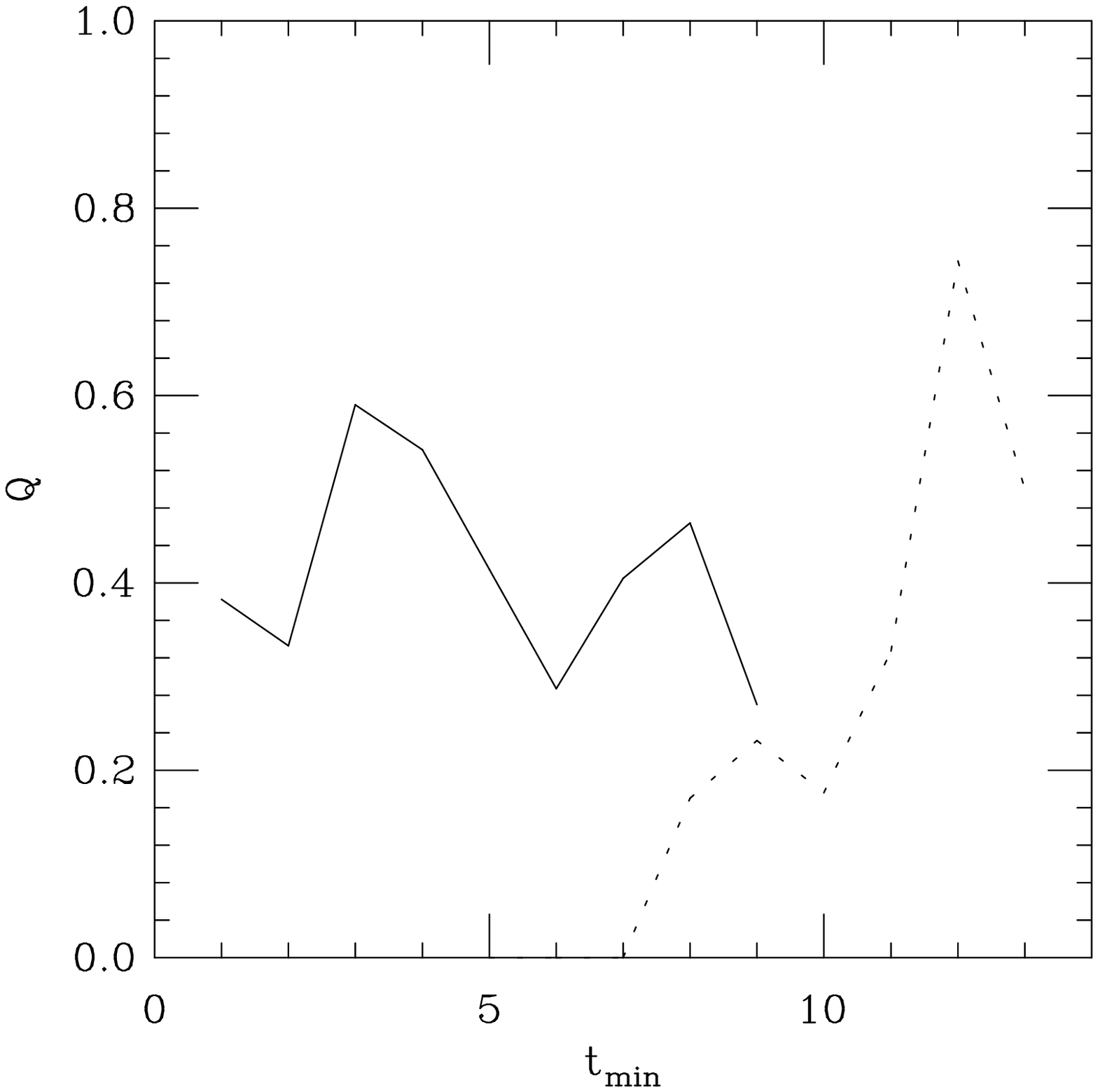}{80mm}
}
\caption{Single exponential fits to the $^1S_0$ $C_{11}$
meson correlator~(circles) and simultaneously to the $C_{1l}$ and
$C_{2l}$ correlators~(crosses) for $aM_0 = \infty$ and
$\kappa_l=0.1585$. $t_{max}=11$ for the fits to $C_{11}$, and $15$ for
the fits to $C_{1l}$ and $C_{2l}$. The corresponding values of $Q$
are shown as a solid line for the fits to $C_{11}$ and as a dotted
line for the fits to $C_{1l}$ and $C_{2l}$.\label{fit_static} }
\end{figure}

We were not able to perform such an extensive analysis in the static
limit. Figure~\ref{meff_static} presents the effective masses for the
$C_{1l}$, $C_{2l}$ and $C_{11}$ correlation functions for a static
heavy quark and $\kappa_l=0.1585$. The static results suffer from the
signal disappearing around timeslice 11.  The smeared correlators
$C_{12}$, $C_{21}$ and $C_{22}$ are dominated by noise and are not
useful.  We performed $n_{exp}=1$ correlated fits to $C_{11}$ and
simultaneously to $C_{1l}$ and $C_{2l}$, detailed in
figure~\ref{fit_static}. A $n_{exp}=2$ correlated fit to the vector of
smeared correlators with $Q>0.1$ was not found. The ground state and
first excited state smearing functions are not sufficiently different
with these statistics at early timeslices to resolve the first excited
state.  Consistency is found between the vector fits and the single
exponential fits to $C_{11}$ and we choose the latter with the time
range $3{-}11$ as the best fit.

% plot of fits for static
%
%

The static case is the infinite quark mass limit of NRQCD. Thus, in
the smooth transition from finite $M_0$ to the static limit the
$signal/noise$ problems, associated with $aM_0=\infty$, will also
present a problem at finite $M_0$ when the heavy quark mass is large
enough. However, as seen in figure~\ref{meff_fit_10} we found that
even the highest value of $M_0$ used is not hampered by noise and the
signal for the ground state remained out to timeslices of 18{-}20.
The static limit appears to be a special point with poor $signal/noise$
properties and our conclusions about the static result will actually
rely on the infinite mass extrapolation of the more accurate
NRQCD results.

% effective mass and tmin plots for M0=10.0
%
%

%
% Fitting ranges:
%  kappa_l =0.1585
%  1s0 -   3-20 11  all M0
%  3s1 -   4-20 11  all M0
%
%  1s0 -   3-11 11  static

%
% Fitting ranges for excited state 1s0  1l,2l 2exp
% 0.1585
%  m0  = 1.0      3-20
%        2.0      3-20
%        4.0      3-18
%
% 0.1600
%  m0  = 1.0
%        2.0
%        4.0

\subsection{Variation of $E_{sim}$ with heavy quark mass}
\label{hqsym}
\begin{figure}
\centerline{\ewxy{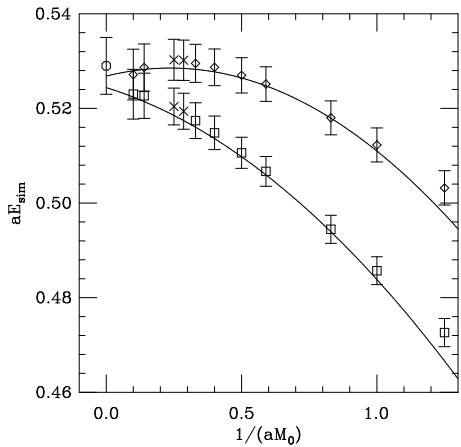}{80mm}
}
\caption{$aE_{sim}$ vs $1/(aM_0)$ for $^1S_0$~(squares) and
$^3S_1$~(diamonds) for $\kappa_l=0.1585$. The static value is
shown as a circle and the solid lines show quadratic fits to the heaviest
six data points. The data points shown as crosses correspond to $aM_0=3.5$
and $4.0$ and are not included in the fits.
\label{esim_all}
 }
\end{figure}

The results for $E_{sim}$ for both $^1S_0$ and $^3S_1$ at finite $M_0$
and in the static limit are plotted against $1/M_0$ in
figure~\ref{esim_all}.  As $M_0$ increases and the hyperfine
interaction becomes smaller the S-states become degenerate and tend
towards the static result where flavour and spin symmetry are
restored. This agreement with the static result coupled with the
multi-state, multi-smearing analysis at finite $M_0$ suggests that the
excited state contribution to $E_{sim}$ for $aM_0=\infty$ is small. In
addition, the deviation from the static limit around the $B$ meson
($aM_0\sim 2.0$ see Section~\ref{mesonmass}) is small and of order a
few percent as expected by naive power counting arguments.

% plot of Esim for 1s0 and 3s1 plus static.
%

In the same way as for the NRQCD action, $E_{sim}$ and other
meson quantities can be parameterised in the heavy quark limit
in terms of an expansion in $1/M$. The coefficients of the
expansion are nonperturbative quantities which can be
extracted from the simulation.
{}From first order perturbation theory in ($H_0+\delta H$) about
the static limit we obtain
\begin{equation}
E_{sim} = E_{sim}^\infty + \hbox{}^{\infty}\hskip-2pt\bra{P}
\bar{Q}(-\vec{D}^2 - c_B\sigma\cdot B)Q\ket{P}^\infty \frac{1}{2M_0}  +
O\left(\frac{1}{M_0^2}\right).
\label{expan}
\end{equation}
where $\ket{P}^\infty$ represents the meson in the limit of infinite
heavy quark mass and the $O(1/M_0)$ coefficient is merely the
expectation value of the kinetic and hyperfine terms appearing in the
NRQCD action. Note that the possibility of using a different choice of
expansion parameter in equation~\ref{expan} introduces an $O(1/M_0)$
uncertainty in the first order coefficient. $E_{sim}^\infty$ is
related to the binding energy of the heavy-light meson,
$\bar{\Lambda}$, in the static limit:
\begin{eqnarray}
\bar{\Lambda} & = & \lim_{m_Q\rightarrow \infty} (M_{hl} - m_Q)\\
              & = & E_{sim}^\infty - E_0^\infty
\label{barlam}
\end{eqnarray}
where $m_Q$ is the pole mass and $E_0^\infty$ is the energy of a heavy quark
in the static theory. Both $\bar{\Lambda}$ and
$\hbox{}^{\infty}\hskip-2pt\bra{P}
\bar{Q}(-\vec{D}^2 - c_B\sigma\cdot B)Q\ket{P}^\infty$  appear in expressions
for $O(1/M)$ corrections to HQET predictions, but cannot be calculated
using HQET alone.

To extract these quantities we performed correlated fits to the data as
a function of $1/M_0$ using the functional form
\begin{equation}
E_{sim} = C_0 + \frac{C_1}{M_0} + \frac{C_2}{M_0^2} + \frac{C_3}{M_0^3}.
\label{fitfunc}
\end{equation}
The fitting procedure was similar to that for the propagator fits;
beginning with a constant fit function the endpoint for the fit is
fixed to the heaviest data point, $aM_0=10.0$, and the initial point is
varied over all finite values of $M_0$. The procedure is repeated for
a linear, quadratic and a cubic fit. The results for the $^1S_0$ meson are
shown in table~\ref{esim_ps}, where, except for the fits to a constant,
only `good' fits to the data are presented~(defined as $Q>0.1$).

The variation of the data with $1/M_0$ is consistent with a constant
only for the heaviest two data points, while a linear term is required
in the region of the $B_s$ meson and a quadratic term for
$1/aM_0\gtaeq0.8$. The lightest two points require a fit function of
higher order than a cubic function to be included in a fit.
Since we have truncated the NRQCD action at $O(1/M_0)$ the coefficients of the
quadratic term are not correct and if the results are to connect with
simulations around the $D$ meson the $O(1/M_0^2)$ terms in the action
are needed.

%
% table of fits to Esim vs 1/M0
%
\begin{table}[htbp]
\begin{center}
\begin{tabular}{|c|c|c|c|c|c|c|}\hline
order & fit range & n.d.o.f & Q & $aC_0$ & $a^2C_1$ & $a^3C_2$  \\\hline
0 & 1-2 & 1 & 0.81     &  0.522(3)       &   -      & - \\\cline{2-7}
  & 1-3 & 2 & 0.04     &  0.515(3)      &   -      & -  \\\hline
1 & 1-3 & 1 & 0.55 &0.527(4) &-0.029(8) & - \\\cline{2-7}
  & 1-4 & 2 & 0.61 &0.529(4) &-0.036(5) & - \\\cline{2-7}
  & 1-5 & 3 & 0.29 &0.533(3) &-0.044(3) & - \\\cline{2-7}
  & 1-6 & 4 & 0.44 &0.532(3) &-0.043(2) & - \\\hline
2  & 1-6 & 3 & 0.42 & 0.528(4) &-0.031(9) & -0.018(7) \\\cline{2-7}
  & 1-7 & 4 & 0.32 & 0.524(4) &-0.018(6) & -0.015(4)  \\\hline
\end{tabular}
\caption{Correlated fits to $E_{sim}^{PS}$ as a function of $1/M_0$ for
$\kappa_l=0.1585$. The fit range $1{-}3$ denotes a fit to
$E_{sim}^{PS}$ at $aM_0=10.0, 7.0$ and $3.0$. The points at $aM_0 =
4.0$ and 3.5 are not included in the fits for reasons explained later.
\label{esim_ps}}
\end{center}
\end{table}

In general, in order to obtain stable values for $C_i$
the fit must include $C_{i+1}$. The results for $C_0$ from both the
linear and quadratic fits are consistent, and also in agreement with
the static result. There are not sufficient data points in the
quadratic region to determine whether the value for $C_1$ is stable
around $\sim-0.03$. Hence, only a rough determination of the
coefficients is possible and the results are detailed in
table~\ref{coeffs} for both $^3S_1$ and $^1S_0$. Quadratic
fits to both $E_{sim}^{PS}$ and $E_{sim}^V$ are shown in
figure~\ref{esim_all}. The spin-average of the energies,
$\bar{E}_{sim}=\frac{1}{4}(E_{sim}^{PS} + 3 E_{sim}^V)$, removes the
spin dependence and the slope of $\bar{E}_{sim}$ depends only on the
kinetic energy of the heavy quark.  The results for $\bar{E}_{sim}$
are also given in table~\ref{coeffs}.

% table of final fits for fitting to spin average etc
%
\begin{table}
\begin{center}
\begin{tabular}{|c|c|c|}\hline
 & $aC_0$ & $a^2C_1$  \\\hline
$E_{sim}^{PS}$ & 0.528(5) & -0.03(1)\\\hline
$E_{sim}^V$ &  0.528(5) & -0.01(1) \\\hline
\raisebox{-0.3ex}{$\bar{E}_{sim}$} &  0.530(5) & -0.02(1)\\\hline
\end{tabular}
\caption{The coefficients extracted from fits to
$E_{sim}$ as a function of $1/M_0$ for $\kappa_l=0.1585$.
The errors include the variation in the coefficients
obtained using different orders in the fit function.
\label{coeffs}}
\end{center}
\end{table}

It is useful to compare with previous results for the slope and
intercept of $E_{sim}$. However, most other calculations have been
performed in the quenched approximation; it is more appropriate to
compare our results at $\beta^{n_f=0}=6.0$~(using the tree-level
Clover action for the light quarks as detailed in
reference~\cite{arifa}) with those of other groups. In particular,
Hashimoto~\cite{hash} using NRQCD for the heavy quark and the
Wilson action for the light quarks and Crisafulli~et~al~\cite{sach} using the
static approximation and the tree-level Clover action have results at
this $\beta$ value. Note that both groups use naive operators i.e. do
not use tadpole improvement. In addition, taking into account the
different systematic errors inherent in each of the simulations only a
fairly rough comparison can be made.

%arifa = E_sim^infty = 0.461(15) (k_c)  slope = -0.02 (2 or 1) for both kap_l
%hashimoto = 0.60 + 0.33/M_0  (k_c)
%sach = 0.61(1) -actually  from ape collab lambda_1 =0.72(14)
% for kappa_l=?

Considering the intercept initially,
\begin{eqnarray}
aE_{sim}^\infty & = & 0.60\mbox{\footnotemark[1]}\hspace{2cm}\mbox{Hashimoto}\\
               & = & 0.61(1)\hspace{1.65cm}\mbox{APE
Collaboration~\cite{ape}\footnotemark[2]}
\end{eqnarray}
where $\kappa_l=\kappa_c$. This is in good agreement with our results at
the same $\beta$ value of $aE_{sim}^{\infty}=0.461(15)$ taking into
account the difference of $\ln u_0=-0.13$~(where $u_0=0.878$ measured
from the plaquette) since we use tadpole improved operators.
\footnotetext[1]{The error in this quantity is not given.}
\footnotetext[2]{Crisafulli et al use the APE results generated on the same
configurations for this quantity.}

\hspace{3mm}In order to calculate $\bar{\Lambda}$ from this quantity we use the
tadpole-improved perturbative value of $E_0^\infty=0.29(8)$ at
$\beta^{n_f=0}=6.0$~\cite{colin,eichten}.  $E_0^\infty$ is an $O(\alpha_s)$
quantity and the corresponding error from omitting 2-loop corrections
may be large, dominating the error in $\bar{\Lambda}$. Here the error
in $E_0^\infty$ is estimated by assuming the 2-loop corrections are roughly
the square of the first order terms. Crisafulli et al suggest that as
a linearly divergent quantity, nonperturbative effects may make a
significant contribution to $E_0^\infty$ and using a perturbative value will
give rise to an ambiguity in the corresponding value for
$\bar{\Lambda}$. Using a nonperturbative renormalisation procedure
which aims to remove all mixing with lower dimensional operators they
obtain the naive~(non-tadpole improved) $E_0^\infty =
0.521(6)(10)$~\cite{sach}. However, the disagreement with the
corresponding naive perturbative result, $E_0^\infty=0.29(8)-\ln
u_0=0.42(8)$, is only at the $1\sigma$ level. In addition, the
nonperturbative result appears to be gauge
dependent~\cite{chris}.

{}From equation~\ref{barlam} we obtain,
\begin{equation}
a\bar{\Lambda} = 0.17(10)  \hspace{1cm}\kappa = \kappa_c
\hspace{1cm}\beta^{n_f=0}=6.0.
\end{equation}
for our quenched results. The error is large enough to provide
consistency with $a\bar{\Lambda}= 0.09(1)$ from
Crisafulli et al.

For this study we find,
\begin{eqnarray}
a\bar{\Lambda} & = & 0.17(14) \hspace{1cm} \kappa = 0.1585,\nonumber\\
              & = & 0.13(14) \hspace{1cm} \kappa = \kappa_c.
\label{lam_bar}
\end{eqnarray}
using $E_0^\infty=0.36(13)$ from perturbation theory. The results for the
meson binding energy at finite values of $M_0$,
$\bar{\Lambda}_{M_0}\equiv E_{sim}{-}E_0$, are given in
table~\ref{lam} and will be referred to below.  It is clear that the
uncertainty in the perturbative value for $E_0$ must be reduced before
a meaningful determination of the meson binding energy is possible
with this approach.

Consider the slope of $\bar{E}_{sim}$ with $M_0$: in
reference~\cite{arifa} we found results for $C_1$ similar to
those presented in table~\ref{coeffs},
\begin{equation}
a^2C_1 = a^2<-\vec{D}^2/2>^{tadpole-improved}_{bare} = -0.02(1)
\hspace{0.5cm}\beta^{n_f=0}=6.0
\end{equation}
independent of $\kappa_l$. The slope is small and negative.
The naive, bare kinetic energy operator, however, is positive definite; in
mean-field theory the two matrix elements are related by
\begin{equation}
<-\vec{D}^2/2>^{tadpole-improved}_{bare} =
<-\vec{D}^2/2>^{naive}_{bare} - 3\left(1-u_0\right).
\end{equation}
Thus, we obtain an estimate for
%\begin{equation}
$a^2<-\vec{D}^2/2>^{naive}_{bare}\sim
+0.35.$
%\mbox{; similarly } a^2<-\vec{D}^2 - c_B\sigma\cdot B
%>^{naive}_{bare}\sim +0.34.
%\end{equation}
Hashimoto extracted the slope of $\bar{E}_{sim}$~(the hyperfine term
is omitted from the NRQCD action), at $\kappa=\kappa_c$, and found
$a^2<-\vec{D}^2/2>^{naive}_{bare}=+0.33$
in good agreement with our
results. Furthermore, Crisafulli et al, using their nonperturbative
procedure calculate the matrix element explicitly and find
$a^2<-\vec{D}^2/2>^{naive}_{bare}= +0.36(7)$ for $\kappa_l$
around $\kappa_s$. The overall agreement between different groups is
encouraging. However, the dominance of the tadpole contributions to
\mbox{$<-\vec{D}^2/2>^{naive}_{bare}$} suggest it is better to quote the
tadpole-improved matrix elements.

The physical kinetic energy of the heavy quark within the $B$ meson,\\
\mbox{$<-\vec{D}^2>_{renorm}$}, is a positive definite quantity and expected to
be very small, of size \mbox{$O(a^2\Lambda_{QCD}^2)\sim 0.04$}.  The
renormalised matrix element corresponds to the slope of
$\bar{\Lambda}_{M_0}$ with respect to $M_0$.  From
table~\ref{pert}, $E_0$ has a negative slope with $M_0$ and is of a
magnitude that indicates a positive result for $<-\vec{D}^2>_{renorm}$
that is not inconsistent with $O(\Lambda_{QCD}^2)$. However, due to
the large perturbative error our results for $\bar{\Lambda}_{M_0}$,
shown in table~\ref{lam}, are consistent with a zero slope.

\begin{table}
\begin{center}
\begin{tabular}{|c|c|c|c|c|c|c|c|c|c|}\hline
$aM_0$  & 3.0 & 3.5 & 4.0 & 7.0 & 10.0 & $\infty$\\\hline
$\kappa_l=0.1585$ & 0.168 & 0.171 & 0.174 & 0.172 & 0.171 & 0.172 \\\hline
$\kappa_l=\kappa_c$& 0.129 & 0.133 & 0.135 & 0.132 & 0.134 & 0.131 \\\hline
\end{tabular}
\caption{$a\bar{\Lambda}_{M_0}$ for the heavier quark mass values.
The corresponding errors are of the same magnitude as those in
equation~\protect\ref{lam_bar}.\label{lam}}
\end{center}
\end{table}

Also shown in figure~\ref{esim_all} are the results for $E_{sim}^{PS}$
and $E_{sim}^{V}$ at $aM_0=3.5$ and $4.0$. The stabilising parameter,
$n=1$, was used for both these values of $M_0$. While this value
satisfies $n\gtaeq3/M_0$ the results for $E_{sim}$ do not smoothly
interpolate between the heavier data points with $n=1$ and the lighter
points which use $n=2$.  In particular, it was not possible to perform
a correlated fit~(even including higher powers in $1/M_0$ in the fit
function) that includes $aM_0=3.5$ and $4.0$ and the data at
$aM_0=3.0$, $2.5$, $\ldots$ for which $n=2$. The problem is remedied
by increasing $n$ for these values of $aM_0$ to $2$~\cite{clover}; the
shift in $E_{sim}$ is much less than $1\sigma$ in the statistical
errors and it is only the high correlations between results at
different $M_0$ and our statistical accuracy that allows us to see
this problem. Using $n=1$ is sufficient for the heaviest data points
at $aM_0=7.0$ and $10.0$ and a correlated fit to the data can be
performed omitting $aM_0=3.5$ and $4.0$.  These data points are
omitted from all further fits to quantities as a function of $1/M$.

\subsubsection{Meson Mass}
\label{mesonmass}
The heavy-light meson mass, $M_2$, can be calculated both
perturbatively and nonperturbatively. It is important to
find consistency between methods and investigate their range
of validity, not least to confirm that Lorentz invariance can
be restored at this order through a mass shift to the
heavy quark. We calculated $M_2$ in three ways:
\begin{enumerate}
\item Directly, from the dispersion relation. $M_2$ is the mass
which appears in the kinetic term of a non-relativistic meson,
\begin{equation}
E(\vec{p}) = M_1 + \frac{|\vec{p}|^2}{2M_2} + \ldots,
\end{equation}
where $M_1=E_{sim}$. For comparison with later results we define the
heavy quark mass shift as $\Delta_{hl} = M_2{-}M_1$.
We extract $\Delta E= E(\vec{p}) - M_1$ by
fitting the ratio of finite momentum to zero momentum $^1S_0$
correlators for $C_{1l}$ and $C_{2l}$ simultaneously to
multiple exponentials for $|\vec{p}|=1$ and $\sqrt{2}$~(for a fit
involving more than one exponential this is not quite the right
ansatz; however, it should allow most of our excited state
contamination to be removed from $\Delta E$). The next order term in
the dispersion relation, $|\vec{p}|^4/8M_3^3$, is not correctly
reproduced by our $O(1/M_0)$ action and thus $M_3\not=M_2$. However,
for lighter meson masses this term may be significant and we solve for
$M_2$ using
\begin{equation}
\Delta E  = \frac{|\vec{p}|^2}{2M_2} - \frac{|\vec{p}|^4}{8M_2^3},
\label{dispn}
\end{equation}
assuming $M_2-M_3$ is small. Errors on the meson mass are obtained
simply by solving for $M_2$ using $\Delta E \pm 1\sigma$.

%
% plot of fits to finite momentum states. for M0=1.0
%
%
\begin{figure}
\centerline{\ewxy{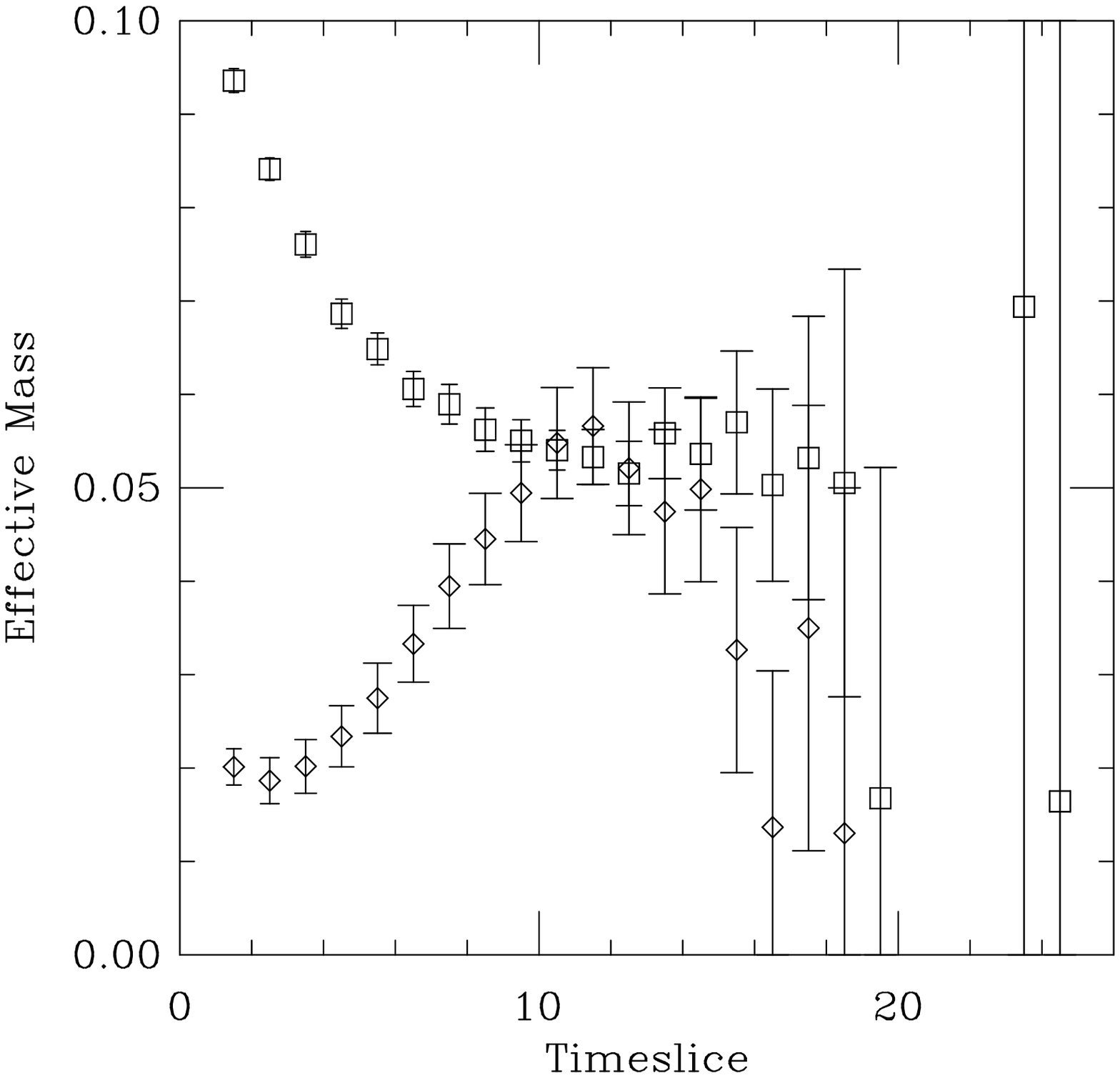}{80mm}
\ewxy{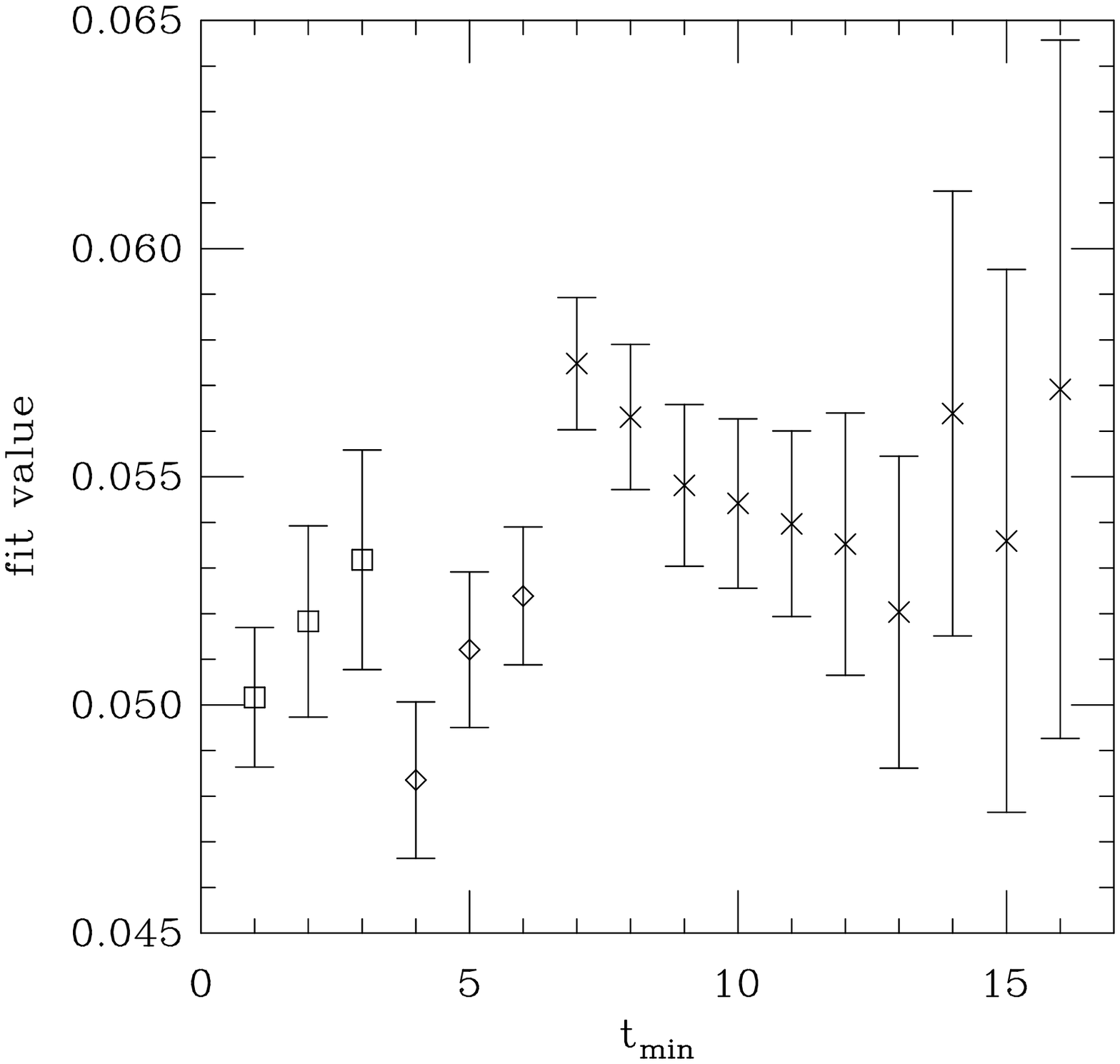}{80mm}
}
\vspace{0.5cm}
\centerline{\ewxy{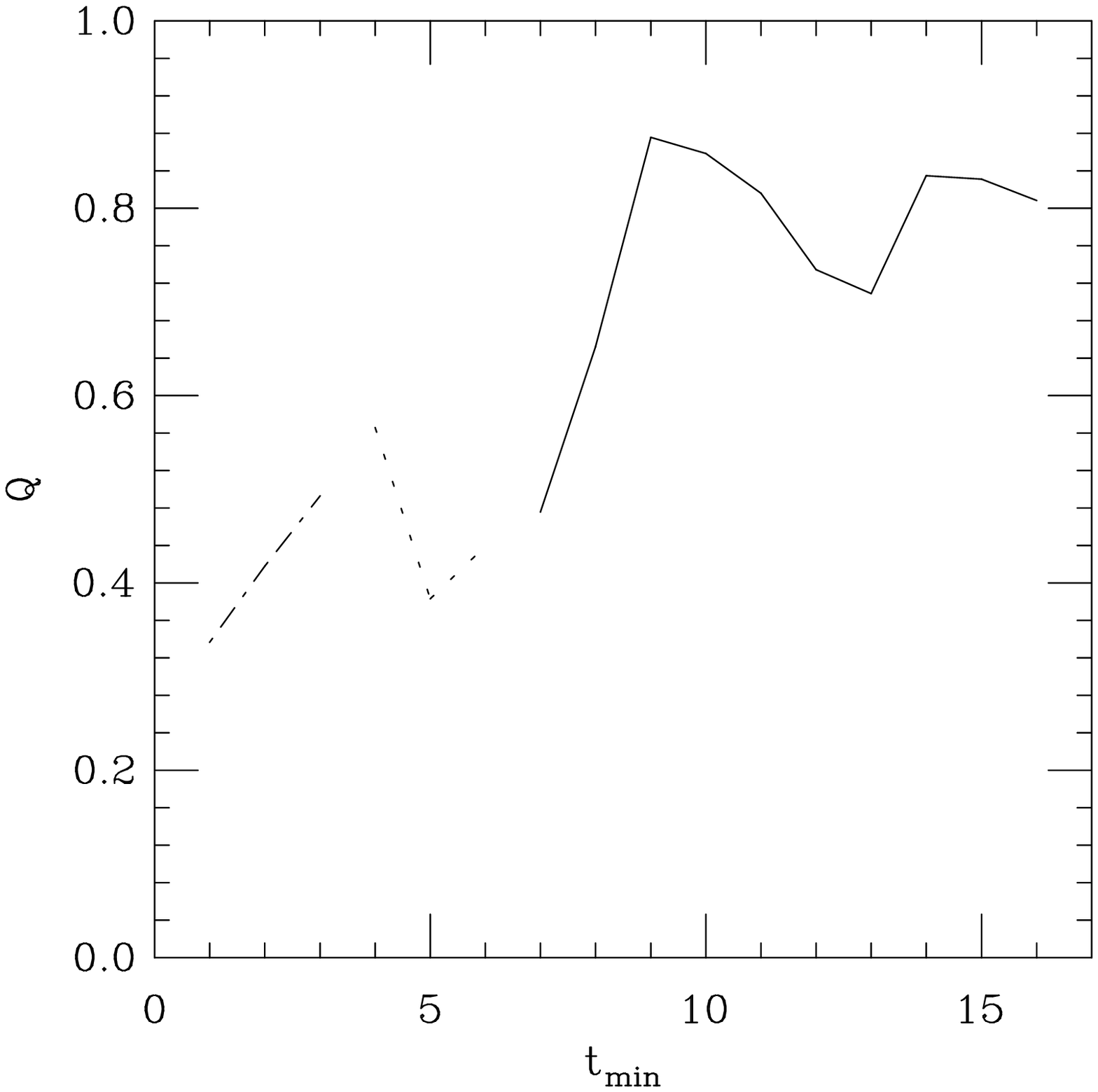}{80mm}
}
\caption{The effective masses of
$C_{1l}(|\vec{p}|=1)/C_{1l}(|\vec{p}|=0)$~(squares) and
$C_{2l}(|\vec{p}|=1)/C_{2l}(|\vec{p}|=0)$~(diamonds) for $^1S_0$ at
$aM_0=1.0$ and $\kappa_l=0.1585$. The fit parameter $\Delta E(\vec{p})$
extracted from 1~(crosses), 2~(diamonds) and 3~(squares) simultaneous
exponential fits and the corresponding values of Q are also shown.
$t_{max}$ is fixed to $20$.}
\label{dispnmom1}
\end{figure}

As an example the results for $^1S_0$ at $aM_0=1.0$ and
$\kappa_l=0.1585$ for $|\vec{p}|=1$ are shown in
figure~\ref{dispnmom1}. The effective mass illustrates a difficulty in
extracting $M_2$ from the dispersion relation; the smearing functions
which are adequate for zero momentum have a reduced overlap with the
ground state at finite momentum~\cite{jim}. The effective mass does
not plateau for $C_{1l}(|\vec{p}|=1)/C_{1l}(|\vec{p}|=0)$ until
$t\approx12$~(compared to $t\approx5$ at zero-momentum), when the
statistical errors are much larger than at earlier timeslices.
However, the onset of the plateau is clear from a comparison of the
effective mass using the excited state smeared source, which has a
very different overlap with the ground and excited states.  The
variation of $\Delta E$ with $t_{min}$ is stable from around
$t_{min}=13$ for $n_{exp}=1$ and consistency is found with the results
obtained with additional exponentials.  A fitting range of $5-20$ with
$n_{exp}=2$ is chosen for $aM_0=0.8$ to $2.0$ and $13-20$ with
$n_{exp}=1$ for $aM_0=2.5$ to $4.0$; the corresponding values of
$\Delta E$ are given in table~\ref{delE}. Finite momentum correlators
were not generated for $aM_0=7.0$ and $10.0$.

%
% Fitting ranges for M2 from hl dispersion relation
% for M0=1.0 p=1
\begin{table}
\begin{center}
\begin{tabular}{|c|c|c|c|c|c|}\hline
$aM_0$ & $n_{exp}$ & fit range & Q & $a\Delta E(\vec{p})$ & $a\Delta
E'(\vec{p})$ \\\hline
0.8 & 2 & 5-20 & 0.6 & 0.060(2) & 0.37(7) \\\hline
1.0 & 2 & 5-20 & 0.5 & 0.051(2) & 0.33(6) \\\hline
1.2 & 2 & 5-20 & 0.4 & 0.045(2) & 0.30(5) \\\hline
1.7 & 2 & 5-20 & 0.2 & 0.034(2) & 0.26(5) \\\hline
2.0 & 2 & 5-20 & 0.2 & 0.029(2) & 0.25(5) \\\hline
2.5 & 1 & 13-20 & 0.2 & 0.025(2) & - \\\hline
3.0 & 1 & 13-20 & 0.2 & 0.021(2) & - \\\hline
3.5 & 1 & 13-20 & 0.1 & 0.019(2) & - \\\hline
4.0 & 1 & 13-20 & 0.1 & 0.016(2) & - \\\hline
\end{tabular}
\caption{Fit parameters and values of
$\Delta E(\vec{p})$ extracted from performing a simultaneous
correlated fit to the ratios $C_{1l}(|\vec{p}|=1)/C_{1l}(|\vec{p}|=0)$
and $C_{2l}(|\vec{p}|=1)/C_{2l}(|\vec{p}|=0)$ for
$\kappa_l=0.1585$.\label{delE}}
\end{center}
\end{table}

As the heavy quark mass increases, $\Delta E$ decreases, and
correspondingly the determination of $M_2$ is statistically more uncertain.
Table~\ref{m2} details the results for $M_2$ from both $|\vec{p}|=1$
and $\sqrt{2}$. Agreement between the meson mass from both momenta
is found to be less than one standard deviation for $aM_0>1.2$, and
confirms the dispersion relation in equation~\ref{dispn}. For
the lighter meson masses the disagreement is less than $2\sigma$
but indicates the increase in significance of the $|\vec{p}|^4/8M_3^3$
contribution and $M_2\not=M_3$. Hence, we take $M_2$ obtained from the lower
value of momentum as more reliable. However, for  meson masses
around the $B$ meson and less the statistical uncertainty in
$M_2$ is becoming prohibitively high.

%
%  M2 from hl dispersion relation
% foR p=2 and p=1
\begin{table}
\begin{center}
\begin{tabular}{|c|c|c|}\hline
$aM_0$ & $aM_2(|\vec{p}|=1)$ & $aM_2(|\vec{p}|=\protect\sqrt{2})$ \\\hline
0.8 & 1.25(5) & 1.16(4)\\\hline
1.0 & 1.49(6) & 1.40(4)\\\hline
1.2 & 1.69(8) & 1.61(6)\\\hline
1.7 & 2.25(14)& 2.20(10)\\\hline
2.0 & 2.64(20) & 2.63(20)\\\hline
2.5 & 3.07(27)& 3.19(29)\\\hline
3.0 & 3.66(39)& 3.74(41)\\\hline
3.5 & 4.05(48)& 4.27(53)\\\hline
4.0 & 4.81(79)& 4.96(69)\\\hline
\end{tabular}
\caption{$M_2$ obtained from $a|\vec{p}|=1$ and $a|\vec{p}|=\protect\sqrt{2}$
for $\kappa_l=0.1585$.\label{m2}}
\end{center}
\end{table}

\item Perturbatively. Since the inequality of $M_1$ and $M_2$ is due to
the omission of the constant term in the NRQCD action, Lorentz invariance
can be restored by a shift to the heavy quark mass,
\begin{equation}
M_2 - M_1 = Z_mM_0 -E_0.
\end{equation}
The mass renormalisation, $Z_m$, energy of a heavy quark in NRQCD,
$E_0$, and the corresponding mass shifts, $\Delta_{pert}=M_2-M_1$, as
calculated in perturbation theory are shown in
table~\ref{pert}~\cite{colin}.  The characteristic scales, $q^*_{Z_m}$
and $q^*_{E_0}$, of $Z_m$ and $E_0$ respectively are also given and
these indicate the reliability of the perturbative result; if the
characteristic scale is very soft perturbation theory is not a
reliable method of calculating the quantity. For $aM_0$ around $2.0$
and $aM_0=7.0$ and $10.0$ the characteristic scales are hard and
perturbation theory works well.  However, as $aM_0\rightarrow 5$,
$q^*_{Z_m}$ becomes softer.  This defect is due to the method of
calculating the characteristic scale and does not relate to any
physical effect. We use the suggestion of Morningstar~\cite{colin} to
guess a characteristic scale $\bar{q}$ to smooth over the defect. The
result $\Delta_{pert}^\dagger$ is also shown in the table. We also
used this method of estimating the perturbative result for the lighter
heavy quark masses where $q^*$ is soft because NRQCD breaks down.

The mass shift is also shown in table~\ref{compdiff} along with an
error estimating the higher loop corrections omitted in this
calculation.  The uncertainty associated with $E_0$ is computed as
detailed in section~\ref{hqsym}. For the quantity $M_0Z_m$ we
estimate the two loop contributions to be roughly $(M_0(Z_m-1))^2$;
this is a more conservative choice than the naive estimate of
$M_0(Z_m-1)^2$.

% table of perturbative results
%
\begin{table}
\begin{center}
\begin{tabular}{|c|c|c|c|c|c||c|c|}\hline
$aM_0$ & n & $aq^*_{Z_m}$ & $Z_m$ & $aq^*_{E_0}$ & $aE_0$ & $a\Delta_{pert}$ &
$a\Delta_{pert}^{\dagger}$\\\hline
0.8 & 4 &0.04 & --     & 0.03 & --     & -     & 0.886 \\\hline
1.0 & 4 &0.19 & 0.668 & 0.14 & --     & -     & 1.063 \\\hline
1.2 & 3 &0.31 & 1.755 & 0.27 & 1.481 & 0.625 & 1.254 \\\hline
1.7 & 2 &0.49 & 1.253 & 0.48 & 0.402 & 1.728 &   -   \\\hline
2.0 & 2 &0.54 & 1.197 & 0.55 & 0.372 & 2.022 &   -   \\\hline
2.5 & 2 &0.56 & 1.146 & 0.62 & 0.359 & 2.506 &   -   \\\hline
3.0 & 2 &0.52 & 1.119 & 0.68 & 0.349 & 3.008 &   -   \\\hline
3.5 & 1 &0.34 & 1.155 & 0.71 & 0.348 & 3.695 &   -   \\\hline
4.0 & 1 &0.23 & -- & 0.74 & 0.346 & - & 3.944 \\\hline
7.0 & 1 &25.10 & 0.994 &0.79 & 0.350 & 6.608 &   -   \\\hline
10.0 & 1 &4.11 & 0.978 &0.81 & 0.351 & 9.430 &   -   \\\hline
\end{tabular}
\caption{Perturbative results for $M_2-M_1$. $\Delta_{pert}^{\dagger}$
is obtained using $\alpha_V(q^*,\bar{q})$~\protect\cite{colin}, where
$\bar{q} = 0.6$ and $0.8$, and the resulting two values for the mass
shift are averaged. For $aM_0=0.8$, $1.0$ and $4.0$, `--' indicates
$\alpha_S$ is ill-defined at the corresponding value of $q^*_{Z_m}$ or
$q^*_{E_0}$.\label{pert}}
\end{center}
\end{table}

\item From the dispersion relation of the heavy-heavy meson at
the same $M_0$. As mentioned above $M_2-M_1$ is due to a mass shift of
the heavy quark and therefore the shift for a heavy-light meson should
be half that for a heavy-heavy meson with the same bare heavy quark
mass:
\begin{equation}
\Delta_{hh} = \frac{1}{2}[ M_2(hh) - M_1(hh)]
\end{equation}
We used the NRQCD action including all relativistic corrections
to $O(Mv^4)$ and the same method of extracting $M_2$. The results are
shown in table~\ref{compdiff}.  As mentioned in Section~\ref{simdet}
discretisation errors in the heavy quark momentum become significant
around $aM_0\sim5$ and thus the results for $aM_0=7.0$ and $10.0$ are
not reliable.
\end{enumerate}

All three methods of computing the mass shift are compared in
table~\ref{compdiff}. There is reasonable agreement for $aM_0\leq4.0$,
while for the heavier quark masses agreement is not expected due to
the large systematic errors associated with $\Delta_{hh}$. The
estimate of the perturbative result for $aM_0=0.8-1.2$ and $4.0$
appears to work well. Furthermore, the agreement between obtaining the
heavy-light meson mass directly and methods~2 and~3 which assume a
mass shift can be applied to the heavy quark mass confirms that
Lorentz invariance can be restored in this way.  Since using the
heavy-heavy dispersion relation is the most accurate method we use
$\Delta_{hh}$ for $aM_0\leq4.0$ and the perturbative results for the
two heaviest quark masses.  Using $a^{-1}=1.8{-}2.4$ we find
$aM_0=2.4{-}1.7$ corresponds to the bare $b$ quark mass.

%
% table of M2-M1 for dispersion relation HL and HH, and pert theory.
%
\begin{table}
\begin{center}
\begin{tabular}{|c|c|c|c|}\hline
$aM_0$ & $a\Delta_{hh}$ & $a\Delta_{pert}$ & $a\Delta_{hl}$ \\\hline
0.8 &   0.89(2) & 0.89(34) & 0.78(5) \\\hline
1.0 &   1.09(3) & 1.06(30) & 1.00(6) \\\hline
1.2 &   1.27(2) & 1.25(31) & 1.20(8) \\\hline
1.7 &   1.76(2) & 1.73(35) & 1.74(14) \\\hline
2.0 &   2.07(2) & 2.02(30) & 2.13(20) \\\hline
2.5 &   2.58(2) & 2.51(26) & 2.56(27) \\\hline
3.0 &   3.10(3) & 3.01(25) & 3.14(39) \\\hline
3.5 &   3.64(3) & 3.70(31) & 3.53(48) \\\hline
4.0 &   4.12(4) & 3.94(20) & 4.29(79) \\\hline
7.0 &   8.10(10) & 6.61(12) & - \\\hline
10.0 &  13.80(25) & 9.43(22) & -  \\\hline
\end{tabular}
\caption{Heavy quark mass shifts determined from the
heavy-light dispersion relation, $\Delta_{hl}$, perturbation theory,
$\Delta_{pert}$ and the heavy-heavy dispersion relation
$\Delta_{hh}$.\label{compdiff}}
\end{center}
\end{table}

\subsection{Mass Splittings}
\label{massplit}
\subsubsection{Hyperfine Splitting}
\begin{figure}
\centerline{\ewxy{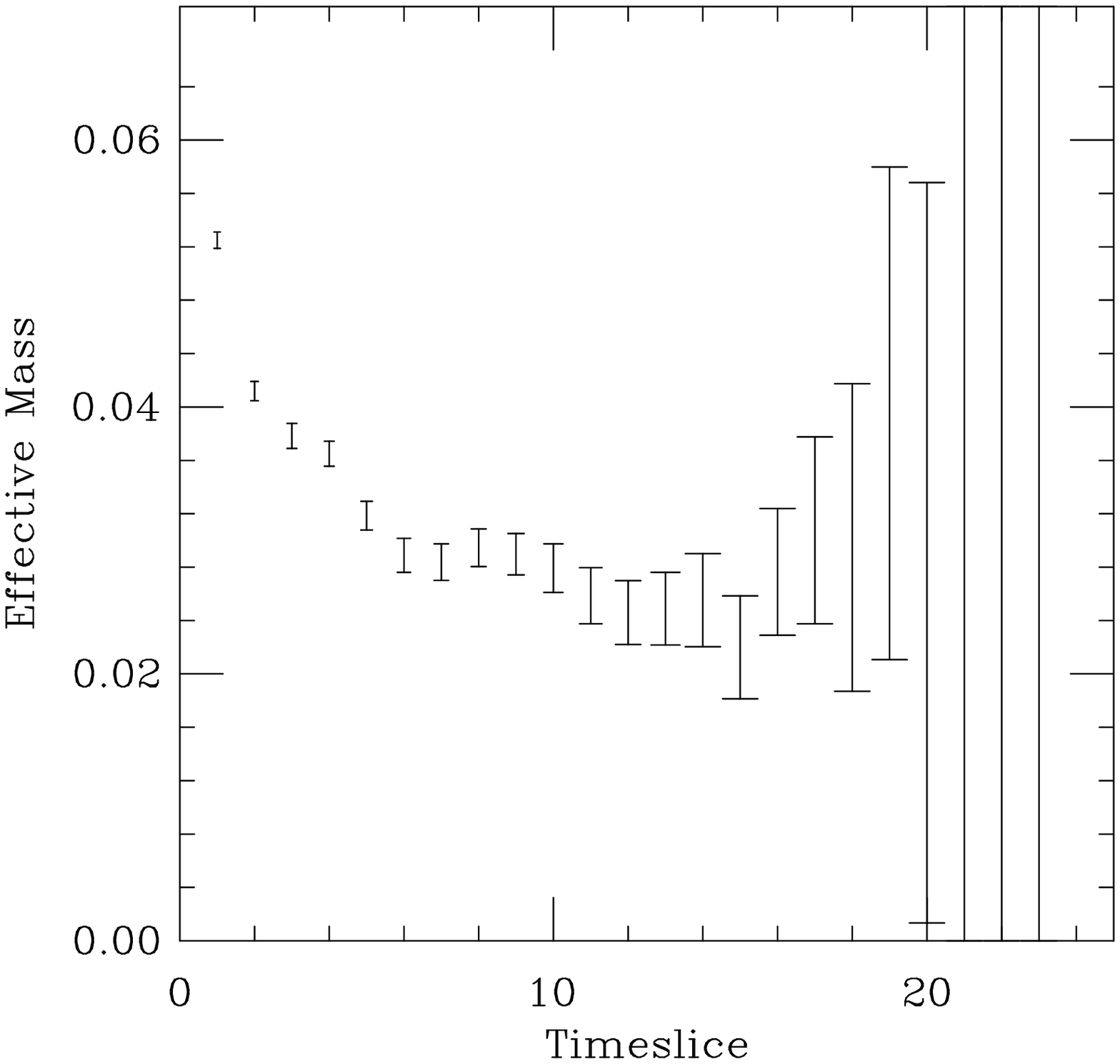}{80mm}
\ewxy{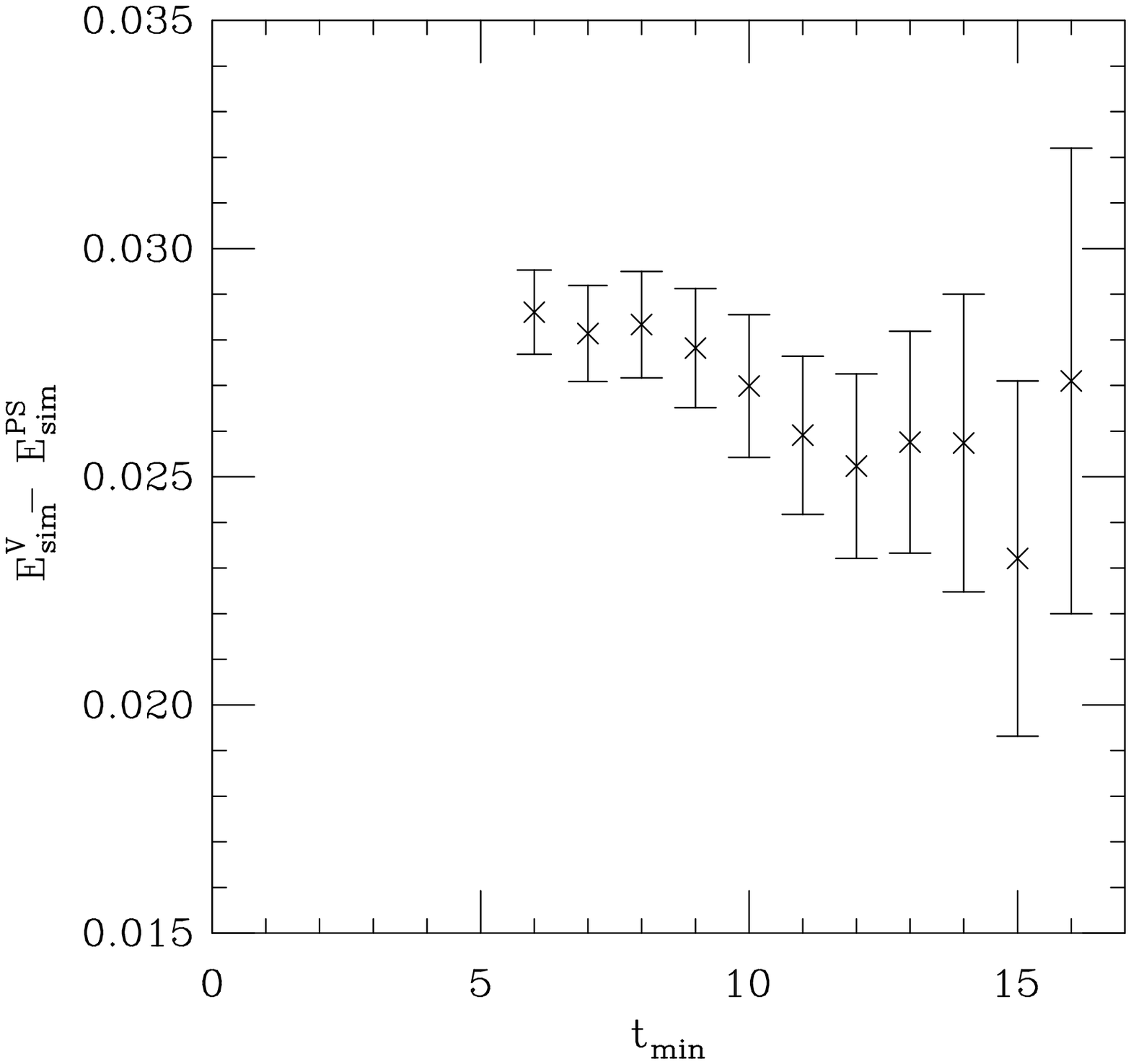}{80mm}}
\vspace{0.5cm}
\centerline{\ewxy{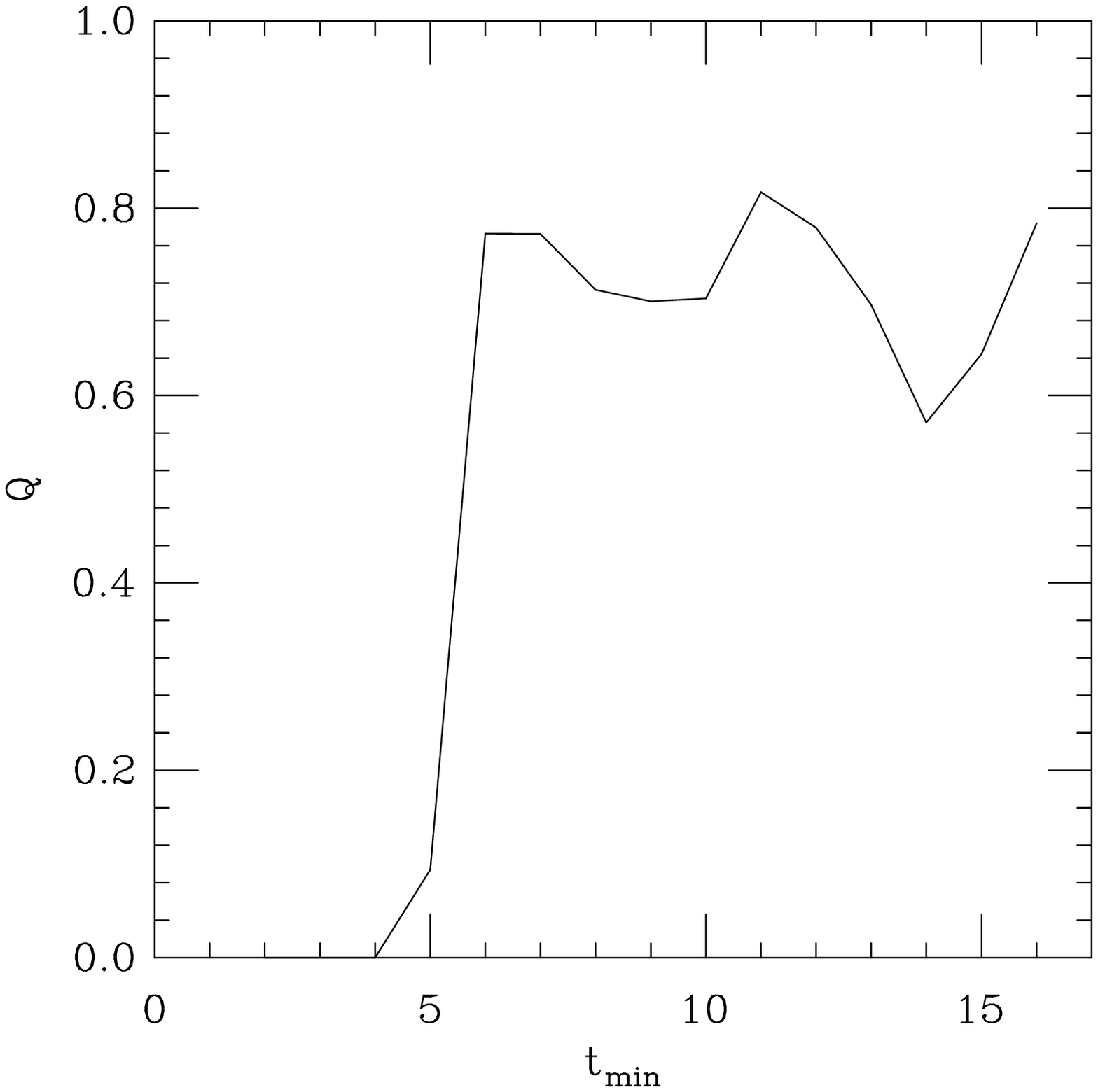}{80mm}}
\caption{Shown are the effective mass of the ratio of $^3S_1$ to
$^1S_0$ $C_{1l}$ correlators for $aM_0=1.0$ and $\kappa_l=0.1585$ and
the mass splitting extracted using $n_{exp}=1$ as a function of
$t_{min}$.  The corresponding values of Q are also shown. $t_{max}$ is
fixed at $20$.
\label{meff_hyp}
}
\end{figure}

We extracted the hyperfine splitting by performing a single
exponential fit to the jacknife ratio of the $^3S_1$ and $^1S_0$
$C_{1l}$ correlators. The effective mass for $aM_0=1.0$ and
$\kappa_l=0.1585$ is shown in figure~\ref{meff_hyp} along with the
hyperfine splitting extracted as $t_{min}$ is varied.  The plateau
begins at $t_{min}\gtaeq 11$, and this is found to be the case for all
$aM_0$.  The splittings extracted using a fitting range of $11-20$ are
shown in table~\ref{tab:e_sim}.

%
% Fitting ranges for hyperfine splitting
%
% k_l = 0.1585
%
%      all M0  11-20
%
%
%
%
% effective mass and tmin/Q plot for hyperfine M0=1.0
%
%

\begin{figure}
\centerline{\ewxy{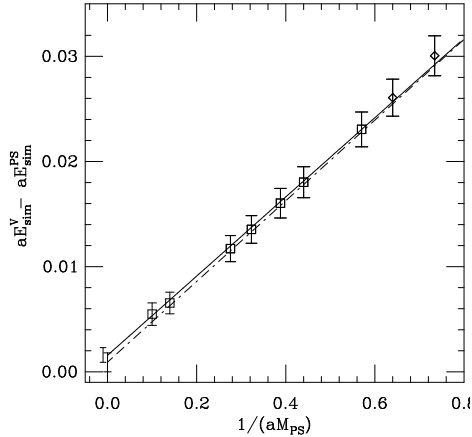}{80mm}}
\caption{The hyperfine splitting vs $1/(aM_{PS})$
for $\kappa_l=0.1585$. The solid line indicates a linear fit to the
data shown as squares.  The dashed line indicates a linear fit to the
data~(squares) omitting the heaviest two points.
\label{fig:hyp}
}
\end{figure}

Figure~\ref{fig:hyp} presents the hyperfine splitting as a function
of $1/M_{PS}$, where $M_{PS}$ is the $^1S_0$ meson mass.
Performing correlated fits to $E_{sim}^V-E_{sim}^{PS}$ using
a fit function of the form in equation~\ref{fitfunc} and the same fitting
procedure we found linear dependence on $1/M_{PS}$ around
the $B$ meson and down to $aM_0=1.2$. The extrapolation to the static
limit is consistent, within $2\sigma$, with zero; the uncertainty in the
meson mass has not been included in the fit. This is largest for
the heaviest two points. If these are omitted the extrapolation
to infinite heavy quark mass is closer to zero. The
coefficients extracted were stable over the fitting range
and we find
\begin{equation}
aC_0 = 0.016(7) \hspace{1cm} a^2C_1 = 0.038(2).
\end{equation}
where the coefficient of the slope corresponds to $\frac{a^2}{2}
<c_B\sigma\cdot B >^{tadpole-improved}_{renorm}$.  This
is consistent with $O(a^2\Lambda_{QCD}^2)\sim 0.04$ expected from
naive power counting. In addition, agreement is found with our results
on quenched configurations, $a^2C_1 = 0.043(6)$~\cite{arifa}. Converting
to physical units,
\begin{eqnarray}
\hspace{-0.4cm}<c_B\sigma\cdot B >^{tadpole-improved}_{renorm}
& = & 0.24-0.44 \mbox{ GeV}^2\hspace{1cm}\beta^{(n_f=2)}=5.6\\
& = & 0.362\pm50\pm36\mbox{ GeV}^2\hspace{0.2cm}\beta^{(n_f=0)}=6.0
\end{eqnarray}
where for the quenched results the first error is statistical and the
second is due to the uncertainty in $a^{-1}$.  This compares reasonably
well with an explicit measurement of the naive matrix element
obtained by Ewing~et~al~\cite{staticukqcd} on quenched
configurations using tree-level Clover fermions at $\kappa_l=\kappa_s$ and
$\beta^{(n_f=0)}=6.2$: \mbox{$Z_\sigma<\sigma\cdot B
>=0.465\mbox{\err{25}{25}}\mbox{(stat)}\mbox{\err{65}{60}}\mbox{(syst)
GeV}^2$}, where $Z_\sigma$ is the renormalisation factor required in
the static theory.
%The approximate $20\%$ difference
%between the NRQCD and static quenched results may be accounted for by
%the one loop corrections to $c_B$.

We find no dependence of the splitting on the light quark mass, and
this is also found in experiment: $B^*{-}B = 46(1)$~MeV and
$B^*_s{-}B_s=47(3)$~MeV. We find a splitting of $25{-}40$~MeV; the
large uncertainty in the splitting reflects the particular sensitivity
to the lattice spacing arising from the $1/M_{PS}$ dependence. While
the hyperfine splitting is uncertain it does appear to lie below the
experimental value. This may be due to residual quenching effects, the
correct number of quark loops being $n_f\sim3-4$. The splitting is
proportional to the wavefunction at the origin and is dominated by a
harder physical scale than those which characterise the quantities
used to calculate the lattice spacing. Hence, it is a quantity which
is sensitive to quenching effects.
%
%
%
% hyperfine plot splitting vs 1/M
%
\subsubsection{P-S Splitting}
The first measurements of the P-states of the $B$ mesons have recently
been made at DELPHI and OPAL~\cite{delphi,opal}. This provides a
further test of the lattice simulation of $B$ mesons through comparison
with experiment, and also a ideal quantity with which to extract a
value for the lattice spacing from the $B$ system itself; the $P-S$
splitting is expected to be very weakly dependent on the heavy quark
mass. So far the spin splittings of the P-states have not been clearly
resolved and the experiments above find
$B^{**}{-}\bar{B}=419(25)$~MeV, where a cross-section weighted mean is
used for $B^{**}$.  A more detailed discussion of the comparison of
our quenched results to experiment is given in~\cite{arifa}.

\begin{figure}
\centerline{\ewxy{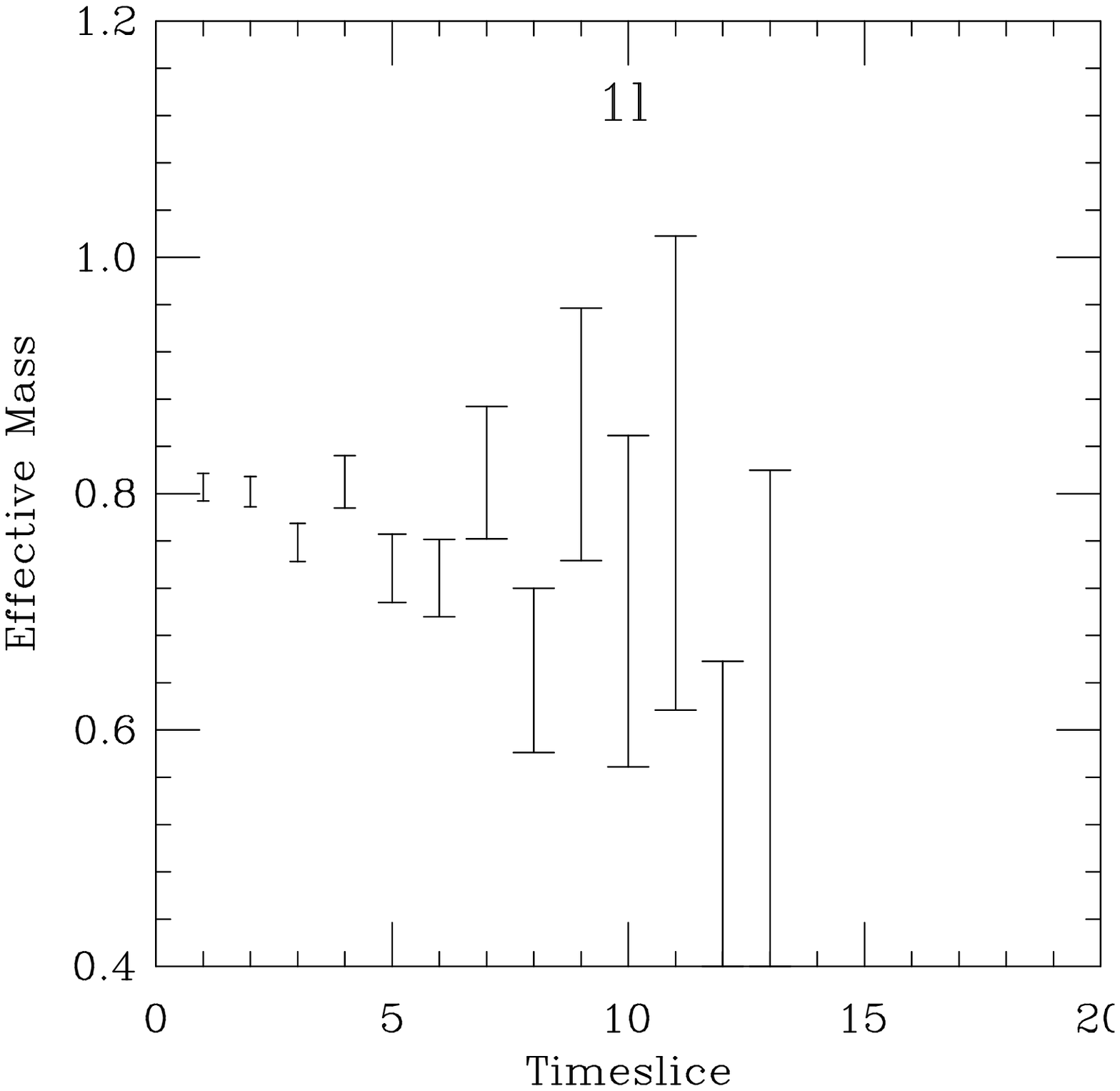}{80mm}
\ewxy{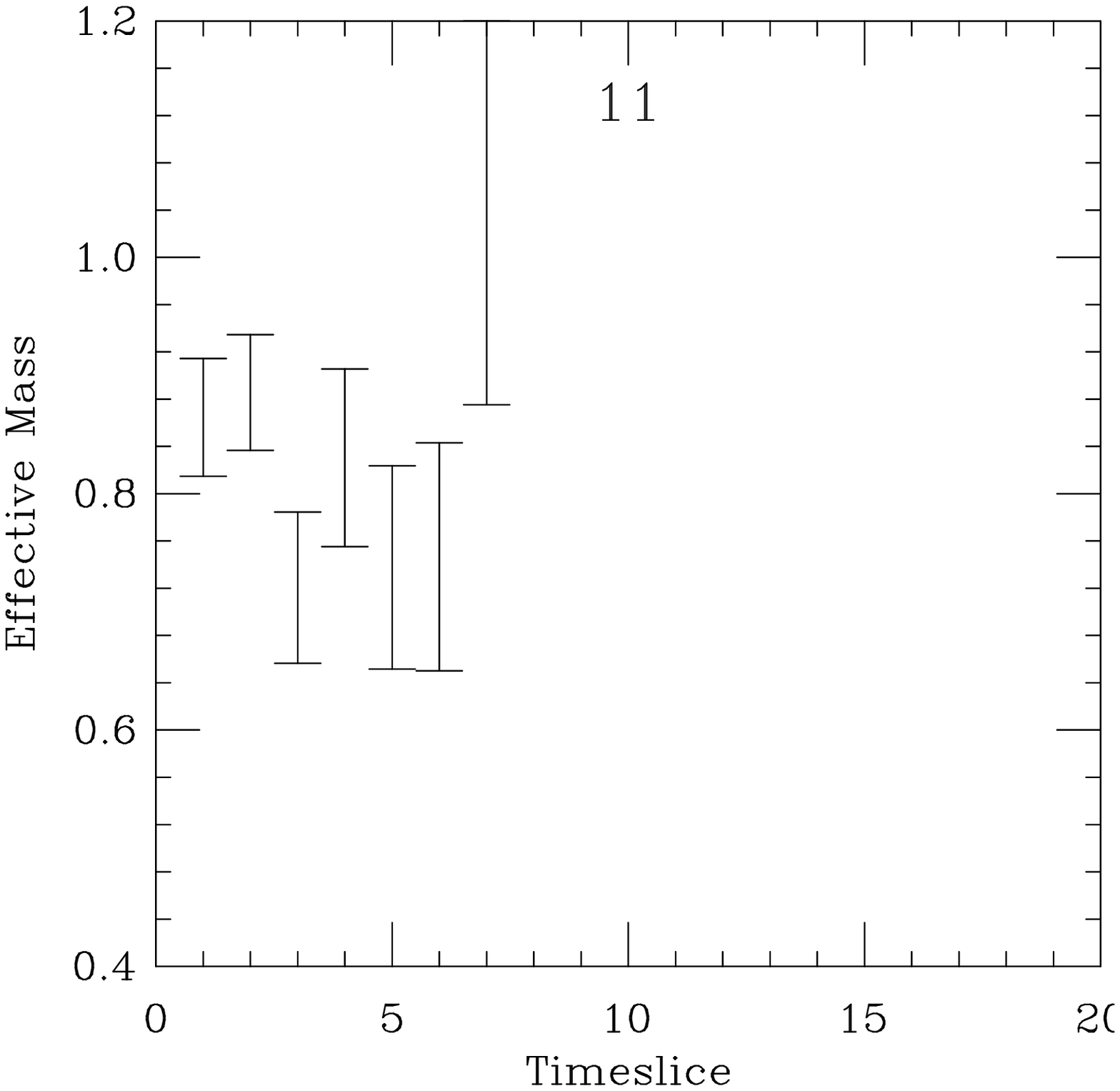}{80mm}}
\centerline{\ewxy{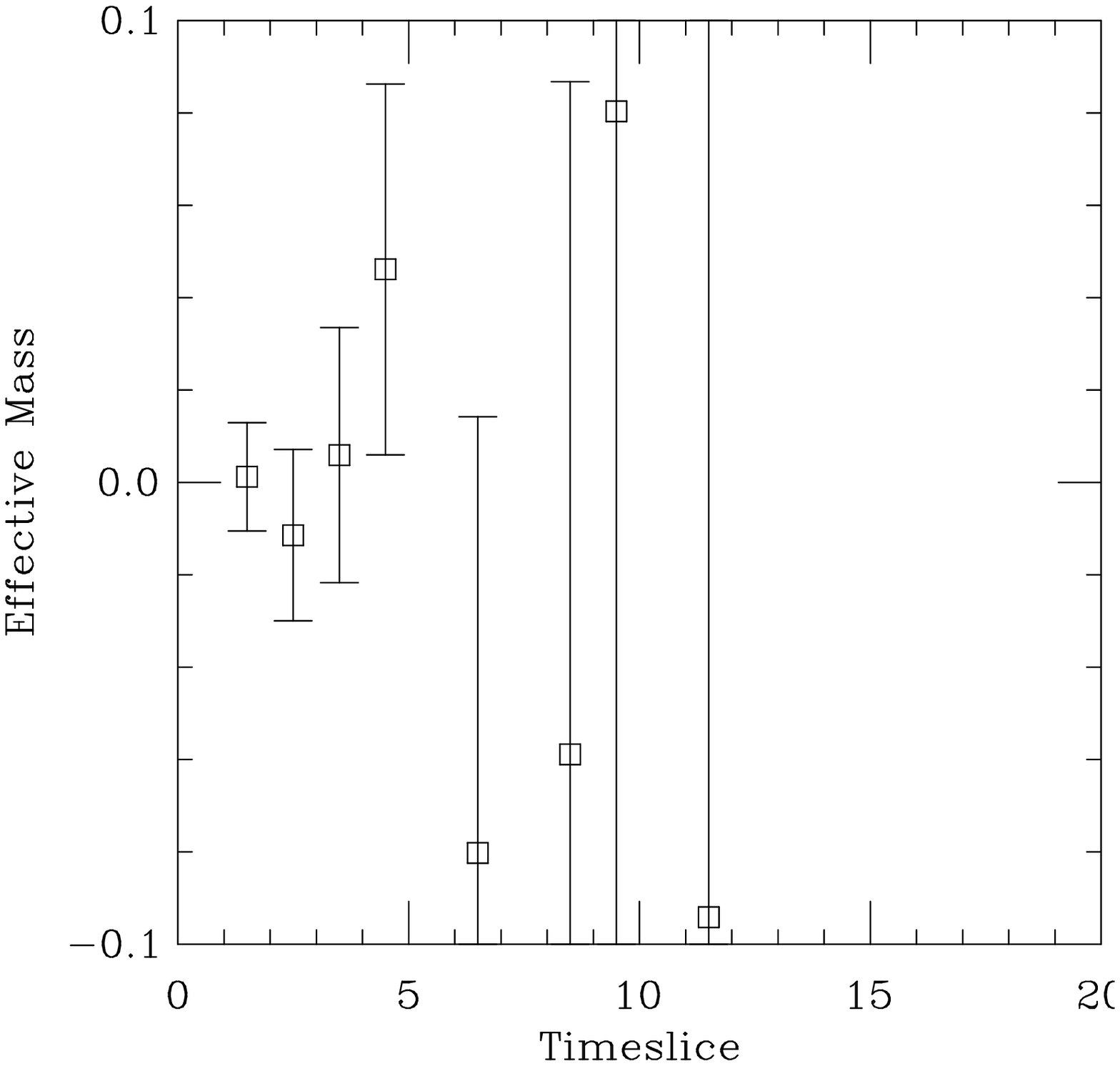}{80mm}
\ewxy{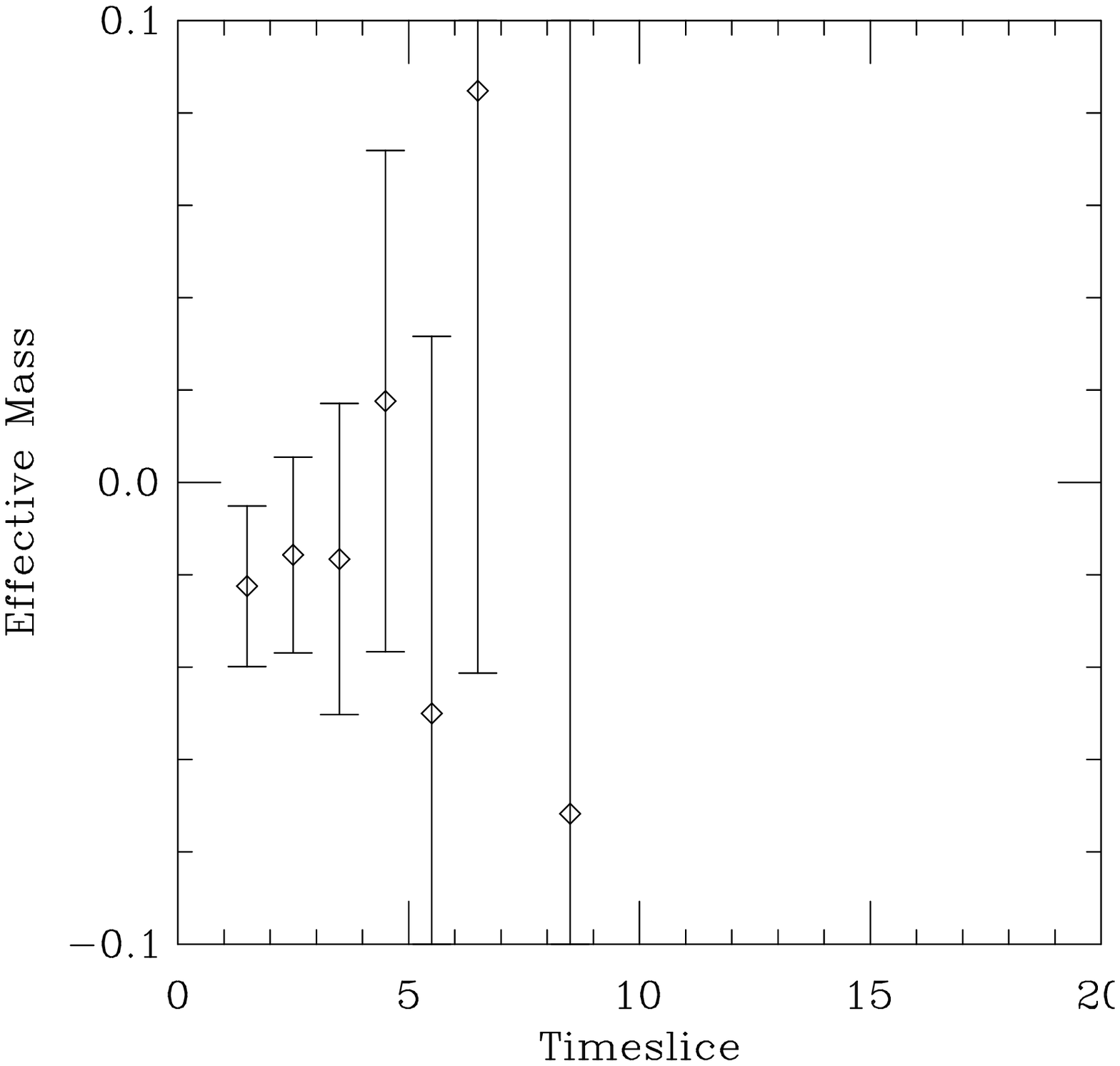}{80mm}}
\caption{The effective masses of the $^1P_1$ $C_{1l}$ and $C_{11}$ correlators,
and the ratios of the $^3P_1$ and $^1P_1$ $C_{1l}$
correlators~(squares), and $^3P_0$ and $^1P_1$ $C_{1l}$
correlators~(diamonds).}
\label{pstate}
\end{figure}

We obtain a signal for the $^1P_1$, $^3P_0$ and $^3P_1$ correlators;
the $^3P_2$ is too noisy to extract an estimate of the mass. However,
the spin-splittings between the P-states are consistent with zero.
This is shown in figure~\ref{pstate} which displays the effective
masses of the $^1P_1$ $C_{1l}$ and $C_{11}$ correlators and the
$^3P_1{-} ^1P_1$ and $^3P_0{-} ^1P_1$ splittings for $aM_0=1.0$ and
$\kappa_l=0.1585$.  The spin-orbit interaction of the light quark,
which we expect to split the $j_l=3/2$ and $j_l=1/2$ states by
$50-100$~MeV~\cite{dunno}, is below the level of our statistical errors.
A multi-smearing analysis is not possible for the P-states, the first
excited state smearing function used for the P states~(see
equation~\ref{firstp}) had almost identical overlap with the ground
and excited states as the ground state smearing function.

%
% effective mass plot of 1p1 1l and 11 for M0 = 1.0
%
%

\begin{figure}
\centerline{\ewxy{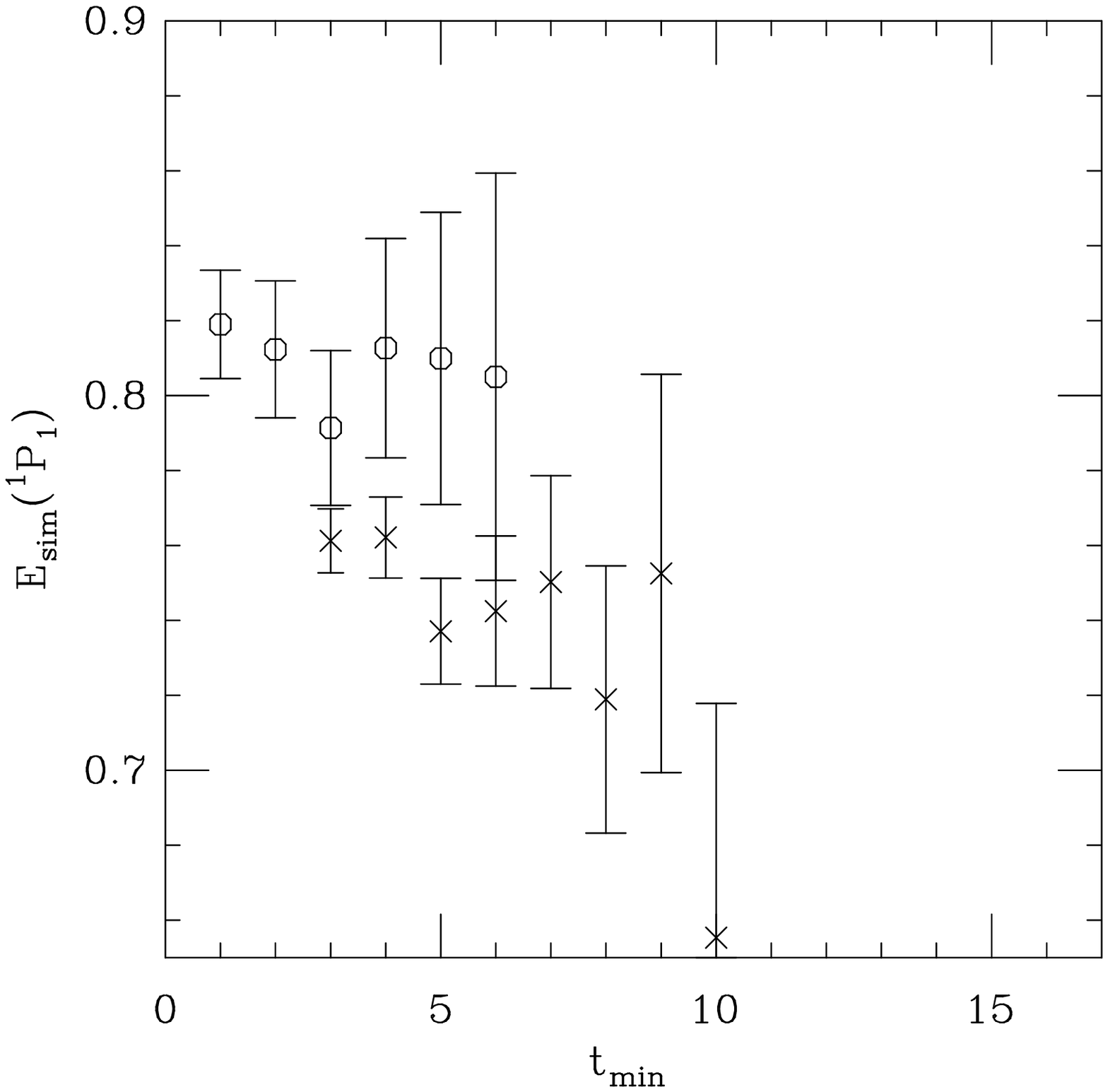}{80mm}
\ewxy{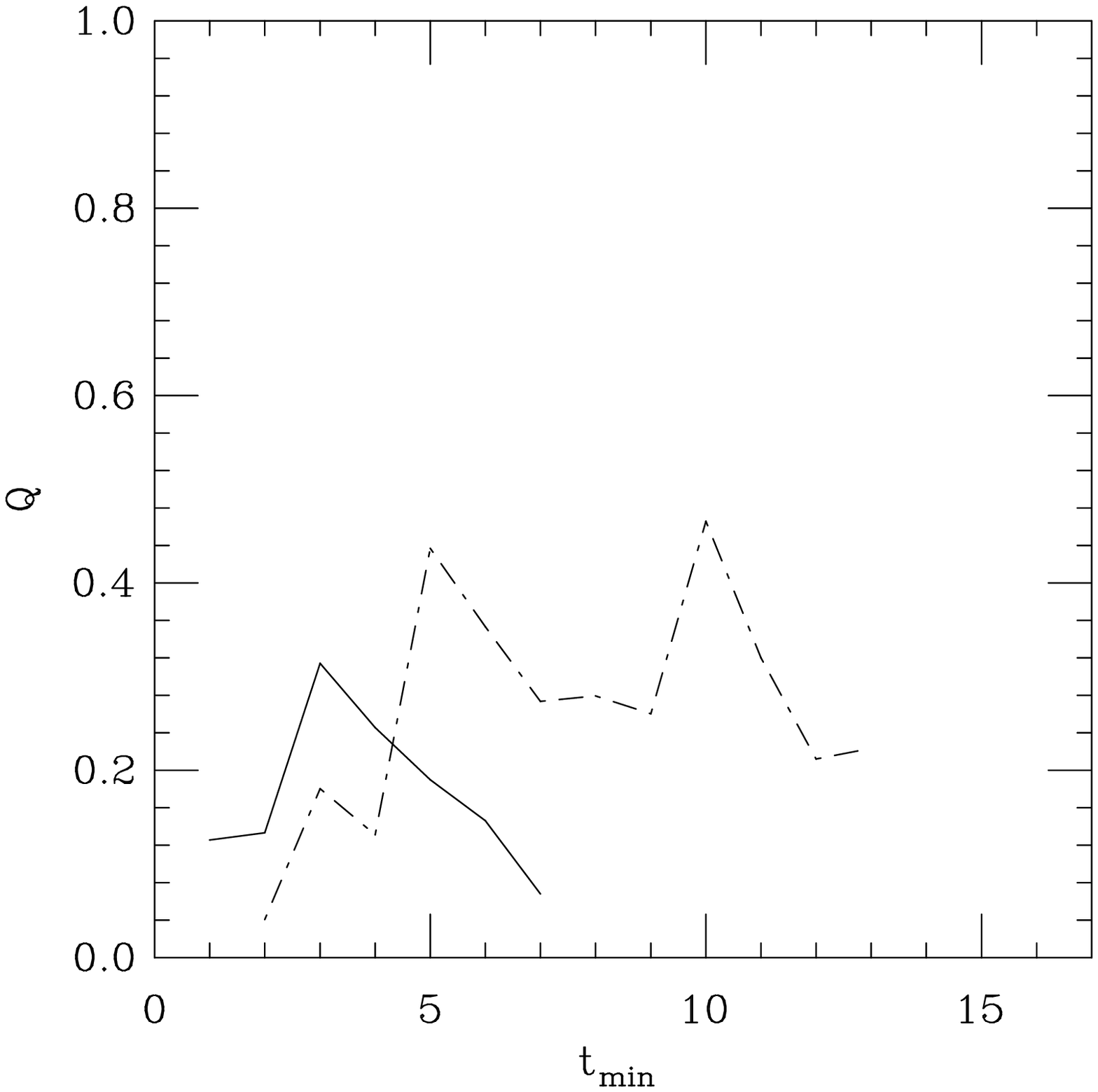}{80mm}}
\caption{The $^1P_1$ energy extracted from a $n_{exp}=1$ fit to
$C_{1l}$~(crosses) and $C_{11}$~(circles) correlators separately and
the corresponding values of Q~(dashed and solid line for the fits to
$C_{1l}$ and $C_{11}$ respectively) for $aM_0=1.0$ and $\kappa_l=0.1585$.
$t_{max}$ is fixed to $12$ for $C_{1l}$ and $10$ for $C_{11}$.}
\label{pfit}
\end{figure}

As we are unable to resolve spin-splittings, we considered only the
$^1P_1$ state and performed $n_{exp}=1$ fits to $C_{1l}$ and $C_{11}$
separately. A simultaneous fit to both correlators was not possible
since the $C_{11}$ correlator becomes dominated by noise as $C_{1l}$
plateaus. The results, given in figure~\ref{pfit}, show a fairly
significant discrepancy between the energies, stable in $t_{min}$,
from the two fits. Unusually, the values from $C_{11}$ lie above those
for $C_{1l}$, which does not approach a plateau from below. Since
noise dominates the $^1P_1$ $C_{11}$ correlator around $t\sim 10$ it
is possible that this correlator has not plateaued and the
results from the $C_{1l}$ correlator, which can be checked over a
larger range of $t_{min}$, are more reliable.

\begin{figure}
\centerline{\ewxy{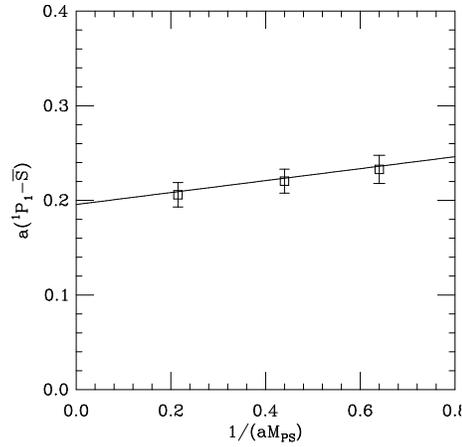}{80mm}}
\caption{The $a(^1P_1-\bar{S})$ splitting obtained using fits to the $^1P_1$
$C_{1l}$ correlator vs $1/(aM_{PS})$
for $\kappa_l=0.1585$.}
\label{splitfig}
\end{figure}

In order to investigate the dependence of the $P{-}S$ splitting on the
heavy quark mass we calculated the splitting at $aM_0=1.0$, $1.7$ and
$4.0$. From table~\ref{splitfit} the $^1P_1-\bar{S}$ splittings, where
$\bar{S}$ is the spin-averaged S-state, are $O(\Lambda_{QCD})$ as
expected for the excitation of a light quark and only mildly dependent
on the heavy quark mass. The results are plotted against $1/M_{PS}$ in
figure~\ref{splitfig}. Considering the discrepancy of $\sim3\sigma$
between the fits for $C_{1l}$ and $C_{11}$ the dependence on $M_0$ is not
significant. We find no dependence of $^1P_1-\bar{S}$ on the light quark
mass and using $aM_0=1.7$ we obtain a splitting of $400-700$~MeV.  The
characteristic scale for the $^1P_1-\bar{S}$ splitting is expected to
be similar to that for light spectroscopy. However, using the
experimental result to determine the lattice spacing,
$a^{-1}=1.6(5)$~GeV, where the uncertainty due to fitting is included
in the error. With such a large error this determination of the scale
is not particularly useful and is consistent with all the other
estimates of $a^{-1}$.  An improvement in the smearing functions and a
resolution of the spin-splittings both in the simulation and in
experiment is needed in order to provide a useful determination of the
lattice spacing.
% table of Pstates for 1s0 for 1.0,1.7,4.0
\begin{table}
\begin{center}
\begin{tabular}{|c|c|c|c|}\hline
$aM_0$ & $C_{1l}$ & $C_{11}$ & $a(^1P_1-\bar{S})$ \\\hline
1.0  &  0.74(1)  & 0.81(2) & 0.28(6) \\\hline   %0.23(1) - 1l  % 31(2) 11
1.7  &  0.74(1)  & 0.80(2)  & 0.26(5) \\\hline  %0.22(1)       % 29(2) 11
4.0  &  0.73(1) & 0.80(2) & 0.25(5) \\\hline    %0.21(1) - 1l  % 27(2) 11
\end{tabular}
\caption{Energy of the $^1P_1$ state obtained from $n_{exp}=1$ fits
to $C_{1l}$ using the fitting range $5-15$ and $C_{11}$ using
$3-10$ for $\kappa_l=0.1585$. Also shown is the $^1P_1-\bar{S}$
splitting, where the error bars take into account the uncertainty due
to fitting.\label{splitfit}}
\end{center}
\end{table}
%
%
% final fitting ranges chosen
%
% 4-16 for M0=1.0,2.0,4.0 using 11
%

%
\subsubsection{$B_s{-}B_d$ Splitting}
\begin{figure}
\centerline{\ewxy{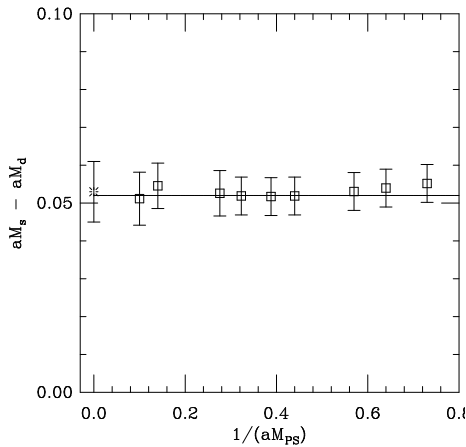}{80mm}}
\caption{The $aM_s{-}aM_d$ splitting vs $1/(aM_{PS})$,
where $\kappa_s=0.1577$ is obtained using the ratio
$M_\phi/M_\rho$. The burst indicates the static result which is
not included in the fit.}
\label{sdmassdiff}
\end{figure}

%
% Fitting range for 0.1600 1s0?
%
%
% same as for k_l = 0.1585
% 1s0 - 3-18
% 3s1 - 4-20
% through the removal of the spin-dependence by taking the
% spin-average
Another quantity of interest is the $B_s{-}B_d$ splitting, which is
expected to be very weakly dependent on the heavy quark mass at the
$O(m_s/M)$ level. In addition, as the splitting is the difference of
the $B_s$ and $B_d$ binding energies it is expected to be $O(m_s)\sim
100{-}300$~MeV.  Experimentally, $B_s{-}B_d=98(6)$~MeV.  In order to
extract the splitting we chirally extrapolated $E_{sim}^{PS}$ using
the energies from the data at $\kappa_l=0.1600$ which are extracted
using the $n_{exp}=1$ fits to $C_{11}$ correlators with the fitting
range $3-20$.  Figure~\ref{sdmassdiff} presents our results for the
splitting, which are consistent with no heavy quark mass
dependence. Converting to physical units we obtain
$B_s{-}B_d=90-130$~MeV; consistent with experiment within the large
systematic uncertainties arising from both $\kappa_s$ and $a^{-1}$.
%

% s-d B splittings.
%  0.8  0.055(4)
%  1.0  0.053(4)
%  1.2  0.052(4)
%  1.7  0.051(5)
%  2.0  0.050(5)
%  2.5  0.050(5)
%  3.0  0.051(5)
%  3.5  0.051(5)
%  4.0  0.051(5)
%  7.0  0.052(6)
%  10.0 0.050(6)
%

\section{Conclusions}
\label{conc}
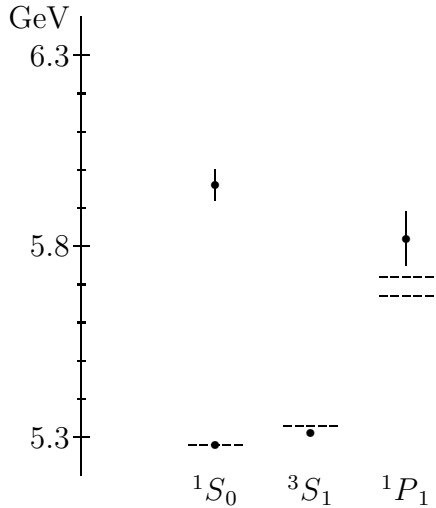
\begin{figure}
\begin{center}
\setlength{\unitlength}{.02in}
\begin{picture}(80,140)(0,930)
% axis
\put(15,940){\line(0,1){120}}
\multiput(13,950)(0,50){3}{\line(1,0){4}}
\multiput(14,950)(0,10){10}{\line(1,0){2}}
\put(12,950){\makebox(0,0)[r]{5.3}}
\put(12,1000){\makebox(0,0)[r]{5.8}}
\put(12,1050){\makebox(0,0)[r]{6.3}}
\put(12,1060){\makebox(0,0)[r]{GeV}}

\put(50,940){\makebox(0,0)[t]{${^1S}_0$}}
\multiput(43,948)(3,0){5}{\line(1,0){2}}
\put(50,948){\circle*{2}}

\put(50,1016){\circle*{2}}
\put(50,1016){\line(0,1){4}}
\put(50,1016){\line(0,-1){4}}

\put(75,940){\makebox(0,0)[t]{${^3S}_1$}}

\multiput(68,953)(3,0){5}{\line(1,0){2}}
\put(75,951){\circle*{2}}

\put(100,940){\makebox(0,0)[t]{${^1P}_1$}}
\multiput(93,992)(3,0){5}{\line(1,0){2}}
\multiput(93,987)(3,0){5}{\line(1,0){2}}
\put(100,1002){\circle*{2}}
\put(100,1002){\line(0,1){7}}
\put(100,1002){\line(0,-1){7}}

\put(125,990){\circle{3}}
\put(125,990){\line(0,1){7}}
\put(125,990){\line(0,-1){7}}

\end{picture}
\end{center}
\caption{The $B$ spectrum. The filled circles denote our results,
where $a^{-1}=2.0$~GeV has been used to convert to physical units and
the error bars do not take into account uncertainties in $a^{-1}$.
The open circle indicates the estimate of the systematic uncertainty
in the $2S-1S$ and $^1P_1-S$ splitting. The dashed lines denote the
upper and lower bounds on the experimental results. The $B$ mass has
been shifted upwards to match the physical value.}
\label{spect}
\end{figure}

We presented a lattice study of the heavy-light spectrum in the region
of the $B$ meson, the first to partially include the effects of
dynamical quarks. A multi-smearing, multi-state fitting analysis was
performed and the ground $S$ and $P$ states as well as the excited $S$
states were extracted.  Figure~\ref{spect} summarizes the results for
the $B$ spectrum.  This is a high statistics calculation and the
systematic errors arising from the Wilson light quarks, indicated in
the figure, are the dominant uncertainty. Within these errors the
results obtained are in agreement with experiment. We are in the
process of repeating the calculation using tadpole-improved Clover
light fermions. This will allow a comparison with our quenched
results, in progress, which use the same light fermions.

We performed a comprehensive calculation of the heavy-light meson
mass, investigating three methods and their range of validity. The
agreement found between these methods confirms that Lorentz invariance
can be restored at this order in NRQCD by a constant shift to all
masses.

We calculated the spectrum over a wide range of heavy quark masses
enabling a detailed investigation of heavy quark symmetry. For the
binding energy and mass splittings the corrections to the heavy quark
limit were found to be in agreement with naive expectations based on a
hydrogen-like picture of the heavy-light meson. However, there is an
indication from the mass dependence of the ground $S$-state energy
that for quark masses around $M_b$ the next order in the NRQCD
expansion is required.  Using our tadpole-improved approach the
nonperturbative coefficients $\bar{\Lambda}$ and $<-\vec{D}^2>$ were
extracted. We showed that previous lattice calculations have large
tadpole contributions and when we correct for this using mean field
theory we find agreement with our results.  The major error remaining
is a perturbative one.

\section{Acknowledgements}
The heavy-light computations were performed on the CM-2 at SCRI.  The
heavy-heavy simulations used in section~\ref{mesonmass} were performed
on the CRAY-YMP at the Ohio Supercomputer Center.  We thank the
HEMCGC collaboration for use of their configurations and light quark
propagators. We are grateful to C.~Morningstar and G.~P.~Lepage for
useful discussions. J.~Shigemitsu thanks the members of the lattice
group at SCRI for their hospitality during a long term visit when part
of this research was carried out.  This work was supported by the
U.S.~DOE under grants DE-FG02-91ER40690, DE-FG05-85ER250000 and
DE-FG05-92ER40742, and the UK PPARC and SHEFC. We acknowledge support
by NATO under grant CRG~941259 and the EU under contract
CHRX-CT92-0051.

%SCRI DOE numbers DE-FG05-85ER250000 and DE-FG05-92ER40742
%

\end{document}